\pdfoutput=1
\documentclass[a4paper,10pt,onecolumn,unpublished,noarxiv]{quantumarticle}

\usepackage[utf8]{inputenc}
\usepackage[english]{babel}
\usepackage[T1]{fontenc}
\usepackage{amsmath,amsfonts,amssymb,amsthm}
\usepackage{booktabs}
\usepackage{longtable}
\usepackage{multirow}
\usepackage{enumitem}
\usepackage{graphicx}
\usepackage{xcolor}
\usepackage{orcidlink}
\usepackage[numbers,sort&compress]{natbib}
\hypersetup{
    colorlinks=true,
    allcolors=blue
}
\graphicspath{{figs/}}

\newcommand{\vv}{\mathbf{v}}
\newcommand{\qq}{\mathbf{q}}
\newcommand{\kk}{\boldsymbol{\kappa}}
\newcommand{\ehat}{\hat{e}}
\newcommand{\ebudget}{\ehat_{\text{budget}}}
\newcommand{\Sel}{\mathcal{S}}
\newcommand{\Var}{\operatorname{Var}}

\begin{document}

\title{Does Data Complexity Predict Quantum Advantage? A framework, a pre-specified test, and an attribution of the measured edge}

\author{Christophe Pere\,\orcidlink{0000-0002-8902-787X}}
\affiliation{\'Ecole de Technologie Sup\'erieure, Universit\'e du Qu\'ebec, 1100 Notre-Dame St W, Montreal, Quebec H3C 1K3, Canada}
\affiliation{Universit\'e Laval, INTRIQ, 1045, avenue de la M\'edecine, Qu\'ebec, Quebec G1V 0A6, Canada}

\begin{abstract}
Quantum machine learning (QML) holds promise for pattern recognition, optimization, and data analysis, but the conditions under which it can outperform classical methods remain unclear. We propose data complexity—the structural, statistical, algorithmic, and topological richness of a dataset—as a central axis for stating such conditions, and test the resulting framework under a pre-specified benchmark.

The central predictive claim fails. A composite complexity measure does not predict the budget-matched quantum edge across 27 datasets in leave-one-dataset-out validation (pooled $R^2=-0.32$, versus a pre-specified threshold of $0.4$). The failure is strongly coupled to the classical baseline: against logistic regression, the selector reaches $R^2=+0.13$ and the parity family wins 17\% of cells, whereas against the best of six standard classical models performance falls to $R^2=-0.94$, with parity winning only 1\%. The apparent signal therefore tracks classical model error rather than a robust quantum contribution. The tested encoding and reduction procedures further leave little nonlinear headroom for the downstream model to exploit; gradient boosting matches a linear probe across the tested widths.

The null is also constrained by the regime: quantum performance is at chance in 81.5\% of cells. Yet a deliberately constructed quantum-signature control produces a $+0.42$ edge under a structure-aware measurement, while generic models required to learn that structure remain near chance. Thus the null result is not evidence that quantum-sensitive structure cannot exist. Rather, in the tested regime, data complexity alone does not predict quantum advantage: the observable edge depends jointly on the classical baseline, accessible data structure, representation and model headroom, and whether quantum-sensitive structure is accessible to the measurement.

\vspace{1ex}

\noindent\emph{Version note.} Version~1 of this preprint presented the framework and its metric catalogue and announced an empirical sequel. This version supersedes it: it carries the v1 material forward, documents the corrections and estimator audit introduced in Sec.~\ref{sec:changes}, and adds the pre-specified test, the attribution analysis and the conclusions drawn from them. Version~1 remains archived and citable as the original presentation of the framework.
\end{abstract}

\maketitle
\tableofcontents


\section{Introduction}
\label{sec:intro}

Debate over when quantum machine learning can outperform classical methods has been conducted largely in terms of the quantum side of the pairing: the circuit family, the encoding, the ansatz, and the hardware. The first version of this work \citep{pere2025datacomplexity}---v1 of this preprint, hereafter simply v1---argued that this leaves out an important property of the learning problem, and proposed \emph{\textbf{data complexity}} as a missing axis: the structural, statistical, algorithmic, and topological richness of the dataset itself. It formalized that intuition as four pillars, entropy, correlation order, compressibility, and topology, combined into a normalized composite $\tilde{C}_{\text{data}}$, and used that construction to formulate three quantitative predictions. Barren-plateau onset should scale as $\Var[g(\theta)] \approx \exp(-\alpha n d\,\tilde{C}_{\text{data}})$, so that gradients vanish faster on more complex data. Generalization error should be governed by the \emph{alignment} between a circuit's expressibility and the data's complexity, penalizing mismatch in either direction. And, integrating both, the achievable quantum edge should be \emph{ranked} by $\tilde{C}_{\text{data}}$, making the composite a practical instrument for deciding, before running anything, whether a given dataset is worth a quantum treatment at all. That last claim is what would make the framework useful rather than merely descriptive, and it is the one this paper sets out to test.

This version is that test. Before testing the empirical predictions, however, preparing the experiments exposed problems in the v1 formalism: dimensionally incoherent terms in the composite, correlated components that make its weights unidentifiable, and a data-free limit that contradicts the circuit-ensemble result of \citet{mcclean2018barren}. These issues were corrected before the confirmatory analysis was run (Sec.~\ref{sec:framework}). We treat this as scientific iteration rather than a change to the motivating question: the core proposition that data complexity is relevant to QML advantage is preserved, while its operationalization and predictive sufficiency are tested empirically.

A failed prediction is informative only if the experiment can distinguish failure of the hypothesis from failure of the measurement mechanism. We therefore fixed a falsification battery with explicit decision rules before collecting confirmatory data. More importantly, we can attribute the measured quantum--classical gap to the part of the pipeline that actually produces it. Any measured gap decomposes exhaustively into a term obtainable by running the preprocessing classically and a residual that is the only quantum claim (Sec.~\ref{sec:encoder}). This decomposition changes how we interpret the negative result: rather than asking only whether the complexity predictor succeeds, we can ask whether any quantum quantity remains for it to predict.

The analysis yields four lessons.

\begin{enumerate}
\item \textbf{The baseline determines the measured advantage}
(Sec.~\ref{sec:baseline}). A fully predictable ``advantage'' appears and disappears depending on the classical model set alone. Against a linear baseline, the selector recovers a signal; against the best of six standard models, that signal largely vanishes.

\item \textbf{The encoding removes the headroom it is meant to predict}
(Sec.~\ref{sec:encoder}). Reducing classical data to few-qubit width removes nonlinear headroom before the quantum model sees it. Gradient boosting matches a linear probe at every width, and where a baseline-robust edge survives, gradient-based training cannot find it. Because this preprocessing is classical, the resulting representation is available to the classical comparator as well and therefore cannot itself constitute evidence of quantum advantage. The experiment instead shows that, for the tested reductions and encoding library, the preprocessing can leave little nonlinear structure for the downstream model to exploit.

\item \textbf{The scaling law is misspecified where experiments live}
(Sec.~\ref{sec:results:h1}). Varying circuit width shows power-law, depth-independent decay rather than the exponential form assumed by the framework, and the observed scaling is inconsistent with the Haar-random rate.

\item \textbf{The test had little quantum variation to detect, and that bounds the null}
(Sec.~\ref{sec:results:degenerate}). The quantum model is at or near chance in $81.5\%$ of cells, With $a_q$ concentrated near chance in this suite, the measured edge is
largely determined by the classical term ($\rho=+0.913$), while the encoding axis carries only $0.5\%$ of the variance. In a regime where the quantum performance has little variation, there is correspondingly little quantum signal for a data-complexity framework to rank. The resulting null is therefore informative about the tested regime, but does not establish that data complexity can never predict a quantum advantage.
\end{enumerate}

Hypotheses, decision rules, the complexity and interaction vectors, the predictor architecture, the encoding library, the cost model, and the cross-validation scheme were fixed in a written pre-specification before any analysis was run. Each result is therefore a measurement of one of the two terms in the attribution: Sec.~\ref{sec:baseline} shows how much of an apparent edge the classical term accounts for; Sec.~\ref{sec:encoder} shows what the encoding destroys and costs before a quantum residual can exist; Sec.~\ref{sec:results:h1} tests whether the proposed scaling law describes the regime in which the experiments operate; and Sec.~\ref{sec:results:degenerate} identifies the regime in which the quantum term is pinned and therefore difficult to measure.

The question forced by this decomposition is therefore more fundamental than whether one composite score predicts the edge. \emph{Under what experimental conditions is there a genuinely quantum quantity left to predict?} That is the question this paper leaves behind. \\

\noindent\fbox{%
\begin{minipage}{\dimexpr\linewidth-2\fboxsep-2\fboxrule}
    \textbf{Main takeaways}:
    
    \begin{itemize}
        \item \textbf{Baseline dependence}: the apparent quantum edge is strongly baseline-dependent: $(R^2=+0.13)$ against LR, but $(-0.94)$ against the best of six classical models.
        \item \textbf{Attribution}: the measured edge is largely explained by classical baseline error rather than an encoding/data-complexity signal.
        \item \textbf{Variance/headroom}: the quantum model is at chance in 81.5\% of cells, while the encoder reduces the nonlinear headroom available to the quantum model.
        \item \textbf{Control}: a designed quantum-signature dataset produces a $(+0.42)$ edge under structure-aware measurement, while generic models remain at chance.
        \item \textbf{Benchmark and generalization scope.}
        The best-of-six result is a fixed-library sensitivity analysis rather than a tuned classical frontier, and the 27-dataset leave-one-dataset-out test supports conclusions within the tested regime rather than universal generalization across QML tasks.
    \end{itemize}
    These results shift the question from whether a scalar data-complexity measure predicts quantum advantage to whether the relevant quantum structure survives the classical encoding and remains accessible to the measurement.
\end{minipage}%
}



\section{Related work}
\label{sec:related}

\paragraph{Data as a constraint on advantage.}
Whether a quantum model can help on a learning task is constrained by the data, not only by the circuit. \citet{huang2021power}, for example, derive projected-kernel bounds that can rule out advantage under specified conditions and provide a prescreening criterion
for when advantage may be possible from properties of the learning problem, while \citet{caro2022generalization} relate generalization from few training samples to circuit structure. These results motivate treating the dataset as an axis of quantum advantage, but they do not by themselves specify which properties of the data should be predictive in an experimental comparison. Version~1 proposed \emph{complexity}, entropy, correlation order, compressibility, and topology as the operative axis and combined them into a linear composite. The present version instead replaces the fixed composite with a learned predictor and tests whether data complexity can predict the measured quantum--classical edge. In this sense it revises the proposed operationalization rather than extending the framework.

\paragraph{The classical comparator sets the observable edge.}
A quantum--classical comparison does not measure a property of the quantum model alone: it measures a difference between two model classes. Recent benchmark studies consistently find that tuned classical baselines remain competitive when comparisons are made carefully \citep{bowles2024benchmarking,yu2026qmlvscml,leither2026oncology}, while end-to-end accounting can further change the apparent advantage \citep{chandarana2026hsqc,gillani2026attribution}. This makes the choice of classical model part of the quantity being measured rather than a secondary implementation detail. Our analysis therefore treats baseline choice as a first-order experimental variable. The resulting question is not simply whether a quantum model outperforms a classical one, but whether the observed edge survives a stronger classical comparator.

\paragraph{Trainability constrains what advantage can be observed.}
Even when a quantum model has a representational advantage, that advantage is observable only if the relevant structure can be reached by the training procedure. Barren plateaus \citep{mcclean2018barren, Larocca_2025}, cost-function locality \citep{cerezo2021cost}, and kernel concentration \citep{thanasilp2024concentration} characterize regimes in which variational or kernel-based quantum models become difficult to train or distinguish. H1 and H2 therefore test not only whether the scaling proposed in v1 is observed, but whether its functional form describes the regime in which the experiments actually operate. \citet{alcantara2026representation} provide exact cost-variance expressions for arbitrary states and observables, removing the dynamical-Lie-algebra membership assumption that excluded settings such as ours; Sec.~\ref{sec:results:h1} compares the observed scaling with this broader framework. Conversely, \citet{cerezo2025simulability} emphasize that the absence of a barren plateau can itself be evidence for classical simulability. Trainability is therefore not independent of advantage: a model that is easy to train may also be one whose behaviour admits an efficient classical description.

\paragraph{Encoding changes the problem before the quantum model sees it.}
Data must generally be transformed before a few-qubit quantum model can process it, making the encoding inseparable from the comparison. Prior work shows that the embedding choice can dominate performance in few-qubit settings \citep{sammartino2026encoding,kurt2026qsvmencoding}; H4 asks whether this choice can be optimized at the feature level. We instead examine the step that precedes the encoder. If dimensionality reduction or feature selection is performed classically, it can remove precisely the nonlinear or high-order structure that a quantum model might otherwise exploit. Sec.~\ref{sec:encoder} therefore asks what information and nonlinear headroom are lost in the reduction, what resources are required to reproduce it coherently, and how much of the eventual quantum--classical gap can be attributed to this classical term before the quantum model is run.

\paragraph{Data access and dequantization.}
A quantum advantage is also conditional on how the classical and quantum algorithms access the data. This point is central to the dequantization literature. Tang \citep{tang2019quantum, jethwani2019} showed that the quantum recommendation algorithm of Kerenidis and Prakash \citep{kerenidis2016quantumrecommendationsystems} could be matched, up to polynomial factors, by a classical algorithm under an $\ell_2$-sampling data structure, removing the claimed exponential speedup under that input model. Subsequent work generalized this idea to a broader sampling-based framework for low-rank matrix arithmetic and quantum singular value transformation, yielding quantum-inspired classical algorithms for recommendation, principal-component analysis, clustering, support-vector machines, regression, and related tasks \citep{chia2022dequantization}. These results do not imply that quantum speedups are impossible; rather, they show that the input model is part of the speedup claim. A quantum algorithm can only claim an advantage over classical alternatives that are denied no more information or access than the quantum procedure itself. H6 therefore stratifies our benchmark by data-access mode. The present suite contains no Mode-Q pairs, so H6 is a descriptive boundary condition rather than a statistical test. We report this explicitly rather than treating the absence of qualifying instances as evidence for or against dequantization.

Taken together, these considerations place the data-complexity hypothesis inside a broader measurement problem. Data may determine where a quantum advantage is possible, but the \emph{measured} edge also depends on what the classical comparator can learn, what structure survives preprocessing and encoding, whether the quantum model can reach the relevant structure, and how the underlying data are accessed. These constraints motivate the attribution framework tested in the remainder of the paper: before asking whether data complexity predicts a quantum edge, we first identify how much of that edge can exist after the classical parts of the pipeline have been accounted for.


\section{Defining data complexity}
\label{sec:defining}

We define \textit{data complexity} as the structural richness of a dataset, expressed through properties that characterize the resources required to represent, compress, or learn from its distribution. In the classical regime, this includes quantities such as \textit{statistical complexity}~\cite{crutchfield1989inferring}, intrinsic dimensionality, correlation structure, compressibility, kernel spectrum, and topology. In the quantum setting, the same question can be asked after an embedding: \textbf{how much correlation, entanglement, geometric structure, or circuit resource is required to represent or learn from the resulting quantum data?} These two views motivate the distinction, developed below, between properties intrinsic to the dataset and properties induced by a particular quantum embedding.

The purpose of this section is therefore twofold. First, it preserves the metric catalogue introduced in v1 and defines the notions of data complexity on which the framework is based. Second, it identifies which of those notions are operationalized in the present experiment. The confirmatory analysis does not assume that all quantities in the catalogue are interchangeable or that they form a single scalar. Instead, the measured classical quantities are retained as a complexity vector $\vv(D)$ after rank normalization (Eq.~\eqref{eq:ranknorm}), while quantities evaluated after embedding are treated separately through the interaction vector $\qq(D,\phi)$ defined in Sec.~\ref{sec:framework:predictor}. Definitions carried in v1 that this study does not estimate, including quantum interaction complexity, Schmidt and tensor rank, quantum Kolmogorov complexity, and the classical--quantum analogies, are reproduced in App.~\ref{app:v1metrics} rather than treated as empirical variables here.

\begin{table}[t]
\centering
\small
\caption{The nine components of the intrinsic complexity vector $\vv(D)$ used
in the confirmatory analysis. Each component is rank-normalized against the
frozen benchmark before use (Eq.~\eqref{eq:ranknorm}); raw values are in
Table~\ref{tab:v_raw} and normalized values in Table~\ref{tab:v_ranknorm}.
The topological component $T$ is the pre-specified estimator; after the
confirmatory analysis revealed defects in its implementation, the corrected
estimator $T^\star$ is reported separately and used in the corresponding
sensitivity analyses (App.~\ref{app:topofix}).}
\label{tab:components}
\begin{tabular}{@{}l p{0.34\linewidth} p{0.30\linewidth} l@{}}
\toprule
 & Quantity & Estimator & Defined in \\
\midrule
$H$              & differential entropy            & $k$-nearest-neighbour            & Sec.~\ref{sec:m-entropy} \\
$d_{\text{pca}}$ & intrinsic dimension             & covariance spectrum             & Sec.~\ref{sec:m-id} \\
$d_{\text{2nn}}$ & intrinsic dimension             & TwoNN                            & Sec.~\ref{sec:m-id} \\
$I_2$            & second-order mutual information & $k$-nearest-neighbour            & Sec.~\ref{sec:m-corr} \\
$I_h$            & higher-order mutual information & MINE                             & Sec.~\ref{sec:m-corr} \\
$K$              & compressibility                 & gzip compression ratio           & Sec.~\ref{sec:m-kolmo} \\
$M$              & description length              & autoencoder MDL                  & Sec.~\ref{sec:m-kolmo} \\
$R$              & kernel effective rank           & RBF, median heuristic            & Sec.~\ref{sec:m-kernel} \\
$T$              & topological complexity          & Vietoris--Rips persistence       & Sec.~\ref{sec:m-topo} \\
\bottomrule
\addlinespace[2pt]
\multicolumn{4}{@{}p{\linewidth}@{}}{\footnotesize $T$ is the component as pre-specified; it was found to be defective after the confirmatory analysis and is replaced by a corrected estimator $T^{\star}$ (App.~\ref{app:topofix}). Both are reported throughout.}
\end{tabular}
\end{table}

\subsection{Classical Data Metrics}

\subsubsection{Distributional entropy}
\label{sec:m-entropy}

The first component measures how much the data distribution spreads: for a density $p$ on
$\mathbb{R}^d$, the differential entropy
\[
  h(X) = -\int p(x)\log p(x)\,dx ,
\]
estimated from samples with a $k$-nearest-neighbour estimator rather than by binning, which is unreliable at the sample sizes used here. One property matters downstream: differential entropy is \emph{extensive in dimension and scale-dependent}, $h(aX)=h(X)+d\log|a|$, so a raw entropy value is not
comparable across datasets of different dimension or diameter. That is one of the defects that motivates rank normalization in Sec.~\ref{sec:framework:predictor}, where it is measured rather than asserted.

\subsubsection{Intrinsic Dimension / Effective Dimension }
\label{sec:m-id}

Many datasets in high-dimensional spaces are concentrated on a much lower-dimensional manifold~\cite{roweis2000nonlinear, tenenbaum2000global}. The \textit{intrinsic dimension} measures the effective degrees of freedom present in the data~\cite{camastra2016intrinsic}. A common estimator is based on the spectrum of the covariance matrix~\cite{bishop2006pattern}:
\begin{equation}
\dim_{\mathrm{eff}}(\mathcal{D}) = \frac{\left(\sum_{i=1}^d \lambda_i\right)^2}{\sum_{i=1}^d \lambda_i^2},
\end{equation}
where $\lambda_i$ are the eigenvalues of the covariance matrix. A low intrinsic dimension indicates redundancy or concentration in a lower-dimensional representation, whereas a high intrinsic dimension indicates that the dataset explores more independent directions. Such properties may affect learning difficulty, but their predictive relationship to the quantum--classical edge is an empirical question tested below~\cite{pope2021intrinsic}.

\subsubsection{Correlation/Interaction Order}
\label{sec:m-corr}

Many classical learning procedures are particularly effective when the
relevant structure is captured by low-order correlations. Conversely, tasks whose useful signal is concentrated in higher-order interactions can be harder for restricted classical model classes~\cite{linial1989constant}, as illustrated by Boolean
functions such as parity or by spin-glass constructions~\cite{haastad2001some, mezard2009information}. The difficulty, however, depends on the model class and representation, rather than on interaction order alone~\cite{eldan2016power}.

Suppose a dataset is described by a joint probability distribution $P(x_1, x_2, \dots, x_n)$.  
We can expand it into cumulants (generalized correlations)~\cite{kendall1994advanced}:
\begin{equation}
\kappa_{i_1 i_2 \dots i_k} = \mathbb{E}\!\left[ \prod_{j=1}^k (x_{i_j} - \mathbb{E}[x_{i_j}]) \right].
\end{equation}
where $\kappa_{i_j}$ corresponds to pairwise correlations. Higher-order interactions correspond to terms with $k > 2$.

We define the \textit{interaction complexity} of a dataset as the highest order $k$ such that
\[
\max_{i_1, \dots, i_k} |\kappa_{i_1 i_2 \dots i_k}| > \epsilon,
\]
for some threshold $\epsilon$ indicating statistical significance.  
Within this operational definition, datasets whose relevant structure is
captured predominantly by low-order interactions receive lower interaction
complexity, whereas datasets for which higher-order terms are required receive
higher interaction complexity~\cite{scholkopf2021toward}. This is a property of the chosen
representation and estimator, not a model-independent ordering of dataset
difficulty.

\subsubsection{Kolmogorov Complexity}
\label{sec:m-kolmo}

Kolmogorov complexity measures the shortest description length of a dataset~\cite{kolmogorov1965three, chaitin1966length, solomonoff1964formal}.  
In practice, this corresponds to its compressibility~\cite{li2008kolmogorov}.  
Highly regular or redundant datasets have low Kolmogorov complexity, 
since they can be described by short programs. Truly random data (without a generative rule) has high Kolmogorov complexity, because no description is shorter than the data itself. By contrast, pseudorandom datasets appear random statistically but are generated from a short seed and program, and thus have low Kolmogorov complexity despite their high apparent unpredictability. For learning tasks, compressibility can provide a useful proxy for structural regularity, although compressibility and learnability are not equivalent: the difficulty of learning also depends on the target function, model class, and representation~\cite{hutter2005universal}.

The Kolmogorov complexity of a dataset $D$ is defined as:
\[
K(D) = \min_{p : U(p) = D} |p|,
\]
where $U$ is a universal Turing machine, $p$ is a program that outputs $D$, and $|p|$ is its length in bits.
The metric can be interpreted as: Low $K(D)$: The dataset is compressible (e.g., regular patterns). High $K(D)$: The dataset is incompressible (pseudorandom).

In practice, one uses approximations such as compression ratios:
\[
C(D) = \frac{\text{Compressed size}(D)}{\text{Original size}(D)}.
\]
Datasets with $C(D) \ll 1$ are low-complexity, while those with $C(D) \approx 1$ approach maximum complexity.

Two instruments of this kind are carried separately. The compression ratio $K$ above uses a general-purpose compressor (gzip), while $M$ is a minimum-description-length estimate obtained from an autoencoder, i.e.\ description length under a \emph{learned} model rather than a fixed one. Both are proxies for the same uncomputable quantity, they need not agree, and neither is averaged into the other: they enter $\vv(D)$ as two components.

\subsubsection{Spectral Complexity of Kernel Matrix}
\label{sec:m-kernel}

Kernel methods characterize complexity through the eigenvalue spectrum of the kernel Gram matrix~\cite{scholkopf2002learning, shawe2004kernel}.  
The effective dimension is defined as~\cite{bach2017equivalence, rudi2017generalization}:
\[
d_{\mathrm{eff}}(\lambda) = \mathrm{Tr}\!\left( K (K + \lambda I)^{-1} \right),
\]
where $K$ is the kernel matrix and $\lambda$ a regularization parameter.  
If the eigenvalues decay rapidly, the kernel captures most of the data structure in a low-dimensional subspace (low complexity).  
A slow decay indicates that many kernel modes contribute substantially to the
representation. This can increase the effective complexity of kernel-based
learning, although its relationship to the measured quantum--classical edge
remains model- and task-dependent~\cite{bartlett2021deep, steinwart2008support}.

\subsubsection{Topological Data Complexity}
\label{sec:m-topo}

Beyond spectral and statistical measures, the topology of data manifolds provides an additional layer of complexity. Topological Data Analysis (TDA) extracts invariants of the data distribution that are stable under perturbations  and insensitive to parametrization. The most common invariants include Betti numbers $\beta_k$, which count $k$-dimensional holes ($\beta_0$: connected components, $\beta_1$: loops, $\beta_2$: voids), the Euler characteristic, and persistent homology barcodes. 

For a dataset $D \subset \mathbb{R}^d$, we construct a Vietoris–Rips complex at scale $\epsilon$, track the birth and death of homological features across scales, and compute their lifetimes.  The \emph{topological complexity} can be written as
\begin{equation}
C_{\text{top}}(D) = \sum_{k=0}^{K} w_k \, \text{Pers}_k(D),
\end{equation}
where $\text{Pers}_k(D)$ is the sum of lifetimes of $k$-dimensional homological features in the persistent homology 
of $D$, and $w_k \geq 0$ are application-dependent weights. 

Datasets with trivial topology (low $\beta_k$, rapidly collapsing barcodes) are topologically simple, while datasets with rich higher-dimensional homology exhibit large $C_{\text{top}}(D)$, which, under this estimator, is taken as evidence of richer topological structure in the sampled data.

Persistent homology has become the standard tool for extracting 
such invariants in practice \cite{Ghrist2008Barcodes, carlsson_2009, Edelsbrunner_2010}, and recent work demonstrates its growing role in machine learning generalization \cite{andreeva2024topologicalgeneralizationboundsdiscretetime} and even in quantum data analysis \cite{komalan2025quantumbarcodespersistenthomology}.

The topological quantity is therefore best regarded as an estimator of topological structure rather than as a model-independent measure of learning difficulty. The estimator-specific limitations encountered in the confirmatory analysis are documented in App.~\ref{app:topofix}.

\subsection{Quantum data Metrics}
\label{sec:quantum-metrics}
The following quantities extend the v1 catalogue to structure that exists only after data are represented in a quantum state or quantum feature space. They provide the conceptual motivation for the interaction vector $\qq(D,\phi)$ and for the distinction between intrinsic data complexity and embedding-dependent structure. They are not all independent empirical predictors in the present benchmark: the confirmatory analysis uses only the embedding-level quantities specified in Sec.~\ref{sec:framework:predictor} and listed in the pre-specification. The remaining quantities are retained here to preserve the framework's conceptual scope and are not interpreted as separately validated measures of quantum advantage.
\subsubsection{Entanglement Entropy}

Entanglement entropy (von Neumann Entropy) quantifies the degree of quantum correlations in a state~\cite{nielsen00, amico2008entanglement}.  
For a bipartition $A|B$ of a pure state $|\psi\rangle$, the entropy is:

\begin{equation}
\begin{split}
S_{\rho_A} & = - \mathrm{Tr}(\rho_A \log \rho_A) \\ & = - \mathrm{Tr}(\rho_B \log \rho_B) = S_{\rho_B}, \\ \rho_A & = \mathrm{Tr}_B(|\psi\rangle\langle\psi|) = \mathrm{Tr}_B(\rho_{AB}).    
\end{split}
\end{equation}
where $S_{\rho_A}$ is the density matrix of the subsystem $A$, $S(\rho_B)$ is the density matrix of the subsystem $B$.
Low-entanglement states can admit efficient classical descriptions for important
state families~\cite{vidal2003efficient}, whereas high entanglement can be
associated with increased classical simulation cost~\cite{eisert2010colloquium}.
Entanglement entropy is therefore a useful descriptor of quantum-state
structure, but its value alone does not establish a quantum computational
advantage.
Thus, entanglement entropy acts as a natural measure of quantum data complexity.

\subsubsection{Quantum Mutual Information}
Quantum mutual information captures the total correlation (classical and quantum) between subsystems~\cite{nielsen00, horodecki2009quantum}:
\[
I(A:B) = S(\rho_A) + S(\rho_B) - S(\rho_{AB}).
\]
where $S(\rho_A)$ is the density matrix of the subsystem $A$, $S(\rho_B)$ is the density matrix of the subsystem $B$, and $S(\rho_{AB})$ is the density matrix of the system $AB$. 

A high mutual information indicates strong total correlations between the subsystems. In analogy with classical correlation measures, it therefore provides a descriptor of dependency structure in the embedded data, although its relationship to learning difficulty depends on the model and task.

\subsubsection{Expressibility \& Barren Plateau Sensitivity}

Expressibility characterizes the range of states accessible to a parameterized
quantum circuit. Greater expressibility can increase the representational
space available to a model, but it does not by itself imply better
generalization and may coincide with reduced trainability in some circuit
families. The relationship between data structure, expressibility, and
trainability is therefore a hypothesis of the framework rather than an
established monotonic law.~\cite{sim2019expressibility, mcclean2018barren, cerezo2021cost, Cunningham_2025}.

\paragraph{Expressibility: } 
Given a PQC $U(\theta)$, expressibility can be measured by comparing the distribution of states it generates to the uniform Haar distribution. One standard metric is the Kullback--Leibler (KL) divergence~\cite{sim2019expressibility}:
\[
\mathcal{E}(U) = D_{\mathrm{KL}} \!\big( P_{U(\theta)} \;\|\; P_{\mathrm{Haar}} \big),
\]
where $P_{U(\theta)}$ is the distribution of states produced by the PQC and $P_{\mathrm{Haar}}$ is the Haar measure over the Hilbert space. Lower $\mathcal{E}(U)$ corresponds to higher expressibility, and $D_{\mathrm{KL}}$ is the Kullback-Leibler divergence defined as:
\[
D_{\mathrm{KL}}(P \| Q) = \int P(x) \log \frac{P(x)}{Q(x)} \, dx
\]
for continuous distributions $P$ and $Q$.

\paragraph{Trainability (Barren Plateaus):} 
Trainability is typically quantified by the variance of gradients of a cost function $C(\theta)$ with respect to circuit parameters. For many PQCs, it has been shown that~\cite{mcclean2018barren}:
\[
\mathrm{Var}\!\left( \frac{\partial C}{\partial \theta_i} \right) \sim \mathcal{O}\!\left( \frac{1}{2^n} \right),
\]
where $n$ is the number of qubits. Exponential decay of gradient variance with $n$ indicates a barren plateau. The trainability of a dataset--circuit pair decreases rapidly with system size unless special architectures or initialization strategies are employed~\cite{cerezo2021cost}.

\subsubsection{Quantum Topological Complexity}

The quantum-topological quantities below belong to the broader v1 framework
and are included as candidate descriptors of structure induced by quantum
representations. They are not used as a standalone composite predictor in the
confirmatory analysis.

In the quantum regime, topological features emerge in both the structure of quantum states and the geometry of quantum feature maps. Relevant invariants include:

\begin{itemize}
    \item \textbf{Topological entanglement entropy (TEE)}: quantifies long-range entanglement beyond local correlations, acting as a marker of topological order.
    \item \textbf{Chern numbers and winding numbers}: characterize the topology of parameter spaces induced by quantum embeddings $U_\phi(x)$, e.g., through Berry curvature integration.
    \item \textbf{Persistent homology of quantum kernels}: analyzing the topology of the dataset after embedding into Hilbert space via kernel-induced distances.
\end{itemize}

We define a composite measure of \emph{quantum topological complexity} as
\begin{equation}
C_{\text{top}}^{(q)}(E) = \gamma_1 S_{\text{topo}} 
+ \gamma_2 \chi(\mathcal{M}_E) 
+ \gamma_3 \sum_{k=0}^{K} \text{Pers}_k(\mathcal{M}_E),
\end{equation}
where $S_{\text{topo}}$ is the topological entanglement entropy of ensemble $E$, $\chi(\mathcal{M}_E)$ is the Euler characteristic of the quantum state manifold $\mathcal{M}_E$, and $\text{Pers}_k(\mathcal{M}_E)$ are persistent homology contributions of dimension $k$. The coefficients $\gamma_i \geq 0$ control relative importance. 

Within this proposed construction, richer topological order or more
nontrivial embedding geometry would correspond to a larger
$C_{\text{top}}^{(q)}(E)$. Whether such a quantity predicts resource
requirements or learning advantage is left as an empirical question.

Recent work by Komalan~(2025) \cite{komalan2025quantumbarcodespersistenthomology} introduces “quantum barcodes,” which apply persistent homology to observable-based point clouds derived from quantum states, successfully distinguishing topological 
phase transitions via homological features—this motivates using $\sum_k \mathrm{Pers}_k(\mathcal{M}_E)$ in the quantum topological complexity metric.

Li~(2023) \cite{li2023characterizingambiguitytopologicalentanglement} explores ambiguities inherent in defining TEE in lattice and continuum settings, 
highlighting that subtleties in state partitioning, boundary conditions, and measurement can significantly affect the TEE value used in complexity bounds.


\section{The framework, corrected}
\label{sec:framework}

Preparing the experiments required implementing the v1 framework, and
implementation exposed several defects in its functional forms: terms that
are not dimensionally commensurable, components whose statistical dependence
makes their weights unidentifiable, and a data-free limit that contradicts an
established circuit-ensemble result. Several of these issues were identified
before confirmatory data were collected; a further issue, the topological
estimator, was identified during the subsequent analysis and required a
separate estimator audit. The status of each change is recorded explicitly
below as withdrawn, corrected, superseded, retained untested, or added.

This section states the framework in its corrected operational form. It is the
form against which the hypotheses, experiments, and results in the remainder
of the paper are evaluated.

\subsection{Changes from v1}
\label{sec:changes}

Table~\ref{tab:changes} registers the changes from v1, their motivation, and
where they are implemented. \emph{Withdrawn} means that the original
quantity was removed before the corresponding confirmatory test and therefore
does not enter its result. \emph{Corrected} means that a replacement
formulation was specified and subsequently tested. \emph{Retained
untested} identifies elements preserved from the conceptual framework but not
validated by the present benchmark.

{\small
\begin{longtable}{@{}p{0.235\linewidth}p{0.455\linewidth}p{0.21\linewidth}@{}}
\caption{Registry of changes from v1 of this preprint. Equation numbers in the first column are those of v1 as archived; the cross-references are to their reproduction in App.~\ref{app:v1} of this version.}
\label{tab:changes}\\
\toprule
\textbf{v1 element} & \textbf{Change, and why} & \textbf{Where} \\
\midrule
\endfirsthead
\multicolumn{3}{@{}l}{\emph{Table~\ref{tab:changes}, continued}}\\
\toprule
\textbf{v1 element} & \textbf{Change, and why} & \textbf{Where} \\
\midrule
\endhead
\midrule
\multicolumn{3}{r@{}}{\emph{continued on the next page}}\\
\endfoot
\bottomrule
\endlastfoot
Composite scalar $C_{\text{data}}$, v1 Eq.~7 \eqref{eq:classical_comp} &
\textbf{Withdrawn before testing.} The four pillars are not commensurable on their native scales (the topological term carries units of length on the data manifold) and they co-vary, so the weights $\{\lambda_i\}$ are not identifiable from a finite benchmark. Measured: one pillar carries $100\%$ of the composite variance, and \emph{which} pillar depends on the estimator. Replaced by the rank-normalized profile $\vv(D)$, Eq.~\eqref{eq:ranknorm}, which keeps the components separate. &
Sec.~\ref{sec:framework:predictor}, Table~\ref{tab:p17} \\
\addlinespace
Normalization $\tilde{C}_{\text{data}}=C_{\text{data}}/C_{\max}$, v1 Eq.~8 \eqref{eq:v1norm} &
\textbf{Withdrawn} with the composite. Rescaling the sum cannot repair a mismatch internal to it. Rank normalization against a frozen benchmark CDF is used instead. &
Sec.~\ref{sec:framework:predictor} \\
\addlinespace
Data-gated barren plateau $\Var[g]\approx e^{-\alpha nd\tilde{C}_{\text{data}}}$, v1 Eq.~9 \eqref{eq:variance} &
\textbf{Corrected.} The form implies $\Var[g]\to1$ as $\tilde{C}_{\text{data}}\to0$, i.e.\ no plateau on trivial data, contradicting the Haar-random circuit-ensemble result of \citet{mcclean2018barren}. The corrected form separates the circuit baseline from the data modulation, Eq.~\eqref{eq:bp}, and recovers the original as the one-dominant-component special case. Then tested: it fails on measurement too. &
Sec.~\ref{sec:framework:bp}; tested in Sec.~\ref{sec:results_nscaling} \\
\addlinespace
Topology-extended plateau law, v1 Eq.~16 \eqref{eq:variance_topo} &
\textbf{Superseded} by the same correction: it inherits the data-free-limit defect of v1 Eq.~9 and is replaced by the component-wise sensitivities $\kk$ of Eq.~\eqref{eq:bp}. &
Sec.~\ref{sec:framework:bp} \\
\addlinespace
Expressibility--complexity alignment $\mathbb{E}[\epsilon_{\text{gen}}]\approx\epsilon_{\text{emp}}+\lambda|\mathcal{E}(U)-C_{\text{data}}|$, v1 Eq.~15 \eqref{equation_expressivity} &
\textbf{Withdrawn before testing.} $\mathcal{E}(U)$ is a KL divergence against the Haar reference while $C_{\text{data}}$ is normalized on $[0,1]$; their absolute difference has no invariant meaning and the ``sweet spot'' $\mathcal{E}(U)\approx C_{\text{data}}$ is not a testable condition. Replaced by a learned alignment head $a(D,U)\in(0,1)$ trained on observed generalization gaps, Eqs.~\eqref{eq:head-align}--\eqref{eq:loss}. The substantive claim that mismatch between data structure and circuit representation costs generalization is retained. &
Sec.~\ref{sec:framework:align} \\
\addlinespace
Induced quantum complexity $C_{\text{ind}}(D;\phi)$, v1 Eq.~11 \eqref{eq:induced_comp} &
\textbf{Replaced.} Complexity induced by an embedding is not a property of the dataset, so it cannot occupy a per-dataset column. Split into the pair $(\vv(D),\qq(D,\phi))$: intrinsic complexity and embedding interaction, carried separately throughout. &
Secs.~\ref{sec:framework:predictor}, \ref{sec:framework:align}; App.~\ref{app:vtables} \\
\addlinespace
Quantum-native composite $C_{\text{quant}}(\mathcal{E})$, v1 Eq.~10 \eqref{eq:quantum_comp} &
\textbf{Corrected; evaluated exploratorily.} Two components were ill-posed and are respecified: the quantum Fisher information as collective-spin QFI normalized by the Heisenberg bound, and the quantum topological term as Fubini--Study persistence normalized by $\pi/2$. With all six components implemented and validated, the composite collapses as Eq.~7 did: the effective rank carries $99.97\%$ of its variance. Replaced by the rank-normalized vector of its components. Its predictive value is untested, since the quantum-native arm has a single width. &
Sec.~\ref{sec:framework:cquant}; Table~\ref{tab:cquant}; App.~\ref{app:vtables} \\
\addlinespace
Topological component, as deployed &
\textbf{Rebuilt.} Four compounding defects: $H_0$ entered at full weight (median ${\approx}90\%$ of the component), no scale normalization ($r(\log\mathrm{diam},\log C_{\text{top}})=+0.922$), a filtration truncated at $0.5\times$ diameter, and edge collapse applied to that truncated filtration. Together they returned $P_1=0.000$ for a perfect circle. The corrected estimator is
untruncated, $H_0$-free, diameter-normalized and validated against known
shapes. The pre-specified H2 result computed with the old estimator is
\textbf{withdrawn}; analyses using the corrected estimator are reported as
post-hoc estimator sensitivity analyses &
App.~\ref{app:topofix} \\
\addlinespace
Encoding cost and data access &
\textbf{Added.} v1 discussed an encoding--expressivity trade-off but priced no encoding. A cost model and an explicit data-access mode are introduced, so that high-expressivity encodings carry their state-preparation assumptions. &
Sec.~\ref{sec:framework:cost} \\
\addlinespace
Geometric curvature pillar &
\textbf{Added, specified but not implemented.} A fifth pillar for graph- and hypergraph-native data (Ollivier--Ricci, ORCHID). It is untested here by construction: a point-cloud suite cannot exercise it. &
Sec.~\ref{sec:framework:predictor}
\end{longtable}}

\paragraph{A note on what this registry is for.} Version~1 is archived and citable, so a reader who has it can diff the two versions. The registry exists so that this is not necessary, and so that the status of each element, withdrawn before testing, corrected then tested, retained untested, or added, is explicit rather than inferred. The methodological point it records is stated in Sec.~\ref{sec:ledger}: three of these six issues were found by implementing the framework rather than by running it, and implementation is a form of theory criticism.

\subsection{From a scalar complexity score to an attributable framework}
\label{sec:framework:predictor}

Version~1 operationalized data complexity as a scalar quantity,
$C_{\text{data}}$, obtained by combining four pillars: entropy, correlation
order, compressibility, and topology. Before testing whether such a quantity
predicts a quantum edge, we first ask whether the four pillars can
legitimately be combined in this way. Two problems arise: the components are
not commensurate on their native scales, and their statistical dependence
makes the weights of a linear combination non-identifiable.

Equation~\eqref{eq:classical_comp} (Eq.~7 of v1) expressed the composite classical complexity as a linear sum of the four pillars, with weights $\{\lambda_i\}$ calibrated on a benchmark suite. Two structural issues prevent this form from serving as an identifiable quantitative predictor on the present benchmark.

\paragraph{The scalar composite is not empirically identifiable.}
The four components of Eq.~\eqref{eq:classical_comp} have different scales and, in some cases, different units. In particular, the topological estimator used in v1 has units of length on the data manifold \citep{carlsson_2009,Edelsbrunner_2010}. Normalizing the final sum by $C_{\max}$ rescales the output but does not repair the internal mismatch: whichever component has the largest variance over the benchmark dominates the composite, independently of whether that component drives learning difficulty.

The components are also statistically dependent. Entropy and intrinsic
dimension co-vary, while correlation order and topological complexity
co-vary on most natural datasets. Consequently, a linear combination of the
components is not identifiable from a finite benchmark: multiple sets of
weights $\{\lambda_i\}$ can produce essentially indistinguishable values of
$C_{\text{data}}$.

We can test both objections directly because every component of
Eq.~\eqref{eq:classical_comp} was evaluated on the 27 classical-track datasets. This analysis is exploratory rather than a pre-specified hypothesis, but it tests whether the proposed scalar has the empirical structure required for its intended use.

\begin{table}[t]
\centering
\caption{The v1 composite $C_{\text{data}}$ (Eq.~\eqref{eq:classical_comp}, equal weights) evaluated on the 27 classical-track datasets. The four pillars differ by up to six orders of magnitude in range and do not share units. The composite is therefore dominated by whichever pillar has the largest empirical variance, and which pillar dominates depends on the estimator. \emph{Exploratory.}}
\label{tab:p17}
\begin{tabular}{lrrrr}
\toprule
& \multicolumn{2}{c}{with $C_{\text{top}}$ as deployed} & \multicolumn{2}{c}{with $C_{\text{top}}$ corrected} \\
\cmidrule(lr){2-3}\cmidrule(lr){4-5}
Pillar & share of $\mathrm{Var}$ & $\rho(C_{\text{data}},\cdot)$ & share of $\mathrm{Var}$ & $\rho(C_{\text{data}},\cdot)$ \\
\midrule
$S$ (entropy, nats)                       & $0.00\%$   & $+0.573$ & $\mathbf{100.00\%}$ & $\mathbf{+0.993}$ \\
$I_{\text{corr}}$ (nats per order)        & $0.00\%$   & $-0.419$ & $0.00\%$            & $+0.272$ \\
$K$ (compression, dimensionless)          & $0.00\%$   & $-0.539$ & $0.00\%$            & $-0.333$ \\
$C_{\text{top}}$                          & $\mathbf{100.00\%}$ & $\mathbf{+0.998}$ & $0.00\%$ & $-0.645$ \\
\midrule
\multicolumn{5}{l}{$\rho\big(C_{\text{data}}^{\text{deployed}},\, C_{\text{data}}^{\text{corrected}}\big) = 0.579$} \\
\bottomrule
\end{tabular}
\end{table}

The two halves of Table~\ref{tab:p17} expose the problem. With the
topological component as deployed, $C_{\text{top}}$ carries $100\%$ of the
composite variance and $\rho(C_{\text{data}},C_{\text{top}})=0.998$: the composite is, empirically, almost exactly the topology component. Correcting that estimator (Appendix~\ref{app:topofix}) transfers the domination to entropy, which then carries $100\%$ of the composite variance with
$\rho=0.993$. The two resulting rankings agree only at $\rho=0.579$.

This transfer does not establish that entropy is the informative pillar.
Differential entropy is extensive in dimension and scale-dependent,
$h(aX)=h(X)+d\log|a|$, and therefore carries the same kind of scale defect.
Across the suite,
\[
S \sim d\,\log(\mathrm{diam}), \qquad R^2=0.999,
\]
with Pearson $r=+1.000$. The largest value in the suite, $S=4036$ nats for \texttt{real-fashion-2v6}, occurs for the dataset with the largest feature dimension ($d=784$); the second largest occurs for the dataset with the largest diameter ($1.2\times10^6$). Thus two of the four pillars inherit the same fundamental problem: the scalar sum reports whichever uncontrolled scale the current estimators amplify.

\paragraph{Rank normalization separates the pillars without choosing their
weights.}
The appropriate repair is therefore to preserve the components rather than
collapse them into a weighted scalar. For each component $f_i$ and dataset
$D$ in a benchmark suite $\mathcal{B}$, we define
\begin{equation}
  v_i(D) =
  \frac{\big|\{D'\in\mathcal{B}:f_i(D')\le f_i(D)\}\big|}
       {|\mathcal{B}|}
  \in[0,1].
  \label{eq:ranknorm}
\end{equation}
The vector
\[
\vv(D)=(v_1,\ldots,v_K)
\]
is the dataset's complexity profile. Each component is now dimensionless on the benchmark scale and retained separately rather than being assigned an unidentifiable weight. The empirical CDF of each $f_i$ is computed once on $\mathcal{B}$ and frozen; new datasets are placed in $\vv$-space by interpolation against those frozen CDFs.

As a diagnostic, applying Eq.~\eqref{eq:ranknorm} makes the four components
carry approximately equal variance shares, $25.00\%$, $25.00\%$, $24.99\%$, and $25.00\%$, as expected from their common rank scale. This does not make the pillars equally informative; it removes the arbitrary dominance created by their native scales so that their predictive contributions can be tested separately.

Table~\ref{tab:v_ranknorm} gives the resulting vectors for the 27
classical-track datasets, which are the representation consumed by the
selector. The underlying estimator values are reported in Appendix~\ref{app:vtables}; the topology estimator is subject to the caveat in
Appendix~\ref{app:topofix}.

\begin{table}[t]
\centering
\caption{The intrinsic complexity vector $\vv(D)$ after rank normalisation (Eq.~\eqref{eq:ranknorm}), for the $27$ classical-track datasets --- the representation the selector consumes. Each column is the empirical CDF of that component over the frozen suite, so entries are comparable across columns by construction. $H$ differential entropy (kNN--KL); $d_{\mathrm{pca}}$, $d_{\mathrm{2nn}}$ intrinsic dimension (PCA-effective, TwoNN); $K$ compression ratio (gzip); $T$ topological complexity as pre-specified (Vietoris--Rips, defective --- see Appendix~\ref{app:topofix}); $T^{\star}$ its corrected replacement (untruncated Rips, $H_0$ excluded, diameter-normalised $\ell_2$); $I_2$, $I_h$ second- and higher-order mutual information (kNN, MINE); $R$ kernel effective rank (RBF, median heuristic); $M$ MDL (autoencoder).}
\label{tab:v_ranknorm}
\small
\begin{tabular}{lrrrrrrrrrr}
\toprule
Dataset & $H$ & $d_{\mathrm{pca}}$ & $d_{\mathrm{2nn}}$ & $K$ & $T$ & $T^{\star}$ & $I_2$ & $I_h$ & $R$ & $M$ \\
\midrule
adv-easy-looking-hard & 0.70 & 0.22 & 0.85 & 0.96 & 0.70 & 0.44 & 0.26 & 0.30 & 0.48 & 0.56 \\
adv-encoded-random & 0.89 & 0.19 & 0.41 & 0.37 & 0.93 & 0.19 & 0.33 & 0.33 & 0.30 & 0.85 \\
adv-hard-looking-easy & 0.33 & 0.26 & 0.04 & 0.89 & 0.19 & 0.41 & 1.00 & 1.00 & 0.22 & 0.19 \\
adv-quantum-trivial & 0.37 & 0.44 & 0.15 & 0.74 & 0.15 & 0.70 & 0.89 & 0.93 & 0.19 & 0.26 \\
\addlinespace
real-breast-cancer & 0.85 & 0.07 & 0.37 & 0.33 & 0.89 & 0.11 & 0.63 & 0.81 & 0.26 & 0.81 \\
real-credit-default & 0.96 & 0.33 & 0.48 & 0.11 & 1.00 & 0.22 & 0.67 & 0.78 & 0.56 & 0.70 \\
real-diabetes & 0.74 & 0.11 & 0.44 & 0.19 & 0.85 & 0.15 & 0.44 & 0.48 & 0.52 & 0.67 \\
real-digits-3v5 & 0.93 & 0.67 & 0.93 & 0.04 & 0.81 & 0.52 & 0.41 & 0.44 & 0.93 & 0.93 \\
real-fashion-2v6 & 1.00 & 0.59 & 1.00 & 0.07 & 0.96 & 0.37 & 0.59 & 0.63 & 1.00 & 1.00 \\
real-higgs & 0.78 & 1.00 & 0.96 & 0.41 & 0.74 & 0.56 & 0.37 & 0.37 & 0.96 & 0.78 \\
real-ionosphere & 0.41 & 0.70 & 0.89 & 0.30 & 0.37 & 0.30 & 0.85 & 0.74 & 0.81 & 0.89 \\
real-iris & 0.56 & 0.15 & 0.30 & 0.15 & 0.04 & 0.26 & 0.74 & 0.85 & 0.37 & 0.30 \\
real-sonar & 0.04 & 0.63 & 0.74 & 0.26 & 0.30 & 0.33 & 0.48 & 0.52 & 0.89 & 0.96 \\
real-wine & 0.81 & 0.04 & 0.26 & 0.22 & 0.78 & 0.04 & 0.52 & 0.67 & 0.04 & 0.74 \\
\addlinespace
synth-2sphere & 0.48 & 0.41 & 0.11 & 0.70 & 0.11 & 0.67 & 0.81 & 0.70 & 0.15 & 0.07 \\
synth-3sphere-pc & 0.52 & 0.48 & 0.33 & 0.81 & 0.33 & 0.74 & 0.56 & 0.56 & 0.41 & 0.11 \\
synth-anisotropic-gmm & 0.63 & 0.56 & 0.81 & 1.00 & 0.41 & 0.07 & 0.93 & 0.89 & 0.44 & 0.33 \\
synth-concentric-rings & 0.30 & 0.30 & 0.07 & 0.78 & 0.07 & 0.48 & 0.96 & 0.96 & 0.07 & 0.22 \\
synth-ea-spin-glass & 0.26 & 0.78 & 0.59 & 0.52 & 0.52 & 0.93 & 0.19 & 0.22 & 0.63 & 0.41 \\
synth-klein-bottle & 0.44 & 0.52 & 0.19 & 0.85 & 0.22 & 0.78 & 0.78 & 0.59 & 0.33 & 0.15 \\
synth-parity-2 & 0.15 & 0.85 & 0.52 & 0.63 & 0.48 & 0.96 & 0.22 & 0.19 & 0.74 & 0.63 \\
synth-parity-3 & 0.07 & 0.81 & 0.56 & 0.52 & 0.67 & 0.81 & 0.07 & 0.07 & 0.67 & 0.52 \\
synth-parity-4 & 0.19 & 0.96 & 0.70 & 0.56 & 0.59 & 1.00 & 0.30 & 0.04 & 0.78 & 0.48 \\
synth-parity-5 & 0.11 & 0.93 & 0.63 & 0.44 & 0.63 & 0.89 & 0.15 & 0.15 & 0.59 & 0.37 \\
synth-prng & 0.67 & 0.74 & 0.78 & 0.67 & 0.44 & 0.63 & 0.11 & 0.26 & 0.85 & 0.59 \\
synth-sk-spin-glass & 0.22 & 0.89 & 0.67 & 0.59 & 0.56 & 0.85 & 0.04 & 0.11 & 0.70 & 0.44 \\
synth-torus & 0.59 & 0.37 & 0.22 & 0.93 & 0.26 & 0.59 & 0.70 & 0.41 & 0.11 & 0.04 \\
\bottomrule
\end{tabular}
\end{table}

\paragraph{The quantum-native composite fails the same way.}
\label{sec:framework:cquant}
Version~1 proposed a parallel composite for quantum data, $C_{\text{quant}}(\mathcal{E})$ of Eq.~\eqref{eq:quantum_comp} (Eq.~10 of v1), defined on an ensemble $\mathcal{E}=\{(p_i,\rho_i)\}_{i=1}^{N}$ of $n$-qubit states. It is the quantity that was meant to carry the quantum side of the framework, and it inherits the structure of Eq.~\eqref{eq:classical_comp}: six components summed with weights. Before it could be evaluated, two of its components had to be made well-posed.

\emph{Quantum Fisher information.} Version~1 writes $F=\mathrm{Tr}[\rho L^2]$ without stating the parameter with respect to which $L$ is taken, and a bare ensemble has none. A Hamiltonian parameter is undefined for most ensembles; encoding parameters belong to the induced complexity of Eq.~\eqref{eq:induced_comp}, not to Eq.~\eqref{eq:quantum_comp}; and ansatz parameters would make the quantity a property of the (ensemble, ansatz) pair, the category error removed from $C_{\text{ind}}$ in Sec.~\ref{sec:framework:predictor}. We use instead the pure-state identity $F_Q[\psi,G]=4\,\mathrm{Var}_\psi(G)$, which makes the quantum Fisher information intrinsic once a generator set is fixed. With the collective spin generators $J_{\hat n}=\tfrac12\sum_i\hat n\cdot\vec\sigma^{(i)}$,
\begin{equation}
  \overline{\mathcal F}(\mathcal{E})=\Big\langle \max_{\hat n}\,F_Q[\psi,J_{\hat n}]\,/\,n^2\Big\rangle_{\psi\in\mathcal{E}}\in[1/n,1],
  \label{eq:qfi_collective}
\end{equation}
bounded below by the separable value and above by the Heisenberg limit. It measures metrological usefulness rather than entanglement: Haar-random states are close to maximally entangled yet score near the separable floor.

\emph{Quantum topological term.} Version~1's $C_{\text{top}}^{(q)}$ sums a topological entanglement entropy, an Euler characteristic and persistence lifetimes. The first is exponential to estimate and requires a two-dimensional lattice that generic ensembles lack; the second is a signed integer on an unspecified manifold. We retain only the persistence term, under the Fubini--Study metric $d(\psi,\phi)=\arccos\lvert\langle\psi|\phi\rangle\rvert$ and normalized by the canonical maximum distance $\pi/2$ between orthogonal states,
\begin{equation}
  C_{\text{top}}^{(q)}(\mathcal{E})=\sum_{k\ge1}\sum_i\Big(\frac{\ell_{k,i}}{\pi/2}\Big)^{2},
  \label{eq:ctopq_fs}
\end{equation}
where $\ell_{k,i}$ are the Vietoris--Rips persistence lengths in dimension $k$, $H_0$ excluded. Unlike the classical case (Appendix~\ref{app:topofix}), projective Hilbert space has an intrinsic scale, so no empirical-diameter normalization is needed. The price is that Eq.~\eqref{eq:ctopq_fs} measures topology weighted by the extent of state space explored, not homology: a small loop scores near zero although it is topologically a circle.

All six components are implemented and validated against closed forms: half-cut entropy $0$ and $\ln 2$ on product and GHZ states, total correlation $n\ln2$ on GHZ, effective rank $1$ and $N$ on identical and orthogonal ensembles, stabilizer R\'enyi-2 magic $0$ and $\log_2(4/3)$ on stabilizer and $T$ states, Eq.~\eqref{eq:qfi_collective} equal to $1/n$, $(3n-2)/n^2$ and $1$ on product, W and GHZ states, and the Betti signatures of the Bloch sphere and a great circle recovered by Eq.~\eqref{eq:ctopq_fs}. We evaluate the composite with equal weights on the ten quantum-native ensembles of this work (Sec.~\ref{sec:positive_control}): five reference ensembles (TFIM and XXZ ground states, Haar-random, stabilizer-with-magic and random-MPS) and the five shadow-hard parity probes $w\in\{1,2,4,6,8\}$, all at $n=8$ and $N=256$. This analysis is \emph{exploratory}.

\begin{table}[t]
\centering
\caption{The v1 quantum-native composite $C_{\text{quant}}$ (Eq.~\eqref{eq:quantum_comp}, equal weights) on the ten quantum-native ensembles ($n=8$, $N=256$). Share of composite variance on native scales and after the rank normalization of Eq.~\eqref{eq:ranknorm}; pairs separated out of $45$ (values differing by more than $10^{-6}$ of the component's maximum); range across the five parity probes. Components 5 and 6 as specified in Eqs.~\eqref{eq:qfi_collective}--\eqref{eq:ctopq_fs}. \emph{Exploratory.}}
\label{tab:cquant}
\begin{tabular}{lrrrrr}
\toprule
Component & range & native $\%\mathrm{Var}$ & rank-norm.\ $\%\mathrm{Var}$ & pairs sep. & parity range \\
\midrule
$\overline S_{\text{ent}}$ (nats)            & $0.40$--$2.28$   & $0.00$ & $17.81$ & $41$ & $0.446$ \\
$\mathcal I_{\text{multi}}$ (nats)           & $3.23$--$5.55$   & $0.01$ & $15.81$ & $44$ & $0.744$ \\
$\mathrm{rank}_{\text{eff}}(K_{\mathcal E})$ & $1.05$--$254.04$ & $\mathbf{99.97}$ & $17.04$ & $35$ & $0$ \\
$\overline{\mathcal N}$ (bits)               & $1.37$--$6.02$   & $0.02$ & $21.56$ & $35$ & $0$ \\
$\overline{\mathcal F}$                      & $0.002$--$0.536$ & $0.00$ & $15.81$ & $45$ & $0.009$ \\
$C_{\text{top}}^{(q)}$                       & $0$--$1.88$      & $0.00$ & $11.96$ & $34$ & $0$ \\
\bottomrule
\end{tabular}
\end{table}

Table~\ref{tab:cquant} shows that Eq.~\eqref{eq:quantum_comp} collapses exactly as Eq.~\eqref{eq:classical_comp} did. The effective rank carries $99.97\%$ of the composite variance, so the composite is, empirically, the effective rank. That component is a count on $[1,N]$: it saturates near $N$ for any ensemble that spreads over Hilbert space ($254.0$ for Haar-random, $252.6$ for random MPS, $252.28$ for every parity probe) and sits near $1$ for ground-state families. Its dominance therefore records ensemble size and spread, not state complexity, and it is constant across the parity probes, the one family in the suite with a complexity knob. As with the classical pillars, the defect lies in the aggregation, not in the components: rank normalization brings the six shares to between $11.96\%$ and $21.56\%$, and every component separates at least $34$ of the $45$ ensemble pairs. Within the parity family only the entanglement entropy and the total correlation move.

Two limits bound what this shows. First, every ensemble has the same width and size, so the extensivity of the entropies in $n$ and the $N$-dependence of the effective rank, both of which would sharpen the domination, cannot be measured here. Second, the composite is evaluated, not tested as a predictor: the quantum-native arm is excluded from the classical grid of H5 by construction, and five reference ensembles at one width offer no variation to predict. The experiment therefore implicates the aggregation rather than showing that the individual components are uninformative.

\paragraph{A fifth pillar: geometric curvature.}
For data with native graph or hypergraph structure, we additionally specify a
discrete Ricci-curvature summary. For graphs we use mean Ollivier--Ricci
curvature \citep{ollivier2009ricci,ni2019ricciflow},
\begin{equation}
  \bar\kappa_{\text{OR}}(D)
  =\frac{1}{|E|}\sum_{xy\in E}
    \left(1-\frac{W_1(m_x,m_y)}{d(x,y)}\right),
  \label{eq:orc}
\end{equation}
where $m_x$ and $m_y$ are probability measures on the one-step
neighbourhoods, $W_1$ is the Wasserstein-1 distance, and $d$ is the
shortest-path distance. For hypergraphs we specify ORCHID node-transport
curvature \citep{coupette2023orchid}, with the edge-Ricci construction
\citep{hacquard2025hypergraph} on the line graph when hyperedge cardinalities
span more than two orders of magnitude.

The motivation is complementary to persistent homology: topology captures
global connectivity, whereas curvature captures local geometric distortion.
The curvature component is also motivated by emerging quantum subroutines for
its estimation, including approaches based on block-encoded geodesic
distances, QSVT, and quantum power methods
\citep{nghiem2025quantumorc}.

\textbf{Scope note.} The curvature pillar is specified but not implemented in
this work. Every dataset in the frozen suite, classes A, C, and D point clouds
and class B state ensembles, is non-graph. Consequently $v_5$ is dropped
throughout the present analysis. H2's curvature arm is therefore
inapplicable here, and H2 reduces to a test of the topological contribution
alone. The unexercised pillar is recorded explicitly rather than silently
omitted. It becomes testable in a future benchmark containing graph- or
hypergraph-native datasets.

The intrinsic complexity and embedding interaction are different
objects. Equation~\eqref{eq:induced_comp} (Eq.~11 of v1) introduced the induced quantum complexity $C_{\text{ind}}(D;\phi)$ and correctly observed that the perceived complexity of a dataset can depend on the feature map $U_\phi$. The limitation was to treat this induced quantity as a property of the dataset itself. It is instead a property of the \emph{dataset--embedding pair}.

We therefore separate the two. The vector $\vv(D)$ contains embedding-free
properties computed from the raw data. The interaction vector
\begin{equation}
  \qq(D,\phi)=(q_1,\ldots,q_M)\in\mathbb{R}^M
  \label{eq:interaction}
\end{equation}
contains quantities evaluated after embedding, including average and
variance of bipartite entanglement entropy, the embedded-kernel effective rank $r_{\text{eff}}(K_\phi)$, circuit-family expressibility $\mathcal{E}(U_\phi)$,
embedding locality/lightcone width, and the persistent-homology profile of
the embedded point cloud under fidelity-induced distance.

The distinction has three consequences. First, dataset complexity can be
computed once and reused across candidate embeddings. Second, feature-level
analysis can be defined through a marginal signature $\sigma(x_j)\subset\vv$, allowing H4 to ask how individual features interact with candidate embeddings. Third, predictor attributions can distinguish variation associated with the dataset from variation associated with its embedding.

The edge is therefore modelled as
\begin{equation}
  \ehat(D,\phi)=
  \Sel\big(\vv(D),\qq(D,\phi)\big).
  \label{eq:selector}
\end{equation}

Two predictor implementations are used to assess robustness. The primary predictor is the sparse variational multi-head Gaussian process (SVGPMultiHead), which uses inducing points and a linear model of coregionalization to model the edge and alignment targets at the benchmark scale. A gradient-boosted model provides an independent non-GP sensitivity analysis and feature attribution. Agreement between the two is evidence of
robustness; disagreement is itself informative.

The cost of this revision is that no single scalar summarizes dataset complexity. The benefit is that the representation avoids the arbitrary weighting of the original scalar composite and permits component-level predictive and attribution analyses.

\subsection{Correcting the barren-plateau scaling law}
\label{sec:framework:bp}

Equation~\eqref{eq:variance} (Eq.~9 of v1) proposed the data-gated form
$\Var[g(\theta)]\approx\exp(-\alpha nd\,\tilde{C}_{\text{data}})$.
That equation has an undesirable data-free limit:
$\tilde{C}_{\text{data}}\to0$ implies $\Var[g(\theta)]\to1$. This treats the absence of data complexity as removing the barren plateau altogether, whereas the Haar-random circuit result of \citet{mcclean2018barren} establishes exponential gradient suppression as a property of the circuit ensemble itself, independent of a particular data distribution.

The corrected formulation therefore separates the circuit baseline from the
data-dependent modulation:
\begin{equation}
  \log\Var[g(\theta)]
  \approx \log A-\alpha nd-nd\,\kk^\top\vv(D),
  \label{eq:bp}
\end{equation}
where $A$ and $\alpha$ describe the data-free circuit family and
$\kk\in\mathbb{R}_{\ge0}^K$ contains component-wise data sensitivities. At
$\vv\to\mathbf{0}$,
\[
\Var[g]\to Ae^{-\alpha nd},
\]
recovering the circuit-only baseline. The original equation can be recovered formally as a special case in which the data-dependent modulation is represented by a single effective complexity coordinate.

This correction separates two questions that were conflated in v1: whether
the circuit family exhibits the expected baseline suppression, and whether
dataset structure modulates that suppression in a systematic way.

H1 tests whether Eq.~\eqref{eq:bp} describes the observed scaling and whether
any $\kappa_i$ are non-negligible. H2 tests whether topology contributes
beyond entropy, correlation, and compressibility. The empirical scaling
results are reported in Sec.~\ref{sec:results_nscaling}.

\subsection{Replacing scalar expressibility--complexity alignment}
\label{sec:framework:align}

Equation~\eqref{equation_expressivity} (Eq.~15 of v1) proposed
$\mathbb{E}[\epsilon_{\text{gen}}]\approx\epsilon_{\text{emp}}+\lambda\left|\mathcal{E}(U)-C_{\text{data}}\right|$.
That construction cannot serve as a well-defined quantitative alignment
test for two reasons. Expressibility $\mathcal{E}(U)$ is a Kullback--Leibler divergence between fidelity distributions, using the Haar-random distribution as the reference \citep{sim2019expressibility}, while $C_{\text{data}}$ is normalized on $[0,1]$. Their absolute difference therefore has no invariant interpretation, and the proposed ``sweet spot'' $\mathcal{E}(U)\approx C_{\text{data}}$ is not a well-defined testable condition.

We replace the scalar alignment distance with a learned alignment
score. The predictor has two heads sharing a representation
$\Phi(\vv,\qq)$:
\begin{align}
  \ehat(D,\phi)
  &=\Sel_{\text{edge}}\big(\Phi(\vv(D),\qq(D,\phi))\big),
  \label{eq:head-edge}\\
  a(D,U)
  &=\sigma\!\left(
  \Sel_{\text{align}}(\Phi(\vv(D),\qq(D,U)))\right)
  \in(0,1),
  \label{eq:head-align}
\end{align}
where $\sigma$ is the logistic sigmoid.

This formulation retains the substantive hypothesis of v1 that mismatch
between data structure and circuit representation may affect generalization,
while removing the arbitrary requirement that expressibility and data
complexity share a numerical scale. The resulting relationship is therefore
tested rather than imposed. The generalization model becomes
\begin{equation}
  \mathbb{E}[\epsilon_{\text{gen}}]
  \approx\epsilon_{\text{emp}}+\beta(1-a(D,U)),
  \label{eq:gen}
\end{equation}
with $\beta\ge0$ parameterized as $\beta=e^b$.

For each $(D,\phi)$, the observed budget-matched edge $e_{\text{obs}}$ and empirical generalization gap $\mathrm{gap}_{\text{obs}}=\epsilon_{\text{test}}-\epsilon_{\text{train}}$ provide the training targets:
\begin{equation}
  \mathcal{L}
  =\sum_{D,\phi}(e_{\text{obs}}-\ehat)^2
  +\gamma\sum_{D,U}
  \left[
  \mathrm{gap}_{\text{obs}}
  -\epsilon_{\text{emp}}
  -\beta(1-a)
  \right]^2,
  \label{eq:loss}
\end{equation}
with $\gamma$ selected on a held-out fold. Thus the alignment score is not
defined from itself: its training signal comes from observed generalization
gaps, while on held-out datasets $a$ is computed from $(\vv,\qq)$ alone.

For the GP this is implemented as a multi-output GP with a linear model of
coregionalization; for the GBM, as a multi-output regressor with shared
splits. H3 tests $\beta>0$ using a bootstrap confidence interval.

\subsection{Making encoding cost and data access explicit}
\label{sec:framework:cost}

Version~1 described an encoding--expressivity trade-off but did not include
encoding overhead in the comparison.The purpose is not to assign a universal cost to state preparation, but to make the resource assumptions used in the present comparison explicit. Amplitude encoding, for example, can require substantial state-preparation resources in the absence of an appropriate coherent data-loading primitive; QRAM is precisely the assumption that provides coherent address-indexed access to stored classical data in superposition \citep{giovannetti2008quantum}.

For an encoding $\phi$ on $n$ qubits we therefore define
\begin{equation}
  \mathcal{C}_\phi
  =D_\phi W_\phi
  (1-\epsilon_\phi)^{-D_\phi W_\phi},
  \label{eq:cost}
\end{equation}
where $D_\phi$ is state-preparation depth, $W_\phi$ is the average number of
two-qubit gates per layer, and $\epsilon_\phi$ is the reference per-gate
error rate. The exponential factor represents compounding fidelity loss
under the reference noise model.

We report three versions of the edge:
$\ehat_{\text{raw}}$, the unadjusted gap;
$\ebudget$, the gap after matching the reference resource budget; and
$\ehat_{\text{Pareto}}$, the hypervolume of the performance--cost Pareto
frontier. The primary target is $\ebudget$: a large raw edge with a small
budget-matched edge indicates that the apparent advantage is attributable
substantially to resource asymmetry.

\paragraph{Operationalizing the budget-matched edge.}
This work uses noiseless simulation, so the resource penalty is introduced
through the fidelity predicted by a frozen reference noise model,
$\epsilon_2=10^{-3}$. For a cell with $n_{2q}$ two-qubit gates,
\[
F=(1-\epsilon_2)^{n_{2q}}.
\]
The noiseless quantum accuracy
$a_q=1-\epsilon^q_{\text{test}}$ is shrunk toward chance by $F$, giving

\begin{equation}
\begin{split}
  \ebudget
  &=
  \left[
  \frac12+F\left(a_q-\frac12\right)
  \right]-a_c \\
  &=
  \left(\epsilon^c_{\text{test}}-\frac12\right)
  -F\left(\epsilon^q_{\text{test}}-\frac12\right).
  \label{eq:ebudget}
  \end{split}
\end{equation}

This reduces to $\ehat_{\text{raw}}$ at $F=1$ and approaches the classical
chance-referenced term as $F\to0$.

\paragraph{Data-access modes.}
The cost of an encoding cannot be interpreted independently of the data-access
model. We therefore classify each $(D,\phi)$ pair as:

\begin{itemize}
\item \textbf{Mode S (sparse/structured):} the encoding exploits explicit
structure in the data, such as sparsity or a structured sampling primitive;
\item \textbf{Mode Q (QRAM-dependent):} the encoding assumes coherent
random-access loading of otherwise unstructured classical data;
\item \textbf{Mode N (quantum-native):} the input is a physical quantum state
rather than a classically stored dataset.
\end{itemize}

This classification determines which classical comparison is appropriate
under the access assumptions adopted in this study. For a Mode-Q pair, the classical comparator must include an appropriate quantum-inspired algorithm under the corresponding sampling/query access model. Otherwise the comparison gives the quantum procedure an access advantage that has not been charged to the quantum side. The dequantization literature makes this issue explicit: when a classical algorithm is given the corresponding sampling access, apparent quantum speedups for some low-rank
linear-algebra tasks can disappear \citep{tang2019quantum,chia2022dequantization}.

Under this framework, a Mode-Q result can be retained as a candidate quantum
advantage only when no applicable classical counterpart exists under the
matched access model. Such cases receive a \emph{sparse-survivor} tag and are compared
against the best unconstrained classical method only after the access
assumption has been made explicit.

For Mode-Q pairs the resource model includes the active data-loading
overhead,
\begin{equation}
  \mathcal{C}^{(Q)}_\phi
  =(D_\phi+n)W_\phi
  (1-\epsilon_\phi)^{-(D_\phi+n)W_\phi},
  \label{eq:costq}
\end{equation}
whereas Eq.~\eqref{eq:cost} applies to Mode~S and
$\mathcal{C}_\phi=0$ to Mode~N.

The present frozen suite contains no encoding requiring coherent QRAM loading.
All evaluated pairs are therefore Mode~S or Mode~N (Sec.~\ref{sec:results:h6}), and H6 is a scope check rather than an empirical
test of dequantization.


\section{Pre-specified hypotheses}
\label{sec:hypotheses}

The revisions to the framework in Sec.~\ref{sec:framework} define six
pre-specified hypotheses (Tab.~\ref{tab:hypotheses}). These hypotheses were fixed before the experimental analysis, together with their decision rules, predictor architecture, encoding library, data-access classifications, and stopping criteria. The complete protocol, frozen estimator code, and analysis pipelines are released with the accompanying code. We committed in advance
to reporting the outcome of every test independently of its direction.

\begin{table}[t]
\centering
\small
\caption{The six pre-specified hypotheses, decision rules, and outcomes.
H5 is the primary integrative test of selector validity. H6 is a boundary
condition on data-access mode: the frozen suite contains no Mode-Q instances,
so this stratum cannot be evaluated.}
\label{tab:hypotheses}
\begin{tabular}{@{}l p{0.26\linewidth} p{0.33\linewidth} p{0.21\linewidth}@{}}
\toprule
 & Claim & Decision rule & Outcome \\
\midrule
H1 & Barren-plateau scaling, Eq.~\eqref{eq:bp}
   & $R^2 \ge 0.7$; some $\kappa_i$ CI excludes $0$
   & \textbf{FAILS} \\
H2 & Topology/curvature contribute
   & $\Delta\mathrm{AIC} \ge 10$ vs.\ reduced $\vv$
   & \textbf{WITHDRAWN}$^{\dagger}$ \\
H3 & Expressibility--data alignment
   & $\beta > 0$, 95\% CI excludes $0$; $\rho \ge 0.3$
   & \textbf{Primary criterion met; predictive support absent}$^{\dagger}$\\
H4 & Per-feature encoding optimality
   & AUROC gain $\ge 5\%$, CI excludes $0$
   & Untested (downstream of H5) \\
\textbf{H5} & \textbf{Selector validity (primary)}
   & $R^2 \ge 0.4$ on $\ebudget$, ECE $\le 0.1$
   & \textbf{FAILS} \\
H6 & Dequantization-stratified validity
   & conditional on H5
   & Not evaluable \\
\bottomrule
\addlinespace[2pt]
\multicolumn{4}{@{}p{\linewidth}@{}}{\footnotesize
$^{\dagger}$ The pre-specified topology test used an estimator later
found to contain implementation defects. Its original result is
therefore withdrawn rather than interpreted as a clean hypothesis
test. Corrected-estimator analyses are reported separately as
estimator-sensitivity analyses in the topology audit.}
\end{tabular}
\end{table}

The battery is structured hierarchically. H1--H3 test the principal mechanisms introduced by the revised framework: data-dependent trainability, geometric structure, and expressibility--data alignment. H4 tests a downstream consequence of the interaction between data characteristics and encoding choice. H5 is the central integrative test: whether these quantities, taken together, provide a predictive selector for the observed quantum edge on previously unseen datasets. H6 tests the scope of that prediction with respect to the data-access model.

\paragraph{Model comparison: }
Several hypotheses use the Akaike Information Criterion (AIC),
\[
\mathrm{AIC}=2k-2\ln\hat{L},
\]
where $k$ is the number of free parameters and $\hat{L}$ is the maximized
likelihood. Lower AIC indicates a preferred model. We define
\[
\Delta\mathrm{AIC}
=
\mathrm{AIC}_{\mathrm{without}}
-
\mathrm{AIC}_{\mathrm{with}},
\]
so that positive values favour inclusion of the tested component. A
threshold of $\Delta\mathrm{AIC}=10$ is used as the pre-specified criterion
for substantive support~\citep{burnham2002model}. Comparisons are made on
the same observations.

\begin{description}[leftmargin=0pt,labelsep=0.4em]

\item[H1: Barren-plateau scaling.]
The revised scaling law in Eq.~\eqref{eq:bp} is expected to describe the
observed gradient-variance scaling across the suite. The hypothesis is
supported if the cross-validated linear regression of $\log\Var[g(\theta)]$ on $(n,d,nd\cdot\vv)$ achieves $R^2\ge0.7$ and at least one data-dependent coefficient $\kappa_i$ has a 95\% confidence interval excluding zero. This tests whether the revised framework captures a systematic modulation of trainability by dataset complexity beyond the data-independent circuit contribution. Failure of this criterion does not test or exclude barren-plateau phenomena more generally; it tests
the adequacy of the proposed functional form in the finite-width regime studied here.

\item[H2: Topological and curvature contribution.]

Adding the topological and/or curvature components of $\vv$ is expected to
improve the explanatory model for gradient behaviour beyond entropy,
correlation, and compressibility. The hypothesis was pre-specified to be
supported if the augmented model achieved $\Delta\mathrm{AIC}\ge10$
relative to the reduced model. The topology estimator used in the
pre-specified implementation was subsequently found to contain
implementation defects. The corresponding H2 result is therefore
withdrawn and is not interpreted as evidence for or against the
hypothesis. Corrected-estimator analyses are reported separately in
the topology audit.

\item[H3: Expressibility--data alignment.]
The fitted alignment coefficient $\beta$ in Eq.~\eqref{eq:gen} is expected
to be positive, with a 95\% confidence interval excluding zero. As a
secondary criterion, the Spearman correlation between predicted
$a(D,U)$ and observed $-\mathrm{gap}_{\mathrm{obs}}$ on held-out folds is
required to satisfy $\rho\ge0.3$. Confidence intervals are obtained by
bootstrapping over leave-one-dataset-out folds.

\item[H4: Per-feature encoding optimality.]
On datasets containing heterogeneous feature structure, mixed-encoding
ans\"atze that route each feature $x_j$ to its predicted-optimal encoding
$\phi^*(x_j)$ are expected to outperform monolithic encodings. Support
requires a relative AUROC improvement of at least $5\%$ with a 95\%
confidence interval excluding zero. Because this test depends on the
validity of the upstream selector, it is not evaluated when H5 fails.

\item[H5: Selector validity (central test).]
The revised data-centric framework is expected to predict the observed
quantum edge on previously unseen datasets. Support requires the trained
selector $\Sel$ to achieve $R^2\ge0.4$ on $\ebudget$ under leave-one-dataset-out cross-validation, together with expected calibration error~\citep{guo2017calibration} of at most $0.1$.

H5 is the primary test of the framework as an operational predictive model.
Its failure means that the quantities specified by the revised framework,
in their present operational form and under the tested benchmark and
access regime, do not provide a reliable predictor of the measured edge
across datasets.

\item[H6: Dequantization-stratified selector validity.]
The validity of the selector is additionally expected to depend on the
data-access model. Conditional on H5 passing globally, performance is
evaluated separately across the available access modes. The pre-specified
criterion requires $R^2\ge0.4$ within both the Mode-S/Mode-N (including sparse-survivor) regime and the Mode-Q regime, or a significant degradation of at least $\Delta R^2\ge0.2$ in Mode-Q. The latter outcome would indicate that the
selector distinguishes regimes in which classical access models can or
cannot reproduce the relevant computation.

The frozen suite contains no Mode-Q instances, however. H6 is therefore a
scope check rather than an evaluable hypothesis in the present experiment.
This limitation is itself relevant to the interpretation of any apparent
quantum edge, since data-access assumptions determine whether a quantum
procedure has a meaningful classical counterpart.

\end{description}

Together, H1--H6 form the pre-specified falsification battery. The
hierarchy is deliberate: H1--H4 test individual mechanisms or
downstream consequences of the revised framework, whereas H5 asks
whether those mechanisms combine into a predictive model of quantum
edge. H6 then places that prediction on the boundary defined by the
data-access model. The subsequent analysis distinguishes between
hypotheses that fail their pre-specified criteria, hypotheses whose
criteria are formally met but whose predictive interpretation does not
hold, hypotheses that could not be evaluated because an upstream test
failed, and hypotheses whose original implementation was withdrawn
after an estimator defect was identified.

This structure also separates two distinct outcomes. A component
hypothesis may fail because the proposed mechanism is absent or
misspecified, while H5 asks the stronger question of whether the
complete framework succeeds as a predictive model. The primary
conclusion of this work therefore rests on H5, while the preceding tests
identify which mechanisms prevent the proposed data-centric predictor
from becoming a reliable measure of quantum advantage.


\section{Methodology}
\label{sec:methodology}

We evaluate the proposed data-centric predictor under a pre-specified experimental protocol. The confirmatory analysis uses a frozen benchmark registry, a fixed encoding--ansatz grid, repeated random seeds, and leave-one-dataset-out (LODO) validation. The benchmark, circuit families, preprocessing rules, predictor architecture, hyperparameters, and decision criteria were fixed before evaluating held-out predictive performance. Later implementation corrections are reported explicitly as deviations or
sensitivity analyses rather than incorporated silently into the confirmatory protocol.

\subsection{Benchmark suite}
\label{sec:benchmark}

The frozen registry contains four dataset classes, of which Classes A, C, and D comprise the 27-dataset confirmatory Phase~B suite analyzed in this paper. These datasets span heterogeneous binary classification problems while remaining within a common classical-data access model.

For each dataset, the experimental protocol evaluates ten deterministic quantum encodings, six variational ansatz configurations, and five random seeds. The confirmatory grid therefore contains
\begin{equation}
27\times 10\times 6\times 5 = 8100
\end{equation}
dataset--encoding--ansatz--seed cells. All 8100 cells were completed.

The confirmatory quantum circuits operate on eight qubits. The ten encodings
comprise four encoding families: angle encoding, $ZZ$-type encoding,
amplitude encoding, and data re-uploading. The six ansatz configurations
comprise three circuit families evaluated at depths $L\in\{1,3\}$. The same experimental grid and completed cells are used throughout the confirmatory analysis; LODO validation partitions these observations by dataset rather than introducing a new experimental sample.

\subsection{Experimental design}
\label{sec:experimental_design}

The encoding procedure is deterministic: for a given input dataset and encoding configuration, the same data point is mapped to the same quantum representation. The encoding itself is therefore not randomized.

The five seeds randomize four components of the experimental pipeline: the train--test split, the random state of the logistic-regression solver, the subsampling used to estimate the interaction vector $\mathbf q(D,\phi)$, and the parameter initialization of the quantum classifier. Because the train--test split changes with the seed, the set of points presented to the encoding also changes. Encoding results are therefore cached with the seed as part of their key, despite the encoding map itself being deterministic.

For every dataset--encoding--ansatz--seed cell, we record the resulting quantum and classical performance together with the quantities required by the proposed predictor. These measurements form the common input table for both the Gaussian-process and gradient-boosting analyses.

\subsection{Measured quantities and prediction target}
\label{sec:measurements}

For each cell $(D,\phi,s)$, we record the quantum accuracy $a_q$, the classical reference accuracy $a_c$, empirical and generalization errors, gradient statistics, encoding cost, the normalized data representation $\mathbf v(D)$, and the circuit--data interaction vector $\mathbf q(D,\phi)$.

The primary prediction target is the budget-matched quantum edge. We model a fixed two-qubit error rate
\begin{equation}
\epsilon_2 = 10^{-3}
\end{equation}
and define the corresponding survival factor for a circuit containing $n_{2q}$ two-qubit gates as
\begin{equation}
F=(1-\epsilon_2)^{n_{2q}}.
\end{equation}
The budget-matched quantum accuracy is then
\begin{equation}
a_q^{\mathrm{budget}}
=
\frac{1}{2}
+
F\left(a_q-\frac{1}{2}\right),
\end{equation}
and the primary edge is
\begin{equation}
e_{\mathrm{budget}}
=
a_q^{\mathrm{budget}}-a_c.
\label{eq:budget_edge}
\end{equation}
Thus, the resource model modifies the quantum performance according to the specified two-qubit error budget while leaving the classical comparator unchanged. The resulting quantity is a budget-matched comparison target, not a direct estimate of performance on a particular hardware platform.

The raw edge,
\begin{equation}
e_{\mathrm{raw}}=a_q-a_c,
\end{equation}
and the Pareto-based edge are retained as secondary analyses, but the budget-matched edge in Eq.~\eqref{eq:budget_edge} is the target used for the primary integrative prediction test.

The predictor has two output heads. The first predicts the edge,
\begin{equation}
\hat e=\hat e(D,\phi),
\end{equation}
while the second stores the predicted generalization gap directly. The observed gap is defined as
\begin{equation}
g_{\mathrm{obs}}
=
\epsilon_{\mathrm{gen}}-\epsilon_{\mathrm{emp}},
\end{equation}
and the predictor returns a corresponding estimate $\hat g$. No empirical error is reintroduced as an independent predictor target.

\subsection{Predictor representation and architecture}
\label{sec:predictor}

The data representation is constructed from rank-normalized data descriptors and circuit--data interaction quantities. For each descriptor $f_i$, its value is converted to an empirical rank across the datasets in the outer training partition:
\begin{equation}
v_i(D)
=
\frac{
\left|
\left\{
D' \in \mathcal B_{\mathrm{train}} : f_i(D')\leq f_i(D)
\right\}
\right|
}{
|\mathcal B|
},
\qquad
v_i(D)\in[0,1].
\end{equation}
The resulting vector $\mathbf v(D)$ is combined with the interaction vector $\mathbf q(D,\phi)$ to form the joint representation
\begin{equation}
\mathbf z(D,\phi)
=
\left[
\mathbf v(D),\mathbf q(D,\phi)
\right].
\label{eq:joint_representation}
\end{equation}

The interaction vector contains measurements of circuit properties relevant to the proposed mechanism, including entanglement, kernel effective rank, expressibility, locality/lightcone width, and persistent-homology descriptors of the embedded point cloud. The topology quantities used in the original confirmatory implementation are retained for reproducibility; corrected topological constructions are evaluated separately in the topology audit.

The confirmatory predictor is implemented as the sparse variational multi-head Gaussian process (SVGPMultiHead), with 256 inducing points and a linear model of coregionalization (LMC) variational strategy over the two prediction heads. An exact Gaussian process is used only as a frozen reference implementation because an exact fit is computationally prohibitive at the full sample size. On an identical set of 800 training inputs, the sparse implementation reproduces the exact-GP predictions with Pearson correlation $r=0.984$ and RMSE $=0.026$. This comparison is an implementation-fidelity check rather than an independent measure of predictive generalization.

A gradient-boosting model is trained in parallel using the same representation and the same outer validation partitions. This provides a non-Gaussian reference model while holding the predictor inputs and cross-validation structure fixed.

\subsection{Cross-dataset validation}
\label{sec:validation}

Predictive performance is evaluated using leave-one-dataset-out cross-validation over the 27 datasets. In each of the 27 folds, all cells belonging to one dataset are held out, while the remaining 26 datasets provide the training set. The Gaussian-process and gradient-boosting predictors use identical outer train--test partitions.

All transformations that estimate dataset-level ranks are fitted using only the outer training rows and then applied to the held-out rows. The raw descriptor values $f_i(D)$ are computed independently of the fold, while the rank transformation defining $\mathbf v(D)$ is fitted using only the outer-training datasets and then applied to the held-out dataset. Hyperparameters that require model selection, including the loss weighting parameter $\gamma$, are selected within the training portion of each outer fold using inner LODO validation.

The predictor is evaluated only after the complete outer training procedure has
been fitted. Predictive performance is assessed on the held-out dataset in each
LODO fold using the coefficient of determination and expected calibration error
(ECE). For the primary edge prediction, the coefficient of determination is
computed as
\begin{equation}
R^2
=
1-
\frac{
\sum_{j\in\mathcal T_{\mathrm{test}}}
\left(e_j-\hat e_j\right)^2
}{
\sum_{j\in\mathcal T_{\mathrm{test}}}
\left(e_j-\bar e_{\mathrm{test}}\right)^2
},
\label{eq:r2}
\end{equation}
where $e_j$ is the observed budget-matched edge, $\hat e_j$ is the corresponding
prediction, and $\bar e_{\mathrm{test}}$ is the mean observed edge over the
held-out test set.

The ECE measures calibration of the predictor's uncertainty estimates by
comparing predicted and empirical coverage across the held-out observations.
The pre-specified H5 acceptance criterion requires
\begin{equation}
R^2 \geq 0.4
\qquad\text{and}\qquad
\mathrm{ECE}\leq 0.1.
\label{eq:h5_criterion}
\end{equation}

This procedure tests cross-dataset generalization rather than interpolation across cells from the same dataset. Because the 8100 experimental cells are fixed before the LODO procedure, each fold partitions the same completed experimental dataset into training and held-out datasets.

\subsection{Classical baselines and attribution analysis}
\label{sec:baselines}

The primary classical comparator is logistic regression trained on the original classical features using the same train--test partition as the quantum model. Its accuracy defines the pre-specified primary classical reference,
\begin{equation}
a_c=a_{\mathrm{LR}},
\end{equation}
and therefore determines both the raw and budget-matched primary edges.

As a sensitivity analysis, we evaluate a fixed library of six classical classifiers:
\begin{enumerate}
    \item logistic regression, with $C=1.0$ and $\mathrm{max\_iter}=2000$;
    \item $k$-nearest neighbors, with $k=5$;
    \item support-vector classification with an RBF kernel, $C=1.0$, and $\gamma=\mathrm{scale}$;
    \item a multilayer perceptron with hidden layers $(64,64)$, $\mathrm{max\_iter}=800$, and $\alpha=10^{-3}$;
    \item histogram gradient boosting with default hyperparameters;
    \item random forest with 300 trees and a single worker.
\end{enumerate}
All six models operate directly on the original classical features, use the same training and test partitions as the quantum experiment, and retain their fixed hyperparameters throughout the analysis.

Let $\mathcal M_6$ denote this library. For each dataset and seed, we define the best-of-six classical accuracy as
\begin{equation}
a_{\mathrm{6CL}}
=
\max_{m\in\mathcal M_6} a_m.
\end{equation}
The corresponding sensitivity edges are
\begin{equation}
e_{\mathrm{6CL,raw}}
=
a_q-a_{\mathrm{6CL}}
\end{equation}
and
\begin{equation}
e_{\mathrm{6CL,budget}}
=
a_q^{\mathrm{budget}}-a_{\mathrm{6CL}}.
\end{equation}
These quantities are not used to redefine the primary H5 target. Instead, the paired comparison tests whether predictor signal persists when the classical comparator has substantially greater model capacity. A signal that disappears against the best-of-six comparator is consistent with the predictor tracking failure of the logistic-regression baseline rather than a robust quantum advantage.

For binary classification, chance accuracy is
\begin{equation}
a_{\mathrm{chance}}=\frac{1}{2}.
\end{equation}
Chance is used as a reference for determining whether a model learns a predictive signal; it is not a competing classical baseline and is not subtracted from the primary quantum edge.

Classical baselines are trained on the original, unencoded features. This is the primary comparison: the classical comparator receives the raw data, whereas the quantum model receives the data after the specified encoding. Where we analyze the effect of the encoding itself, we additionally compare models under the same encoded representation; these are attribution analyses rather than replacements for the primary baseline.

Finally, the framework specifies an access-model-matched comparison for settings in which the classical and quantum methods do not receive the same effective information. Under sparse or structured access (Mode~S), the classical comparator may use the corresponding sparse representation. Under quantum-access assumptions (Mode~Q), a valid comparison requires a classical or quantum-inspired method with the corresponding sampling or query access. This condition prevents an apparent quantum edge from arising solely from asymmetric data-access assumptions. The confirmatory suite analyzed here does not contain Mode~Q cells; consequently, this comparison is treated as a framework boundary condition rather than as an empirical test in the present study.

\subsection{Reproducibility and analysis freeze}
\label{sec:reproducibility}

The confirmatory circuits, encoding maps, ansatz families, hyperparameters, random seeds, preprocessing operations, predictor architecture, and decision rules were frozen before held-out predictor performance was evaluated. When implementation defects were subsequently identified, the original confirmatory outputs were retained for reproducibility and the corrected analyses were reported separately as deviations or sensitivity analyses. No component of the predictor or benchmark was modified in response to the held-out results.

Three distinct sources of variability are considered separately. First, seed variation captures the complete experimental pipeline, including data splitting, parameter initialization, and interaction-vector estimation. Second, the exact-versus-sparse Gaussian-process comparison quantifies implementation fidelity on identical inputs. Third, LODO cross-validation measures generalization across datasets. These three layers are not treated as interchangeable estimates of uncertainty.

The resulting analysis therefore distinguishes variability within experimental cells, fidelity of the sparse predictor approximation, and cross-dataset predictive generalization. This separation is maintained throughout the confirmatory analysis.


\section{Results}
\label{sec:results}

\subsection{H5: the selector does not predict the edge}
\label{sec:results:h5}

\begin{table}[t]
\centering
\caption{Confirmatory H5 evaluation. Pipeline: logistic-regression
$\ebudget$, SVGP\,$+$\,GBM multi-head, leave-one-dataset-out over 27 folds,
inner $\gamma$-CV, and three-step EM. The pre-specified threshold is
$R^2 \ge 0.4$.}
\label{tab:h5}
\begin{tabular}{lr}
\toprule
Metric (GP head) & Value \\
\midrule
Pooled LODO $R^2$            & $-0.32$ \\
\quad (corrected-gap re-run) & $-0.37$ \\
Per-fold $R^2$, median       & $-1.12$ \\
Per-fold $R^2$, mean         & $-28.54$ \\
Fold-bootstrap 95\% CI       & $(-58.5,\, -4.7)$ \\
GBM head, pooled LODO $R^2$  & $-0.58$ \\
Expected calibration error  & $0.60$ \\
\midrule
\textbf{H5 pass} ($R^2 \ge 0.4$) & \textbf{False} \\
\bottomrule
\end{tabular}
\end{table}

The selector does not predict the budget-matched edge on held-out datasets.
Every aggregation falls far below the pre-specified threshold
(Table~\ref{tab:h5}); the fold-bootstrap upper bound of $-4.7$ is nowhere
near $0.4$. The best single fold is $+0.077$
(\texttt{synth-parity-3}); no fold reaches $+0.1$. The GBM head, an independent
model on the same representation, agrees ($-0.58$).

\begin{figure}[t]
\centering
\includegraphics[width=0.8\columnwidth]{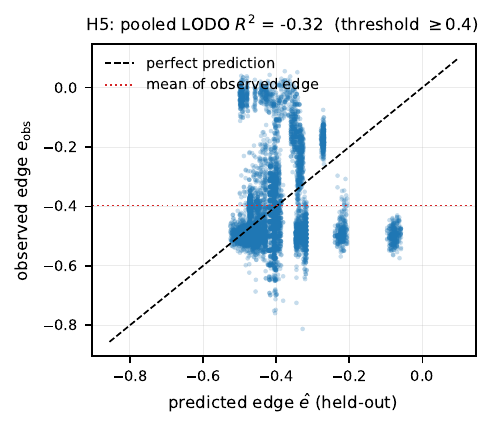}
\caption{H5, the central test. Predicted against observed budget-matched edge
on held-out datasets, pooled over all 27 leave-one-dataset-out folds. A
predictive selector would concentrate along the diagonal; the cloud instead
sits largely around the mean of the observed edge, corresponding to a pooled
$R^2=-0.32$.}
\label{fig:h5}
\end{figure}

We take the \emph{pooled} $R^2$ as the headline metric. Per-fold $R^2$ is
unstable here because several held-out datasets have near-zero edge variance,
on which $R^2$ becomes extremely negative (e.g.
\texttt{synth-klein-bottle} $-271$,
\texttt{synth-3sphere-pc} $-246$,
\texttt{synth-prng} $-150$). These values inflate the per-fold mean to
$-28.5$ without carrying additional information. The pooled statistic,
obtained by predicting every held-out point and scoring once, still shows
that the predictor performs worse than the intercept-only mean predictor (Fig.~\ref{fig:h5}).

\textbf{A negative $R^2$ is not a weak positive result. It means that, on this suite
and for this edge definition, the specified predictor does not provide a
reliable cross-dataset prediction of the measured quantum edge}. Under the
pre-specified criterion, H5 therefore fails. The remaining results decompose
this failure into the dependence of the measured edge on the classical
comparator, the loss of headroom introduced before the quantum model, and
the near-chance performance of the quantum model itself.

\subsection{The classical baseline determines the measured edge}
\label{sec:baseline}

The primary H5 analysis uses logistic regression as the pre-specified classical comparator. We next vary that comparator while leaving the quantum pipeline unchanged. This sensitivity analysis isolates how much of the measured edge and its predictability depends on the capacity of the classical reference.

The primary classical baselines therefore use the raw feature representation; encoded-feature comparisons are used only to isolate the contribution of the encoding.

\begin{figure}[t]
\centering
\includegraphics[width=\textwidth]{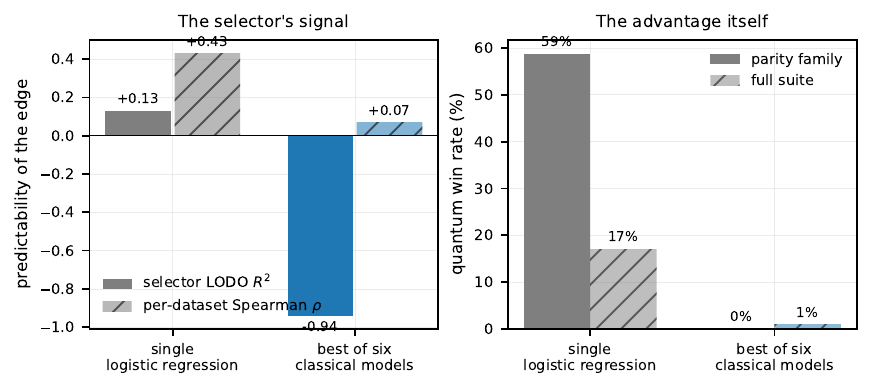}
\caption{The classical baseline determines the measured edge. Left: the selector's leave-one-dataset-out $R^2$ and per-dataset Spearman $\rho$ on the edge. Right: the quantum win rate on the parity family (solid) and across the full suite (hatched). The quantum pipeline is unchanged; only the classical comparator changes.}
\label{fig:baseline}
\end{figure}

Against logistic regression, the complexity selector gives an apparent positive held-out relationship: leave-one-dataset-out $R^2=+0.13$, with per-dataset Spearman $\rho=0.43$ (Fig.~\ref{fig:baseline}). Replacing the comparator with the best
performer among six fixed classical models removes both the apparent advantage and its predictability. The six-model library consists of logistic regression, $k$-nearest neighbors, SVC with an RBF kernel, a multilayer perceptron, histogram gradient boosting, and random forest, with the fixed
hyperparameters specified in Sec.~\ref{sec:baselines}.

All six models are trained on the same train--test partitions as the quantum experiment and are evaluated on the original classical features. They are fixed rather than tuned. The library is therefore a deliberately modest sensitivity comparator, not an estimate of the strongest classical performance achievable with unrestricted model selection.

On the full suite, the quantum win rate falls from $17\%$ against logistic regression to $1\%$ against the best of six. On the parity subset, where the framework predicts an edge, the win rate falls from $59\%$ to $0\%$. Correspondingly, the selector's $R^2$ falls from $+0.13$ to $-0.94$ and per-dataset Spearman $\rho$ from $0.43$ to $0.07$.

The endpoints alone could suggest that only taking the best of six removes
the apparent advantage. The individual-baseline sweep shows otherwise.
Table~\ref{tab:baseline-ladder} reports the win rate against each of the six
models on the same $8{,}100$ cells.

\begin{table}[tbp]
\centering
\small
\caption{Quantum win rate against each classical baseline taken \emph{individually}, over the
same $8{,}100$ Phase~B cells. A win is a strictly lower held-out error at matched
$(\text{dataset},\text{seed})$. Logistic regression is not merely the weakest of the six --- it
is a factor of three below the next weakest, and every other single model already places the
win rate under $6\%$. The headline therefore does not require taking the best of the set to
disappear; any of the other five would have removed it.}
\label{tab:baseline-ladder}
\begin{tabular}{@{}lrr@{}}
\toprule
baseline & full suite & parity family \\
\midrule
logistic regression & 17.0\% & 58.3\% \\
$k$-NN & 5.8\% & 0.0\% \\
MLP & 4.8\% & 0.0\% \\
SVC-RBF & 4.0\% & 0.0\% \\
random forest & 3.6\% & 1.8\% \\
HistGBM & 3.5\% & 6.8\% \\
\midrule
worst of six & 20.9\% & 62.4\% \\
\textbf{best of six} & \textbf{1.1\%} & \textbf{0.0\%} \\
\bottomrule
\end{tabular}
\end{table}

Logistic regression is a clear outlier: every other individual baseline
already reduces the full-suite win rate below $6\%$. On the parity subset,
$k$-NN, MLP, and SVC-RBF each give a win rate of exactly $0\%$ on their own.
The apparent $58$--$59\%$ quantum win rate therefore requires the logistic
baseline specifically.

The mechanism is especially transparent for parity. \textbf{Parity is a canonical
example of a task that a linear model cannot represent directly, while
nearest-neighbour and kernel methods can capture it. A quantum win rate
measured only against logistic regression can therefore reproduce the
linear model's shortfall rather than isolate a quantum contribution}. The
spread across individual baselines is itself a useful diagnostic: \textbf{a large
gap between one baseline and the rest should be treated as a warning before
any aggregate advantage is interpreted.}

\textbf{A quantum advantage reported against a single weak classical baseline does
not isolate a quantum contribution. A baseline sweep is therefore required
to establish whether the measured edge is robust to reasonable variation in
classical model capacity.}

\subsection{Classical preprocessing removes headroom}
\label{sec:encoder}

The baseline analysis identifies one source of the H5 failure. A second
mechanism occurs earlier in the pipeline: the classical preprocessing
required to place high-dimensional data on a few-qubit register can remove
the nonlinear structure that a downstream model could exploit.

\begin{table}[!htbp]
\centering\footnotesize
\setlength{\tabcolsep}{3pt}
\caption{\textbf{The cost of dimensionality reduction to run on a superposition.} Upper
block: garbage of the reduction map, enumerated exactly over the whole quantized input cube at
matched $(p,d,b)$, so both reducers discard the same $p-d$ dimensions. Feature selection attains
the floor $(p-d)b$ \emph{exactly} in every row --- it drops whole registers and touches nothing
else --- while an orthonormal rotation must round, and rounding destroys more than it was asked to.
The gap is small in the Landauer average and $2.1$--$2.3\times$ in the worst-case preimage, which
is what an ancilla count pays for. Lower block: reversible circuit resources at $b=8$. Selection is
a relabelling of registers; a rotation accumulates $p$ constant-multiplies per output, needing
$2b+\lceil\log_2 p\rceil$ bits of accumulator.}
\label{tab:reducercost}
\begin{tabular}{@{}l c c c c@{}}
\toprule
$(p,d,b)$ & floor & selection & rotation & max preim. \\
\midrule
$(5,2,3)$ & 9 & \textbf{9.000} & 9.220 & 1182 \\
$(6,2,3)$ & 12 & \textbf{12.000} & 12.216 & 9709 \\
$(6,3,3)$ & 9 & \textbf{9.000} & 9.352 & 2047 \\
$(7,2,3)$ & 15 & \textbf{15.000} & 15.177 & 68475 \\
\addlinespace
\midrule
dataset ($d{=}4$) & sel. & rot.\ anc. & garbage & overhead \\
\midrule
breast cancer & \textbf{0} & 84 & 52 & $2.6\times$ \\
wine & \textbf{0} & 80 & 48 & $2.5\times$ \\
digits & \textbf{0} & 88 & 56 & $2.8\times$ \\
\bottomrule
\end{tabular}
\end{table}

At matched $(p,d,b)$, both reducers discard the same $p-d$ dimensions, so the
minimum information loss associated with the discarded coordinates is fixed.
Feature selection attains this floor, $(p-d)b$ bits, in every row of
Table~\ref{tab:reducercost}. An orthonormal rotation must additionally
quantize the transformed coordinates, producing greater loss in the measured
preimage statistics. The circuit cost is also substantially different:
selection requires no gates or ancillas, whereas a rotation requires
additional accumulator resources. On a 32-qubit data register, the PCA
implementation requires 84 ancillas before the quantum encoder is applied.

\begin{table}[!htbp]
\centering\footnotesize
\setlength{\tabcolsep}{4pt}
\caption{\textbf{Headroom after reduction, and the controls that make the zeros
readable.} \emph{Headroom} is gradient boosting minus a linear probe on the same reduced features,
the accuracy a non-linear model --- quantum or otherwise --- could still be demonstrating something
on. The maximum is taken over widths $d$ from $2$ to the full feature count, so these are not
small-$d$ artefacts. On the benchmark datasets it is indistinguishable from zero at \emph{every}
width, including no reduction at all. The two positive controls are what license reading that as a
property of the data: the certified $[16,4,5]$ code data, whose label is a $\mathrm{GF}(2)$
combination no linear-over-$\mathbb{R}$ model can see, and parity-8. \textbf{Parity-8 also shows the
metric's limit} --- its headroom reads $0.004$ because gradient boosting fails on parity too, so
headroom near zero means \emph{either} no exploitable structure \emph{or} structure beyond the
non-linear baseline as well.}
\label{tab:headroom}
\begin{tabular}{@{}l l c c c@{}}
\toprule
dataset & kind & $p$ & max headroom & $d^{*}$ \\
\midrule
breast cancer & benchmark & 30 & -0.004 & --- \\
wine & benchmark & 13 & -0.011 & --- \\
digits & benchmark & 64 & +0.077 & 4 \\
certified $[16,4,5]$ & \textbf{control} & 16 & \textbf{+0.127} & 2 \\
parity-8 & \textbf{control} & 16 & \textbf{+0.126} & 3 \\
\bottomrule
\end{tabular}
\end{table}

The reduction leaves little nonlinear headroom.
Table~\ref{tab:headroom} reports headroom as the performance of gradient
boosting minus that of a linear probe on the same reduced features,
maximized over widths from $d=2$ to the full feature count. On breast cancer,
wine, and digits, the measured headroom is indistinguishable from zero at
every width, including with no reduction. PCA improves accuracy relative to
selection ($0.872$ versus $0.812$ averaged over
$d\in\{2,4,8\}$), but this gain is already available to a linear classical
model.

This provides a mechanism for the H5 null. If the classical reduction has
already exposed the task to a low-capacity model, there may be little
nonlinear structure left for the quantum model to exploit. This is
consistent with the baseline result: the apparent edge against logistic
regression is reduced when the comparator has access to nonlinear classical
capacity.

Now, the classical preprocessing versus the quantum feature map. The attribution should be restricted to the classical preprocessing part of the pipeline. A classical reduction, normalization, or coordinate
transformation can be handed to the classical comparator and therefore
cannot itself constitute evidence of quantum advantage. This does not rule
out an advantage arising from the subsequent quantum feature map: an
entangling quantum map may induce a classically difficult kernel. The
experiment here instead finds that, for the tested data and encoding library,
the quantum prediction axis remains near chance.

The distinction can be made explicit by adding and subtracting the
performance of a classical learner operating on the same preprocessed representation. 
The first term is obtainable by a classical party at the cost of applying the same classical preprocessing. Only the residual can support a claim specific to the quantum operation. This decomposition motivates the baseline-sensitivity analysis (see Proposition~1).

Data processing does not create label information.
For data $(X,Y)$ and an encoding channel $\mathcal{E}$ acting on $X$, $I\big(\mathcal{E}(X);Y\big)\leq I(X;Y).$ Thus an encoding cannot create label information absent from the input (see Proposition~2). A quantum feature map can nevertheless alter the computational difficulty of accessing information that is already present. \textbf{The relevant condition for advantage is therefore not quantum structure alone, but quantum structure aligned with label-relevant information.}

\paragraph{The encoder can only raise the classical floor.} The two measurements above, the information floor and the ancilla cost, describe what reduction \emph{does}. There is a shorter
statement of why it cannot do otherwise, and it is structural rather than empirical.

An encoder acting on classical data is itself a classical algorithm. Whatever it computes, a rotation, a projection, a coordinate selection, a normalization, is computation performed on a classical machine, and therefore computation the classical comparator could also have performed. It follows that the encoder cannot lower the quantum ceiling. It can only raise the classical floor.

This is why the effect presents as structure being revealed when it is not. A good encoding makes the data separable, and separability is exactly the property that makes a task classically easy. Linear discriminant analysis is the limiting case and the reason we exclude it below: it projects onto at most $C-1$ directions chosen to maximize between-class scatter, so whatever separation survives is close to trivial and any downstream classifier finds it. The encoder has done the work the model is
then credited with.

Two things follow that we adopt as method. First, a fairness rule: \textbf{the classical baseline must be evaluated on the encoded representation, not on the raw data}. Comparing a quantum model on
encoded features against a classical model on raw features charges the classical side for work the quantum side was given. Second, a scope condition: the argument holds when the encoding is a
classical transformation of classical data, which is the entire setting of this paper. It does not hold when state preparation itself performs something classically hard, or when the state does not originate from classical data, a device's output, an experimental sample, a sensor record. There the classical comparator has nothing to be handed, and the argument has no purchase.

One distinction has to be kept or the argument overreaches. It applies to the \emph{classical preprocessing} half of an encoding pipeline, the reducers, normalizations and coordinate choices
this paper measures, because those are classical maps and can be handed over intact. It does not apply to the quantum feature map itself: making the map more entangling can render the induced kernel harder to evaluate classically, and that is not a classical computation the comparator could have performed. Advantage from a feature map is not ruled out here; advantage from \emph{preprocessing} is.

With that scope, the consequence for the program is stronger than the null results that prompted it, and it is not that these pipelines were badly designed. \textbf{The classical-preprocessing stage of a quantum pipeline cannot place an advantage beyond classical reach}, because it is classical and can always be handed over; improving it improves the baseline it is measured against. Whatever survives a comparison run on the encoded representation is attributable to the quantum map, and whatever does not was never the model's to claim.

The complementary axis is not untested here, and its result is why this paper cannot speak to feature-map advantage either way. The Phase~B sweep varies the encoding over ten configurations in four families, angle, ZZ, amplitude, re-uploading, including genuinely entangling ZZ maps reaching $1.52$ nats of half-cut entropy. The axis \emph{collapsed}: mean $a_q$ spans $0.462$ to $0.497$ across all ten, a range smaller than the within-encoding spread of $0.05$--$0.09$, and the best encoding is scattered over nine of the ten configurations as expected from noise. Nothing distinguishes a product-state angle encoding from an entangling ZZ map because both sit at chance. So the residual term of Eq.~\eqref{eq:attribution} was measured and found indistinguishable from zero \emph{in this regime}; that is a statement about the regime, not about feature maps.

\paragraph{Proposition 1 (attribution decomposition).}
\label{prop:attribution} Let $P$ be a classical preprocessing map
computable in time $T_P$, $E$ an encoding, $Q$ a quantum learner and $C$ a classical learner class, and write $\mathcal{A}(\cdot)$ for performance under a fixed metric and split, abbreviating $QEP=Q\circ E\circ P$. Then

\begin{align}
&\underbrace{\mathcal{A}(Q E P) - \mathcal{A}(C_{\mathrm{raw}})}_{\text{reported gap}} \nonumber\\[2pt]
&\quad=\; \underbrace{\mathcal{A}(C P) - \mathcal{A}(C_{\mathrm{raw}})}_{\text{preprocessing-attributable}}
\;+\; \underbrace{\mathcal{A}(Q E P) - \mathcal{A}(C P)}_{\text{residual}}
\label{eq:attribution}
\end{align}

The identity is trivial, it adds and subtracts $\mathcal{A}(C\circ P)$, and we state it as a proposition because its content is not the algebra but the attribution it forces. The first term is obtainable by any classical party at cost $T_P$ per example, since $P$ is a classical map; it is therefore not available as evidence of quantum advantage. Only the residual is.

\emph{Corollary.} A comparison reporting only the left-hand side does not distinguish the two terms, and so upper-bounds rather than measures the quantum-attributable part. The decomposition is exhaustive, so the question ``which term did you measure?'' has an answer for every published gap,
and a study that cannot answer it has not isolated a quantum effect.

We adopt Eq.~\eqref{eq:attribution} as the reporting standard for the work that follows from this one. It is also the discipline behind Sec.~\ref{sec:baseline}: the best-of-six classical baseline evaluated on
the same reduced representation is precisely $\mathcal{A}(C\circ P)$, and the collapse of the measured edge when it replaces a raw-data logistic regression is the first term being paid back.

\paragraph{Proposition 2 (an encoding cannot manufacture label information).} Let $(X,Y)$ be the data and label random variables and let $\mathcal{E}$ be any encoding channel acting on $X$ alone, classical preprocessing, quantum feature map, or their composition. Then

\begin{equation}
I\big(\mathcal{E}(X);Y\big) \;\le\; I(X;Y),
\label{eq:dpi}
\end{equation}

by the data-processing inequality, which holds in the quantum case as well because $\mathcal{E}$ is a channel on $X$ and $Y$ is untouched. Equality requires $\mathcal{E}$ to be sufficient for $Y$.

The consequence separates two things that are routinely run together. A quantum feature map \emph{can} manufacture computational hardness: an IQP-type map induces a kernel believed hard to evaluate classically, and that hardness is a property of the map, obtainable from any input. What Eq.~\eqref{eq:dpi} forbids is manufacturing \emph{relevance}. The map cannot put label information into the representation that the data did not already carry; it can only make some of what is there harder for a classical party to reach, while, by the reducer measurements above, destroying some of the rest.

An advantage therefore needs both conditions at once: the induced kernel must be classically hard \emph{and} aligned with the structure the labels actually have. Hardness alone buys nothing, because a kernel that is expensive to compute and unrelated to the task is an expensive random kernel.

This is what the collapse of the encoding axis measures. The ZZ maps in the sweep produce genuinely entangled states, $1.52$ nats of half-cut entropy against a possible $2.77$, so quantum structure is present in the \emph{states}. It is absent from the \emph{predictions}, which sit at chance alongside the product-state angle encodings. Structure in the representation is not structure aligned with the task, and only the second is an advantage.

\paragraph{Why LDA is not among the reducers.} Linear discriminant analysis is the standard supervised answer and it is excluded on principle rather than on cost. It projects onto $C-1$ dimensions chosen to make the classes linearly separable, driving the headroom to zero \emph{by construction}: a threshold suffices afterwards, and a quantum model placed on top of it is demonstrating nothing. The measurement above says PCA does the same thing on these datasets, merely less deliberately.

\paragraph{The controls, and the metric's limit.} A headroom instrument that reads zero on everything cannot be told from a broken one, so Table~\ref{tab:headroom} carries two positive controls: the certified $[16,4,5]$ code data of the companion decoding-window work, whose label is
a $\mathrm{GF}(2)$ combination invisible to any model linear over $\mathbb{R}$, and parity-8. Both read large. Parity-8 also exposes the metric's limit and we state it rather than bury it: its headroom is $0.004$, indistinguishable from breast cancer's, because gradient boosting fails on parity as well. \textbf{Headroom near zero therefore means either no exploitable structure or structure beyond the non-linear baseline too}, and only a known Bayes floor separates the two, which is available for generated data and not for these benchmarks.

\subsection{H1 and H2: trainability and topology}
\label{sec:results:h1}

We estimated $\operatorname{Var}[g(\theta)]$ using the two-point
parameter-shift rule on the $Z_0$ observable, with 200 draws of encoded
state, $\theta\sim\mathrm{U}[0,2\pi]^P$, and parameter index per cell.
The confirmatory fixed-width grid contains $27\times8\times6\times5=6{,}480$ cells. The complexity vectors are the Phase~B vectors themselves, so the regressors are identical to those used by the selector. The width sweep reported below is exploratory and is kept separate from the confirmatory H1/H2 analysis (Table~\ref{tab:h1}).

\begin{table}[t]
\centering
\caption{H1 and H2. Fit of Eq.~\eqref{eq:bp} with $\vv$ rank-normalised on
training folds only; $R^2$ is pooled leave-one-dataset-out; coefficient
intervals are cluster bootstrap over datasets ($2{,}000$ resamples).
Reported coefficients are those of the design matrix, so the fitted value
on $nd$ is $-\alpha$ and on $nd\,v_i$ is $-\kappa_i$.}
\label{tab:h1}
\begin{tabular}{lr}
\toprule
Quantity & Value \\
\midrule
Pooled LODO $R^2$, full $\vv$        & $+0.026$ \\
Pre-specified threshold              & $0.70$ \\
$\hat\alpha$ (from $nd$ coefficient) & $-0.075$ \\
$\kappa$ with 95\% CI excluding $0$  & $2$ of $9$ \\
\quad higher-order MI                & $\kappa = +0.052$ \\
\quad MDL (autoencoder)              & $\kappa = +0.053$ \\
\midrule
$\Delta$AIC$, topology in vs.\ out$   & $+0.1$ \\
Pre-specified threshold               & $10$ \\
LODO $R^2$ without topology           & $+0.027$ \\
\midrule
\textbf{H1 pass} & \textbf{False} \\
\textbf{H2 pass} & \textbf{Inconclusive}$^\dagger$ \\
\bottomrule
\end{tabular}
\end{table}

\textbf{H1 fails.} Equation~\eqref{eq:bp} explains essentially none of the
held-out variance in $\log\operatorname{Var}[g]$: pooled LODO
$R^2=+0.026$ against a threshold of $0.70$. Two of nine components,
higher-order mutual information and MDL compression length, have $\kappa$
intervals excluding zero and both have the predicted positive sign. The
data-modulation terms are therefore detectable for these components, but
their magnitude is insufficient to make the pre-specified scaling relation
predictive at fixed circuit width.

The fitted relation is phenomenological rather than an exact consequence
of the circuit family. \textbf{The result therefore does not establish that data
complexity is absent from gradient statistics; it establishes that the
specified multiplicative correction does not provide the required
cross-dataset prediction on this grid}. Exact variance expressions can
depend on the representation channels of the encoded state and observable,
suggesting that a single scalar modulation of an $nd$ decay rate may be
too restrictive for this setting. This interpretation is post hoc and does
not alter the pre-specified H1 verdict.

\textbf{H2 is inconclusive.} The production estimator changes AIC by only
$+0.1$ and leaves held-out $R^2$ essentially unchanged. Audit of the
topological instrument found that its value was dominated by dimension-0
persistence, tracked ambient scale, used a truncated filtration, and
returned zero for a perfect circle (Appendix~\ref{app:topofix}). The
pre-specified test therefore did not measure the intended topological
quantity reliably.

We rebuilt the estimator using untruncated Rips persistence, excluding
$H_0$ and normalizing bars by point-cloud diameter, and validated it on
shapes with known Betti numbers. Re-running H2 with these corrected
summaries gives an estimator-dependent verdict: the threshold-free
summaries fail, whereas a bar-count summary passes only in the narrow
significance-cut range $\theta\in[0.08,0.10]$
(Table~\ref{tab:h2fix}). Because that choice is informed by the observed
outcome, it cannot be promoted to a confirmatory result.

We therefore retain H2 as inconclusive and treat the repaired analysis as
exploratory. Repairing an implementation defect is distinct from selecting
among repaired definitions after observing the outcome. The defensible
conclusion is that this experiment does not determine whether topology
contributes to barren-plateau prediction. Since the frozen $\vv$ contains
no explicit curvature component, H2 reduces on this suite to a test of
topology alone.

\begin{table}[t]
\centering
\caption{H2 re-run on the corrected topological component ($n=8$ grid).
The production estimator reproduces the original values, while corrected
summaries give an estimator-dependent result.}
\label{tab:h2fix}
\begin{tabular}{lccc}
\toprule
Topology summary & $\Delta$AIC & $\kappa_{\text{top}}$ (95\% CI) & H2 \\
\midrule
production (unnormalised, $H_0$ in) & $+0.1$ & $+0.014\ (-0.022,+0.046)$ & fail \\
corrected, $\ell_2$                 & $+0.5$ & $+0.018\ (-0.010,+0.070)$ & fail \\
corrected, max bar                  & $-1.7$ & $-0.004\ (-0.027,+0.026)$ & fail \\
corrected, count $\theta=0.08$      & $+11.5$ & $+0.056\ (+0.025,+0.093)$ & pass \\
corrected, count $\theta=0.10$      & $+15.4$ & $+0.073\ (+0.042,+0.115)$ & pass \\
corrected, count $\theta=0.15$      & $-1.7$ & $-0.016\ (-0.142,+0.054)$ & fail \\
\midrule
\multicolumn{4}{l}{\emph{Threshold for a pass:} $\Delta$AIC $\ge 10$.} \\
\bottomrule
\end{tabular}
\end{table}

\subsection{Why the primary test had little quantum variation to detect}
\label{sec:results:degenerate}

The H5 target is informative only to the extent that the quantum model
produces variation in performance beyond the classical comparator. Across
the $8{,}100$ cells, quantum test accuracy is
$0.481\pm0.069$, with median $0.493$. The model is at or near chance
($a_q\geq0.45$) in $81.5\%$ of cells and within one percentage point of
chance ($a_q\geq0.49$) in $55.2\%$. In contrast, the best of the six
classical models reaches $a_c\leq0.05$ error in $68.9\%$ of cells.

With $a_q$ concentrated near chance, the measured edge is largely determined
by the classical term:
\begin{equation}
e \approx \frac{1}{2}-a_c.
\end{equation}
Consistently, the edge correlates strongly with classical accuracy,
$\rho(e,a_c)=+0.913$, whereas
$\rho(e,a_q)=-0.337$. Approximately $82\%$ of the variance in the measured
edge is attributable to the classical term under this decomposition. \textbf{The
result explains why a data-complexity vector can appear predictive against
a weak linear baseline while failing against stronger classical models.}

The null is nevertheless narrow. None of the tested alternatives predicts
the edge robustly: the nine complexity components give $R^2=-0.111$,
sample size $-0.138$, dimension $-0.192$, their combination $-0.266$, and
best-classical error $-0.078$. The best individual complexity component
reaches only $-0.003$ and the best pair $-0.016$. Predicting $a_q$ itself
also fails ($R^2=-0.746$ for the linear model and $-0.078$ for boosting),
with only $\operatorname{sd}(a_q)=0.020$ across datasets.

The appropriate conclusion is therefore conditional rather than universal:
on classical datasets for which the tested quantum model remains at chance,
there is no robust quantum edge for the complexity selector to rank. \textbf{This
does not establish that data complexity can never predict a quantum
advantage; it establishes that the present quantities do not do so in a
regime where the quantum performance itself has little variation.}

\subsection{Exploratory circuit-width scaling}
\label{sec:results_nscaling}

Because the fixed-width Phase~B grid cannot identify the width dependence
of Eq.~\eqref{eq:bp}, we collected an exploratory width-scaling arm at
$n\in\{4,6,8,10,12,14\}$ on a stratified nine-dataset subset, all five
seeds, for $12{,}960$ cells. This arm tests the functional form of the
scaling relation rather than a pre-specified hypothesis.

\begin{table}[t]
\centering
\caption{Functional form of the width dependence on the exploratory
$n$-scaling arm. Lower AIC is better.}
\label{tab:form}
\begin{tabular}{lrr}
\toprule
Model for $\log\operatorname{Var}[g]$ & $R^2$ & AIC \\
\midrule
$1+nd$          & $+0.049$ & $19837$ \\
$1+n+d$         & $+0.164$ & $18175$ \\
$1+n$           & $+0.164$ & $18174$ \\
\textbf{$1+\log n$} & $\mathbf{+0.171}$ & $\mathbf{18054}$ \\
$1+n+n^2$       & $+0.171$ & $18068$ \\
\bottomrule
\end{tabular}
\end{table}

Replacing $nd$ by $\log n$ improves AIC by $1783$, strongly favouring the
log-width form within this exploratory arm. Adding depth to the width-only
model does not improve AIC; at $n=12$, mean
$\log\operatorname{Var}[g]$ is $-5.85$ for both $L=1$ and $L=3$.

\begin{figure}[t]
\centering
\includegraphics[width=0.8\columnwidth]{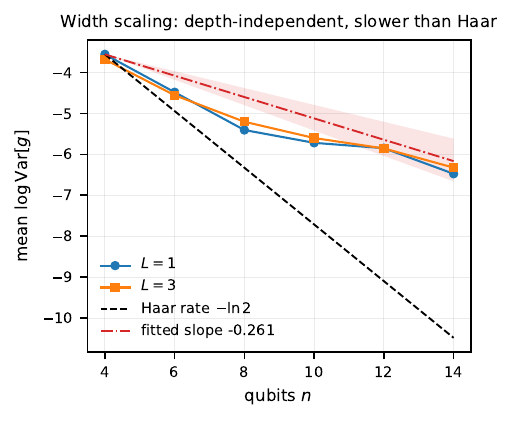}
\caption{Mean $\log\operatorname{Var}[g]$ against circuit width, by depth.
The measured decay is substantially slower than the Haar-random reference.}
\label{fig:width}
\end{figure}

The corresponding decay is approximately
\[
\operatorname{Var}[g]\sim n^{-2.15}.
\]
For comparison, forcing a linear-in-$n$ form gives a slope of $-0.261$ with
cluster-bootstrap 95\% CI $(-0.311,-0.206)$, excluding the Haar-random rate
$-\ln2=-0.693$. Mean $\log\operatorname{Var}[g]$ falls from $-3.55$ at
$n=4$ to $-6.47$ at $n=14$ for $L=1$ (Fig.~\ref{fig:width}).

These observations do not contradict asymptotic barren-plateau results
for deep, sufficiently random circuits. They instead show that the
finite-width regime tested here is not described by the particular
$nd$ phenomenological form used in H1. \textbf{For shallow circuits with local
observables and encoded real data, the leading dependence can differ from
the asymptotic Haar-random scaling.}

\subsection{Refitting the complexity correction on the corrected baseline}
\label{sec:results:h1refit}

The exploratory width-scaling result raises a direct question: \textbf{does the
complexity correction become predictive once the leading width dependence
is corrected?} We therefore refit the correction on the $n$-scaling arm,
using the $1+\log n$ baseline. This analysis is exploratory (Table~\ref{tab:form}).

Adding $\kk^\top\vv$ to $1+\log n$ improves AIC by $492.5$. The apparent
in-sample strength is misleading because $\vv$ is a dataset-level covariate:
the nine-dataset arm contains only nine distinct values of $\vv$ across
$12{,}960$ records. AIC therefore counts thousands of repeated circuit
observations as independent evidence for a predictor whose effective
sample size is nine datasets.

The consequence is visible under LODO validation. The LODO $R^2$ falls from
$+0.158$ for the $1+\log n$ baseline to $-32.9$ with the full complexity
vector. A one-component model gives only $+0.0115$ and the best pair
$+0.0143$, both selected post hoc. Randomly reassigned complexity vectors
give a null best-of-nine gain with mean $+0.0067$ and 95th percentile
$+0.0224$, placing the observed one-component gain at $p=0.235$.

Thus the H1 null is not explained solely by the original width baseline.
Even after correcting the leading scaling form, the complexity vector does
not add robust held-out predictive information in this nine-dataset arm.
More generally, dataset-level predictors are constrained by the number of
independent datasets, not by the number of circuit and seed observations.

\subsection{Two axes that carry no variance in \texorpdfstring{$a_q$}{aq}}
\label{sec:results:aqwidth}

The degeneracy of Sec.~\ref{sec:results:degenerate} is the paper's central mechanism, and the
obvious objection to it is that $n=8$ was a pre-specified choice made before anything was
known. If quantum test error were width-dependent, the whole H5 analysis would be a statement
about one circuit size. We therefore repeated the Phase~B protocol at $n \in \{4,6,10\}$ on the
nine-dataset subset, restricted to the eight non-reuploading encodings, and recomputed the
$n=8$ baseline the same way so the arms are comparable ($6{,}480$ additional cells). This arm
is \emph{exploratory} in the specification's sense, it was collected after the fixed-width
result was known.

\begin{table}[tbp]
\centering
\small
\caption{Quantum test error does not unpin with circuit width. All four arms use the same
nine-dataset subset and the same eight non-reuploading encodings, so the rows are
like-for-like; $n=8$ is the frozen grid restricted the same way. \emph{Exploratory}: these
data were collected after the fixed-width result was known and are reported in the register of
Sec.~\ref{sec:results_nscaling}, not as a confirmatory test. The pre-declared reading was that
$a_q$ counts as unpinned if the across-dataset spread rises above about $0.05$ with the
near-chance fraction falling materially below $0.8$; no width approaches either.}
\label{tab:aq-width}
\setlength{\tabcolsep}{4pt}
\begin{tabular}{@{}lrrrr@{}}
\toprule
width & cells & mean $a_q$ & \shortstack[r]{sd across\\datasets} & \shortstack[r]{fraction\\$a_q \ge 0.45$} \\
\midrule
$n=4$ & 2160 & 0.492 & 0.014 & 0.850 \\
$n=6$ & 2160 & 0.493 & 0.016 & 0.856 \\
\textbf{$n=8$} & 2160 & 0.484 & 0.019 & 0.827 \\
$n=10$ & 2160 & 0.484 & 0.019 & 0.825 \\
\bottomrule
\end{tabular}
\end{table}

The pre-specified reading was that $a_q$ counts as unpinned if the across-dataset spread rises
above about $0.05$ with the near-chance fraction falling materially below $0.8$.
\textbf{No width approaches either threshold} (Table~\ref{tab:aq-width}). The spread is $0.014$--$0.019$ throughout and
$n=10$ lands exactly on the $n=8$ baseline; the near-chance fraction never drops below $0.825$.
Quantum test error is pinned at every width tested.

This matters more in combination than alone. We already reported that the \emph{encoding} axis collapses: mean $a_q$ spans only $0.462$ to $0.497$ across all ten encodings, a range smaller than the within-encoding spread, with no encoding winning systematically. The width sweep adds a second, independent axis with the same outcome. \textbf{Neither the representation chosen nor the circuit size moves $a_q$ off chance on this suite.}

We also checked a third candidate explanation and closed it. \citet{appel2026fractal} shows that angle-encoded quantum kernels collapse once the feature map is \emph{wider than the intrinsic dimension of the data}, with the correlation fractal dimension $D_2$ giving the threshold. Several suite datasets are low-dimensional manifolds in higher ambient space (\texttt{synth-2sphere} measures $D_2 = 2.00$, \texttt{synth-torus} $2.14$), so at $n=8$ they are encoded three to four times wider than their intrinsic dimension. But the suite is not uniformly over-encoded, \texttt{real-higgs} ($D_2 = 10.08$) and \texttt{adv-encoded-random} ($11.03$) are \emph{under}-encoded; and the two groups are indistinguishable in $a_q$ ($0.480$ over-encoded, $0.492$ under-encoded), with $\mathrm{corr}(D_2, \overline{a_q}) = +0.18$ across 27 datasets. Encoding width relative to intrinsic dimension does not explain the pinning either.

It licenses the generalization that the fixed-width design could not support: \textbf{the degeneracy is a property of the data and the model class, not of the $n=8$ choice}, and the H5 verdict is not an artifact of a single circuit size.

It does \emph{not} license the stronger reading that representation is irrelevant. \citet{appel2026fractal} finds feature-map geometry determines whether a quantum kernel retains any information at all, and \citet{peng2026interaction}
finds a map aligned to the data's interaction structure beats classical kernels outright. The
defensible statement is narrower and more useful: \textbf{for the tested classical preprocessing and encoding library, the experiments provide no evidence that changing representation creates a quantum edge when the resulting structure remains accessible to the classical comparator.} On the spectral reading these encodings all deliver the same low-degree content to the model, and a quantum model given low-degree structure has nothing a classical model lacks. \textbf{That is a claim about the interaction of representation with} \emph{this data}\textbf{, and it predicts its own reversal: change the data so the signal sits only at high degree, and the picture should invert. The next study does exactly that.}

\subsection{H3: expressibility--data alignment}
\label{sec:results:h3}

\emph{The expressibility--data alignment result is similarly narrower than its primary criterion suggests.} The corrected analysis satisfies the pre-specified coefficient criterion: the fitted slope is positive with a confidence interval excluding zero. However, the alignment head has no held-out predictive power, with mean $R^2=-0.013$ for the GP and $-0.007$ for GBM. The positive coefficient therefore establishes an association with the level of the observed generalization gap, not a useful cross-dataset predictor of its variation. The secondary rank-correlation criterion could not be evaluated because the required per-point alignment quantities were not serialized. This is an instrumentation limitation rather than a null result, and it prevents a stronger interpretation of H3.

H3's secondary criterion, Spearman $\rho\geq0.3$ between predicted alignment and observed negative gap, could not be computed from the stored confirmatory outputs because the per-point alignment scores and observed gaps were not serialized. We record this as an instrumentation failure, not as a null result. The serializer has since been fixed. The negative gap $R^2$ nevertheless provides direct evidence that the alignment head did not generalize.

\paragraph{Correction of the gap convention.}
The initial confirmatory run reported $\beta=0$ on every fold and a gap $R^2\approx-300$. The discrepancy arose because the implementation treated the already-computed observed gap as though empirical error still had to be subtracted. The corrected implementation returns the gap directly, matching the stored $g_{\mathrm{obs}}$. A synthetic validation recovered the expected
relationship with gap $R^2=+0.86$.

The correction does not alter the H5 conclusion. The GBM edge is bit-identical between the two runs (pooled $R^2=-0.579$), while the jointly fitted GP edge moves from $-0.322$ to $-0.370$; both remain far below the H5 threshold of $0.4$.

H3 therefore passes its literal primary criterion but not its intended predictive interpretation. \textbf{This illustrates the distinction between a coefficient having the expected sign and a model whose corresponding quantity generalizes across datasets.}

\subsection{H4: per-feature encoding optimality}
\label{sec:results:h4}

H4 is not tested because it is conditionally downstream of H5. The
hypothesis asks whether ansatz configurations that route each feature to its predicted-optimal encoding outperform monolithic encodings. The routing rule itself is produced by the selector. With pooled held-out $R^2=-0.32$, the selector provides no demonstrated basis for choosing those encodings on unseen datasets.

Running H4 despite this failure would conflate two questions: whether mixed encoding can help at all, and whether the selector can identify when it should help. The former can be tested with an empirical oracle routing, which establishes the achievable ceiling; the latter should only be tested against that ceiling once a valid selector exists. We therefore leave H4 untested under the pre-specified hierarchy.

\subsection{H6: dequantization stratification}
\label{sec:results:h6}

H6 is not evaluable because the confirmatory suite contains no Mode-Q cells. The access-model condition is nevertheless important: an apparent quantum advantage is not meaningful if the quantum method receives a stronger data-access primitive than its classical comparator. Mode-Q would therefore require a classical or quantum-inspired comparator under the corresponding sampling or query access model. The present suite uses Mode-S or Mode-N assumptions, so H6 remains a framework boundary condition
rather than an empirical result.

\subsection{A designed quantum-signature control}
\label{sec:positive_control}

The H5 null could have two interpretations: either the proposed predictor fails even when a quantum-sensitive signal exists, or the confirmatory suite contains little quantum structure for the predictor to rank. We distinguish these possibilities with a designed quantum-signature control.

\begin{figure}[t]
\centering
\includegraphics[width=0.8\columnwidth]{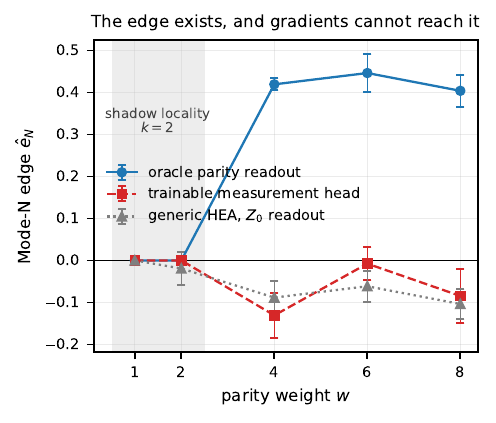}
\caption{Designed quantum-signature control as a function of parity weight $w$. The structure-aware measurement turns the edge on once the parity exceeds the $k=2$ shadow locality, while the generic and trainable readouts remain near chance. The control demonstrates that a baseline-robust quantum-sensitive signal can exist without being discoverable by a generic gradient-trained model.}
\label{fig:positive_control}
\end{figure}

We construct a shadow-hard parity probe in which states are labelled by a weight-$w$ global parity and sampled within the corresponding eigenspace. The classical comparator uses the best of the same six models applied to low-order ($k\leq2$) classical-shadow features. For $w\leq2$ the edge is approximately zero; for $w\geq4$ a structure-aware measurement produces an edge of approximately $+0.42$, with $a_q\approx0$ against $a_c\approx0.42$ (Fig.~\ref{fig:positive_control}).

The control separates the existence of a quantum-sensitive signal from the ability of a generic trainable model to discover it. A generic hardware-efficient ansatz with $L=3$ and a $Z_0$ readout remains near chance. A trainable measurement-selection head that must discover the parity support also fails for $w\geq3$. At $w\geq4$, neither trainable model captures the available $+0.42$ edge.

This control has a deliberately narrow purpose.  It shows that the
constructed instance contains a quantum-signature structure for which a structure-aware measurement yields an observable edge, while the generic models used here do not learn that structure.  It is not a classical-versus-quantum separation result.  In particular, the present comparison does not equip a classical method with explicitly constructed higher-order parity features, the same prior knowledge of the planted structure, or a matched access and measurement/query budget.

A stronger parity benchmark would pre-specify these matching choices: the classical feature construction and its order, the structural information supplied to each procedure, the raw-data access model, and a comparable measurement or query budget.  Such a study would answer a different question from the present control and is left for future work; we therefore make no claim that the observed edge persists under those stronger classical comparisons.

\textbf{The result should be read as a control, not as a general theorem about
variational quantum algorithms. Here the quantum edge is accessible when
the relevant measurement is specified and disappears when the same
structure must be inferred by the trainable models.} This distinguishes two
questions that are otherwise conflated: whether quantum-sensitive
structure exists in the data, and whether a generic learning procedure can
discover the operation that exploits it.

Taken together, the confirmatory and control experiments support a narrower
conclusion than either a universal null or a universal advantage claim.
\textbf{On the classical benchmark suite, the quantum model is near chance and the
measured edge is strongly determined by the classical comparator. A
baseline-robust edge can nevertheless be constructed when the data contain
a deliberately inaccessible global structure. The central limitation is
therefore not the logical possibility of a quantum edge, but the absence of
a robust, learnable quantum quantity for the proposed data-complexity
selector to predict in the tested regime.}



\section{Discussion}
\label{sec:discussion}

\subsection{Analysis of the v1 approach}
\label{sec:ledger}

This part is intended to provide information about what v1 proved to be correct and what was discarded.  

\paragraph{What survived.}
\emph{The quantities are real and measurable.} All nine complexity components were estimated across the broader $32$-dataset frozen registry with reported uncertainties, and the estimators were validated against closed-form testbeds before the confirmatory experiment. The confirmatory predictor, however, was evaluated on the $27$-dataset A/C/D suite defined in Sec.~\ref{sec:methodology}. v1's operational claim, that one can assign quantitative descriptors to the structural, statistical, algorithmic, and topological richness of a dataset, therefore holds up, even though the validity of any particular descriptor as a predictor of quantum advantage remains an empirical question.

\emph{The framework was falsifiable, and that is not a small thing.} v1 made claims sharp enough to be tested and potentially rejected, which is why this paper can distinguish between quantities that survived implementation and those that did not~\citep{bowles2024benchmarking, yu2026qmlvscml,leither2026oncology}. The resulting analysis also makes explicit where the failure arises: from the functional form, from the estimator, from the predictive model, or from the absence of useful quantum variation in the tested regime.

\emph{The embedding-dependence observation was correct.} v1 noted that ``the perceived complexity of the same classical dataset is not intrinsic, but depends on the choice of feature map.'' That observation is right, and it motivates the $\vv$/$\qq$ separation of Sec.~\ref{sec:framework}; only its formalization as a property of $D$ alone was mistaken. The present results reinforce the distinction: $\vv(D)$ describes dataset-level structure, whereas $\qq(D,\phi)$ captures properties that arise only from the interaction between a dataset and a particular encoding--circuit configuration.

\emph{The direction of the data-modulation claim has support.} Two of nine complexity sensitivities are significantly non-zero \emph{with the sign v1 predicted}, higher-order mutual information and MDL compression
length, both with $\kappa_i > 0$, meaning higher complexity implies a stronger plateau. v1's intuition that data modulates trainability is not empty.

\paragraph{What did not survive.}
\emph{The scalar composite} (Eq.~\eqref{eq:classical_comp}, Eq.~7 of v1) is dimensionally incoherent and unidentifiable, and \emph{we withdrew it before testing it} (Sec.~\ref{sec:framework}). The same applies to the generalization model of Eq.~\eqref{equation_expressivity} (Eq.~15 of v1), whose absolute difference between an unbounded KL divergence and a normalized scalar was not a testable quantity. These were corrected by preparing the experiment, not by running it.

The quantum-native counterpart is different. Unlike the classical composite, it was made well-posed, implemented, and evaluated exploratorily on ten quantum-native ensembles (Sec.~\ref{sec:framework:cquant}). The resulting equal-weight composite is nevertheless dominated by a single component: effective rank accounts for $99.97\%$ of its variance. The individual components retain substantial discriminatory information after rank normalization, so the experiment does not show that the underlying quantum descriptors are uninformative. Rather, it shows that their equal-weight scalar aggregation is not a stable measure of quantum data complexity. We consequently replace $C_{\mathrm{quant}}$ by the rank-normalized vector $\vv_N(\mathcal{E})$, while leaving its predictive relationship to quantum advantage untested because the quantum-native arm was not part of the H5 benchmark.

\emph{The data-gated barren-plateau formulation did not survive.} The v1 expression in Eq.~\eqref{eq:variance} fails theoretically because its data-dependent form does not recover the data-independent circuit-ensemble limit on trivial data. The revised expression in Eq.~\eqref{eq:bp} also fails its empirical criterion: over the exploratory width range $n=4$ to $14$, gradient variance is better described by a power law than by the proposed exponential-in-width form, with little additional explanatory contribution from depth in the tested regime. This does not test or exclude barren-plateau phenomena generally. Rather, it shows that the particular finite-width functional form proposed here is not an adequate description of the measured regime.

\emph{The original topology result is withdrawn; the underlying question remains open.} Version~1 gave topological complexity a load-bearing role, extended in Eq.~\eqref{eq:variance_topo}. The pre-specified test appeared to refute that contribution, but the estimator behind that number was subsequently found to contain implementation defects: it was sensitive to ambient scale and dispersion, used a truncated filtration, and combined edge collapse with that truncation. We therefore withdraw the original inferential result rather than treating it as evidence against topology.

The corrected estimator removes these defects and gives a genuinely estimator-dependent picture. Threshold-free summaries do not provide a stable contribution, whereas significance-thresholded bar counts pass the AIC criterion for some thresholds. Because the choice among these summaries cannot be made from the present data without introducing post-hoc selection, the corrected analyses are reported as sensitivity analyses rather than as a replacement pre-specified H2 test. The curvature component introduced in Sec.~\ref{sec:framework} is likewise untested here by construction. Thus the broader question of whether geometric structure provides a stable contribution to trainability or quantum advantage remains open, and requires a dedicated study with the geometric estimators and hypotheses fixed in advance.

We flag it for a \emph{dedicated} study rather than for the next one, because the two halves fail for different reasons and neither is repaired by more of what this suite contains. The topological half is a \textbf{specification} problem: on the corrected estimator the verdict depends on
which summary is chosen, threshold-free or significance-thresholded, and that choice must be fixed in advance, it cannot be adjudicated post hoc by us or by anyone reading these data. The curvature half is a \textbf{data} problem: $v_5$ needs graph- or hypergraph-native inputs, which a point-cloud suite structurally cannot supply, so no reanalysis of the frozen grid will ever exercise it. A study that fixes both needs a different instrument and a different corpus from the one
built here, which is why we do not fold it into the roadmap below as an increment.

It is nonetheless committed work rather than a gesture: the program this paper belongs to is developing it as a dedicated geometry-and-topology study, carrying forward the corrected estimator and the shape validation established here (Appendix~\ref{app:topofix}) and adding the graph-native corpus this suite lacks.
Whether it appears as a direct continuation of this work or as a standalone paper is not yet fixed, and we prefer to leave that open rather than announce a sequence we may not keep: what comes next is driven by which instrument the preceding result demands, not by a plan settled in advance.

\emph{The central predictive claim does not survive the confirmatory test.} The proposed complexity representation does not predict the budget-matched quantum edge across the $27$-dataset leave-one-dataset-out benchmark. The pre-specified pooled $R^2$ is $-0.32$, far below the required $0.40$, and the independent gradient-boosting sensitivity model gives $R^2=-0.58$. Replacing the defective topological component with corrected alternatives does not rescue the result; the corrected full rerun gives pooled $R^2=-0.451$ for the GP and $-0.674$ for GBM.

The failure is informative because the edge itself is highly structured. Quantum performance is near chance in most cells, while the measured edge correlates strongly with the classical comparator, $\rho(e,a_c)=+0.913$, and approximately $82\%$ of its variance is attributable to the classical term under the reported decomposition. The result therefore does not show that data complexity is irrelevant to quantum learning. It shows that, in the present benchmark and access regime, the proposed complexity quantities do not provide a robust cross-dataset signal from which to rank quantum edge.

\paragraph{The pattern, stated plainly.}
The v1 \emph{observations} were largely sound; its \emph{functional forms} were not. Every specific equation proposing how complexity enters, the linear composite, the exponential data-gated plateau, the absolute-difference alignment, was either dimensionally incoherent, contradicted by established theory, or empirically unsupported, while the underlying observations that motivated them survive. The framework's failure mode was
that its quantities were real and its equations were guessed.

There is a methodological corollary worth recording. Three of the six issues in Sec.~\ref{sec:framework} were found by \emph{implementing} the framework, before any data was collected: dimensional analysis of the composite, correlation structure among components, and the data-free limit. Implementation is a form of theory criticism.

\subsection{The four lessons, as recommendations}
Each lesson converts into a concrete requirement on future work.

\paragraph{Lesson 1 $\Rightarrow$ report a baseline sweep, not a baseline.}
Any advantage measured against a single classical model should be treated as unquantified until the baseline is varied. We recommend reporting the edge against a \emph{declared baseline sweep}, including a strong comparator such as the best of a fixed standard set, and publishing the per-baseline breakdown, so that a reader can see how much of the claimed edge is model-selection slack. \textbf{Had we not varied the baseline, this paper would have reported a positive, pre-specified, reproducible, and spurious result.}

\paragraph{Lesson 2 $\Rightarrow$ separate the existence of an edge from its reachability.} Our positive control shows these come apart cleanly: the edge is present at the level of the measurement but absent from the gradient-trained models tested here. \textbf{A selector trained on gradient-trained outcomes therefore has a target that is systematically missing exactly the advantages worth having.} Advantage claims should state which of the two they concern.

\paragraph{Lesson 3 $\Rightarrow$ vary circuit width, and fit the regime you are in.} Benchmark suites at a single width cannot test the functional form of any scaling law in $n$; ours could not, until we added an arm for it.
Doing so overturned the form: decay is power-law in width, depth-independent, and slower than Haar-random. \textbf{Any framework predicting advantage from circuit-scaling arguments should be fit in the finite regime rather than
extrapolated from asymptotic results}, and should report the width sweep that licenses its form.

\paragraph{Lesson 4 $\Rightarrow$ check that your axes carry variance before fitting a model on them.} A framework is only tested on the axes along which its target actually varies. Ours did not vary: the quantum model sat at chance in $81.5\%$ of cells, so the edge collapsed onto classical error
($\rho(\hat e, a_c) = +0.913$), and the encoding axis carried $0.5\%$ of the
variance against $83.8\%$ between datasets. The consequence is that $\qq(D,\phi)$ was never given a chance to matter, and the null on $\vv(D)$ is a statement about a regime rather than about complexity. Before fitting a
predictor, report the variance decomposition of the target over every axis the model uses,  and it distinguishes a hypothesis that failed from one that could not have been tested.

\paragraph{And a corollary on where to look next.} The only baseline-robust edge we found was engineered to be non-local in the measured observables, and its onset tracked that locality precisely. This suggests the informative axis is the locality structure of the label--observable relationship, not
aggregate data complexity, a sharper and more testable target than the one v1 proposed.

\subsection{Another interpretation, and what it would change}
\label{sec:interpretation}

The results above are reported as a falsification, and that reading stands:
the selector does not predict the edge, and none of the hypothesis outcomes is rescued by a more careful analysis. But there is a second reading of the same evidence which does not weaken the null, and which points somewhere
different. We set it out here rather than in the conclusion, because it is an interpretation rather than a result.

\textit{The quantity we chose to predict is a comparison}. The primary edge
\begin{equation*}
  e_{\mathrm{budget}}
  =
  a_q^{\mathrm{budget}}-a_{\mathrm{LR}}
\end{equation*}
is, by construction, a difference between two rival systems. Section~\ref{sec:baseline} shows what follows: widening the classical model set moved the selector's leave-one-dataset-out $R^2$ from $+0.13$ to $-0.94$ without changing a single quantum measurement. The predictand was not a property of the data, or of the circuit, or even of the pair, it was a property of the comparison, and the comparison moved.

The six-model comparison is a fixed-library sensitivity analysis, not a search for the best attainable classical benchmark.  Its purpose is to test whether the apparent selector signal survives a broader, standard set of classical references under the specified protocol.  It does not establish the result for a hyperparameter-tuned classical frontier. Conversely, because no modest tuning study was run, we do not infer the direction or size of any further change in the measured edge.

A useful follow-up would pre-specify a modest and equal tuning budget for the classical library and report it as an additional sensitivity analysis, without redefining the pre-specified LR-based H5 target.

That is worth separating from the framework's content. \textbf{A target defined against a baseline can be destroyed by improving the baseline, and no amount of better feature engineering on the quantum side prevents it. Whether data complexity predicts} \textit{something} \textbf{is a different question from whether it predicts} \textit{this} \textbf{something.}

\paragraph{Another interpretation of the designed quantum-signature control.} We reported in Sec.~\ref{sec:positive_control} that the only baseline-robust edge we found was unreachable by gradient-based training: the advantage sat at the level of the measurement, not the optimization. Read as a competition result, that is a quantum win that cannot be trained for, and therefore of limited use.

Read structurally, it says something else. It says the two substrates were good at different halves of the same problem, \textbf{the quantum device at preparing and measuring a state whose label is invisible to low-order classical features, the classical optimizer at everything downstream of that measurement}. The failure was in requiring one substrate to do both. On that reading the designed control is not a curiosity attached to a null result but a
worked example of a decomposition, and \textbf{the natural follow-up is not a better selector but a better division of labour.}

Some changes need to be proceed to test this interpretation. A target internal to a single pipeline rather than defined against a rival, for instance the marginal contribution of a quantum component, measured by replacing it with a capacity-matched classical control and holding the rest of the workflow fixed. Such a quantity cannot be moved by widening a baseline set, because there is no second system in the comparison. We do not test it here, and we note that it carries its own failure mode: \textbf{a marginal-contribution measure is only as strong as the control it is measured against}, which is Sec.~\ref{sec:baseline}'s lesson in a new place rather than an escape from it. Sec.~\ref{sec:scope} sketches what a study built on this footing would look like.

\subsection{Limitations}
\label{sec:limitations}

A falsification is only valuable if the case that the null is a fact about the framework rather than an artifact of the setup. We therefore separate the limitations by what they threaten.

Here are the limitations that do not plausibly flip the verdict. $N = 27$ datasets is small for leave-one-dataset-out generalization, and it does limit effect-size precision: we can say the selector carries no held-out signal, not how much signal a better one would carry. But the pre-specified threshold is $R^2 \ge 0.4$ and the measured value is $-0.32$ pooled, with the GBM head at $-0.58$, worse than predicting the mean, not marginally short of a bar. Recovering H5 from this sample would require the selector to have been unlucky in a consistent direction across 27 independent folds. The same applies to H1, where the observed $R^2 = +0.026$ sits against a pre-specified $0.7$. Both verdicts are also invariant to the $\beta$-convention correction
of Sec.~\ref{sec:results}: the GBM edge path is bit-identical across runs and the GP edge moves only from $-0.322$ to $-0.370$.

Now, the limitations that bound what the null covers. These do not weaken the result but do restrict its scope, and each is a concrete way the framework could survive in a regime we did not test.
\begin{itemize}
  \item \emph{Circuit regime.} All pre-specified circuits are $n = 8$ qubits,
        simulated, noiseless, NISQ-scoped and fault-tolerance-free. Width
        dependence is no longer untested: gradient variance was measured at
        $n \in \{4,\dots,14\}$ (Sec.~\ref{sec:results_nscaling}) and \emph{quantum test error} at $n \in \{4,6,8,10\}$ (Sec.~\ref{sec:results:aqwidth}), the
        latter closing a gap the earlier arm left open, the scaling sweep measured gradients only, so $a_q$ had been untested at \emph{any} width other than $8$. Both are exploratory. Behaviour above $n = 14$ remains
        untested, as does hardware noise.
  \item \emph{The classical software layer is a possible and undiagnosed
        source of loss.} Every number here comes from an 8{,}100-cell simulator
        sweep, and \citet{panambalom2026postprocessing} show that postprocessing
        routines developed against simulators can encode assumptions that fail at
        device scale, producing data loss \emph{not detectable from high-level
        model outputs}. Our pipeline is simulator-native throughout, so the
        device-scale form of that failure does not arise; but the general point,
        that a plumbing defect can be invisible in the outputs one inspects, is one this paper has already been bitten by once, in the $C_{\text{top}}$
        estimator of Appendix~\ref{app:topofix}. We have not audited the postprocessing layer with the same care we eventually applied to that estimator, and we do not claim to have done so.
  \item \emph{Encoding and ansatz libraries are finite.} Four encoding
        families in ten configurations, and a hardware-efficient ansatz family.
        A selector that fails over this library might succeed over another.
        Re-uploading encodings are excluded from H1/H2 outright, because they
        admit no fixed encoded state to compute $\qq$ on.
  \item \emph{The fifth pillar was never exercised.} Every dataset in the
        frozen suite is non-graph, so the curvature component $v_5$ is dropped
        throughout and H2 reduces to a test of the topological contribution
        alone. Curvature is scoped out, not falsified.
  \item \emph{Mode~Q is empty.} No encoding in the suite assumes coherent QRAM
        loading, so H6 has no stratum to test and the framework's claim to
        discriminate dequantizable from non-dequantizable regimes is
        unexamined.
  \item \emph{The two tracks are parallel, not unified.} The designed
        quantum-native control (Sec.~\ref{sec:positive_control}) is a positive
        control, not an input to the selector; nothing here shows what a selector
        trained on quantum-native data would do.
  \item \emph{The target variable is cell-dependent, and we tested only one.}
        Sec.~\ref{sec:scope} argues the framework's claim is properly stated
        per (problem type, algorithm, data-access mode) cell. The same applies
        to the \emph{predictand}: $\hat{e}$ compares a quantum method against a
        classical rival, which presumes the two are substitutes. In cells where
        they are complementary rather than rival, a competitive target may be
        the wrong instrument regardless of whether data complexity is the right
        axis (Sec.~\ref{sec:interpretation}). The result of this study is a null for this target in this cell.
\end{itemize}

The embedding-interaction half was never given a chance to matter.
The framework has two halves, intrinsic complexity $\vv(D)$ and embedding
interaction $\qq(D,\phi)$, and only the first was tested against a target with structure to predict. Decomposing the variance of the budget-matched edge over all $8{,}100$ cells: \textbf{$83.8\%$ lies between datasets, $0.5\%$
between encodings and $0.5\%$ between ans\"atze.} So $\vv(D)$ faced a real target, while $\qq(D,\phi)$ faced an axis carrying half a percent of the variation.

The cause is the degeneracy of Sec.~\ref{sec:results:degenerate} acting on a
second axis. Because no configuration escapes chance, mean $a_q$ spans only
$0.462$ to $0.497$ across all ten encodings, a range \emph{smaller} than the within-encoding spread of $0.05$--$0.09$, the encoding dimension collapses. Nothing distinguishes a product-state angle encoding from a \textit{ZZ}-map producing
genuinely entangled states (mean half-cut entropy $1.52$ nats of a possible $2.77$), because both land at chance. Consistently, no encoding wins systematically: across the 27 datasets the best-performing encoding is
scattered over nine of the ten configurations, as expected from noise.

This is not a defect in any encoding. It is that an interaction term cannot be tested when the quantity it modulates is pinned. \textbf{We therefore report $\qq(D,\phi)$ as untested rather than unsupported}, and note that any future design must establish that its encoding axis carries variance \emph{before}
fitting a model that depends on it.

We also report limitations in the instruments themselves.
Three of the nine complexity components are measured more weakly than the framework assumes. Topological metrics are relative rankings across the suite, not absolute Betti detectors, so $C_{\text{top}}$ orders datasets but does not characterize them, and in the form used for the pre-specified tests it did
not even order them by topology (Appendix~\ref{app:topofix}). The corrected
estimator is scale-invariant and shape-validated, but it normalizes by the global diameter, so widely separated clusters dilute the measure and multiplicity is lost: three well-separated circles score \emph{below} one.
Local-scale normalization repairs that at the cost of sample-size invariance, which matters more when comparing datasets of differing $n$; we took the trade deliberately and record it as a limitation of the corrected instrument. Higher-order mutual-information estimation is brittle at $k \ge 3$, which is precisely where the correlation-order pillar should carry its distinctive content, so that pillar is weakest exactly where it matters most. And the H3 secondary criterion, specified to catch the failure mode we ultimately diagnosed by other means, was lost to a serialization defect (Sec.~\ref{sec:results:h3}); we report the gap $R^2$ as a proxy and the serializer is fixed, but the pre-specified statistic itself is unavailable for these runs.

The latest is the limitation that cuts against our own positive result. Lesson~1 (Sec.~\ref{sec:baseline}) says the classical baseline determines the measured edge, and that applies to the designed quantum-signature control as much as to anything else. The control edge is measured against the best of six models on $k \le 2$ shadow features; it is baseline-robust with respect to \emph{that} set, and the construction gives a principled reason to expect robustness. The label is provably invisible to any $k < w$ Pauli feature, but a classical learner given higher-order shadow features, or a different representation entirely, has not been ruled out empirically. We report the edge as surviving the sweep we ran, not as a proof that no classical method reaches it.

The parity control also has a limited comparative scope
(Sec.~\ref{sec:positive_control}).  Its positive result establishes recoverable signal under the specified structure-aware measurement; it does not benchmark against classical procedures supplied with matched higher-order features, prior structural knowledge, or a matched measurement/query budget.

\subsection{Scope of the algorithm suite, and the roadmap for the next study}
\label{sec:scope}

The validation instantiates a \emph{single} quantum-algorithm family: parameterized-circuit classification, comprising an embedding $\phi$, a variational ansatz $U(\theta)$ and a single-Pauli $\langle Z_0\rangle$ readout, trained by parameter-shift descent. Every task is supervised binary
classification on classical feature vectors, which places all pairs in Mode~S. The selector is defined for any (algorithm, mode) cell, but only this one cell is populated.

The primary result must therefore be read conditionally. What H5 tests is ``data complexity predicts the edge of variational classification on Mode-S data'', not a claim about quantum machine learning at large. The scoping is not incidental: Mode-S supervised learning on unstructured classical vectors
is precisely where classical surrogates are strongest, and the primary comparator is the pre-specified logistic-regression baseline, supplemented by a fixed six-model sensitivity analysis. The null therefore applies to this specific Mode-S variational-classification comparison, not to quantum machine learning at large.

\begin{table}[t]
\centering
\caption{The honest unit of the framework's claim is a (problem type, algorithm, data-access mode) cell. This work populates one row.}
\label{tab:scope}
\small
\setlength{\tabcolsep}{4pt}
\begin{tabular}{@{}p{0.27\columnwidth}p{0.17\columnwidth}@{\hspace{1.2em}}c@{\hspace{1.2em}}p{0.29\columnwidth}@{}}
\toprule
\raggedright Problem type & \raggedright Algorithm & Mode & \raggedright Separation status \tabularnewline
\midrule
\raggedright Supervised classification \emph{(this work)} & \raggedright variational / kernel & S & \raggedright weakest; kernels often dequantizable \tabularnewline
\raggedright Combinatorial optimization & \raggedright QAOA, annealing & S & \raggedright contested vs.\ classical heuristics \tabularnewline
\raggedright Linear systems, PCA & \raggedright HHL, QSVT & Q & \raggedright dequantizable unless sparse-survivor \tabularnewline
\raggedright Ground states, dynamics & \raggedright VQE, Trotter & N & \raggedright quantum-native; separation depends on problem family \tabularnewline
\raggedright Generative modeling & \raggedright Born machine & S/N & \raggedright partial \tabularnewline
\bottomrule
\end{tabular}
\end{table}

The two cells where separation is theoretically most defensible, \textbf{Mode~N} and \textbf{Mode-Q sparse-survivor}, are exactly the ones this suite does not populate (Table~\ref{tab:scope}). Extending to a full
algorithm $\times$ mode matrix is the principal open direction, and the quantum-native cell, for which this work already supplies an honest classical-shadow baseline and a positive control, is the natural first
expansion. The next study develops it, on the quantum-native probe ensembles drawn
from this corpus; that study does not carry the mode taxonomy of Table~\ref{tab:scope}
across, so the correspondence is by construction rather than by name.

Also outstanding, in the order we would rank them. \textbf{First, geometry}: the topological verdict is summary-dependent and the curvature pillar was never exercised (Sec.~\ref{sec:ledger}), and of everything listed here it is the item whose resolution would most change the framework, since on this suite the composite \emph{is} the topology component ($\rho = 0.998$). It needs a
pre-specified topological summary and a graph-native corpus, so it is a separate study rather than an extension of this one, and one the program is developing (Sec.~\ref{sec:ledger}). Then: H4, which needs the mixed-encoding implementation; an $n$-scaling arm promoted from exploratory to
pre-specified; a QRAM-dependent encoding to make H6 evaluable; and hardware.


\subsection{Open questions carried over from v1}
\label{sec:v1open}

\emph{The open questions below were stated in v1, before any of the measurements reported here. They are retained unchanged; Sec.~\ref{sec:ledger} states which of them this work closes, narrows, or leaves open.}

The preceding sections show that any empirical assessment of QML advantage must account for the interaction between data structure, representation, model class, and data-access assumptions. The present study does not resolve the broader question of when quantum advantage should occur. Instead, it narrows several of the conditions that a useful complexity-based benchmark would need to distinguish: intrinsic dataset descriptors from embedding-dependent quantities, quantum performance from classical baseline variation, and generic trainability from the presence of a quantum-sensitive structure.

First, it motivates the development of complexity-based benchmarks for deciding when QML should be deployed. Rather than comparing quantum and classical algorithms only in terms of runtime or hardware requirements, benchmarks should explicitly incorporate measures of dataset complexity, including encoding cost, structural diversity, and manifold dimensionality. Such benchmarks would allow researchers to identify realistic use cases where QML provides value beyond classical alternatives~\cite{cerezo2021variational,abbas2021power}.

Second, it suggests a need for complexity-aware dataset design in demonstrations of quantum advantage. Synthetic and real-world datasets could be crafted to emphasize correlations or structures that are provably difficult for classical learners but naturally embedded within quantum Hilbert spaces. This would parallel the role of hardness assumptions in computational complexity theory, where carefully chosen problem families expose separation between classical and quantum capabilities~\cite{aaronson2015read,huang2022quantum}.

Third, complexity interacts intimately with noise and trainability in QML models. Noisy quantum hardware can blur the fine-grained structures that confer advantage, while barren plateaus in variational circuits limit the trainability of highly expressive models~\cite{mcclean2018barren,wang2021noise, Larocca_2025}. The challenge lies in determining whether the intrinsic complexity of data amplifies or mitigates these issues: does richer structure improve the signal-to-noise ratio in optimization, or does it exacerbate the fragility of quantum models? Addressing this interplay requires a joint framework that unifies data complexity with hardware limitations.

Finally, an open and fundamental question is whether one can derive a complexity-theoretic threshold for QML advantage, akin to a “quantum PAC-bound.”\footnote{Quantum PAC-bounds establish the theoretical limits on the number of quantum superposition samples (weighted by square root probabilities) required to learn a target concept within specified accuracy and confidence parameters \cite{Hadiashar_2024, nayak2024propervsimproperquantum}.} Such a result would formalize the boundary at which the representational and generalization capacity of quantum models provably exceeds that of classical models under standard learning-theoretic assumptions~\cite{valiant1984theory,arunachalam2017survey}. Establishing such thresholds would not only sharpen the theoretical foundations of QML but also guide practical strategies for identifying domains where quantum advantage is most likely to emerge.

In summary, the implications of data complexity extend beyond algorithmic design into benchmarking, dataset engineering, hardware-aware training, and foundational learning theory. Addressing these open questions will be essential for charting a rigorous roadmap toward demonstrable quantum machine learning advantage.

The present study therefore closes the question of whether the current complexity vector is a useful selector for the tested benchmark regime: it is not. It does not close the broader questions of complexity-aware dataset construction, quantum-native access, geometric complexity, or complexity-theoretic thresholds for quantum learning advantage. Those remain the appropriate targets for subsequent experiments.


\section{Conclusion}
\label{sec:conclusion}

We set out to test whether the proposed data-complexity framework could predict quantum machine learning advantage under a frozen benchmark, encoding, circuit, and access regime. It does not. The primary selector fails its pre-specified predictive criterion, and the failure persists under stronger classical baselines and corrected topological estimators. The result therefore rejects the framework's central predictive claim in the tested regime, rather than establishing a universal absence of quantum advantage.

The reason the result is worth reporting is that we can say \emph{why}, and each reason outlives the framework that produced it. The apparent complexity--advantage signal was real but misattributed: it tracked where a linear baseline fails, and a single weak baseline is enough to manufacture a fully predictable, reproducible and spurious quantum advantage. Where an edge did survive a best-of-six classical comparator, a shadow-hard parity
construction, at $+0.42$, it proved unreachable by gradient training, in a way that unifies classical parity hardness, barren plateaus and shadow-hardness as one absence of low-order signal. And varying circuit width, which a fixed-width benchmark cannot do, showed the scaling law to be misspecified in the regime experiments actually occupy: power-law rather than exponential, depth-independent, and slower than the Haar-random rate. A selector trained on gradient-trained outcomes was therefore fitting a small correction on top of a wrong leading term, to a target that systematically excludes the advantages worth having.

What we take from this is narrower and more useful than the hypothesis we started with. Aggregate data complexity is the wrong axis; the locality structure of the label--observable relationship is a sharper one, and it is what the only surviving edge was built from. The claim is also properly conditional: one algorithm family, one data-access mode, simulated circuits of at most $14$ qubits, and the cells where a separation is most defensible are exactly the ones this suite does not populate. The next study takes up the quantum-native of those. In the meantime we would put the methodological result first: report a baseline sweep rather than a baseline, distinguish an edge that exists from an edge that can be found, and vary circuit width before fitting anything that scales with it. Had we done only the first of those, this paper would have reported a positive result, and it would have been wrong.

\subsection*{In the grey zone}

It is worth being exact about the reach of the negative result, because the tempting overstatement is close by and we do not make it. We tested one task formulation, twenty-seven datasets, ten encodings, six ans\"atze and one data-access mode, and on those we found no advantage attributable to the quantum model. That is not evidence that quantum machine learning over classical data cannot work. We have not tested all classical datasets, and we have no theorem that forbids it; Propositions~1 and~2 constrain what an encoding can \emph{do}, not what a model can \emph{achieve}. An encoding cannot manufacture label information and cannot place preprocessing gains beyond classical reach; neither statement closes the residual term, and the residual is where an advantage would live.

The 27 datasets support a pre-specified test within this benchmark suite, not a population-level claim about all tabular learning problems or all QML pipelines.  Leave-one-dataset-out evaluation tests transfer among the sampled datasets, but the suite is not a probability sample of task families, data-generating mechanisms, encodings, or classical competitors.  The central conclusion should therefore be read as conditional on the tested regime: the specified complexity representation does not reliably predict the measured edge across these 27 datasets.

What the propositions do supply is a reason the null was not an accident of our choices. The preprocessing-attributable term is classical by construction, so it was never ours to claim. The residual was measured across ten encodings, including entangling maps carrying $1.52$ nats of half-cut entropy, and found indistinguishable from zero \emph{in a regime where the quantum model sits at chance in $81.5\%$ of cells}. A pinned quantity cannot exhibit an interaction. That is a statement about the regime we could reach, and the honest form of our claim is conditional on it.

\subsection*{The question we would leave behind}

If one thing from this paper outlives the framework it was written to test, we would rather it were a question than a result. Every reported quantum--classical gap decomposes exhaustively, by Eq.~\eqref{eq:attribution}, into a preprocessing-attributable term and a residual. The first is obtainable by any
classical party at the cost of running the preprocessing. Only the second is a
quantum claim.

So: \textbf{which term did you measure?}

It applies to work this paper never touched; it can be answered from a comparison that is already published, and a study that cannot answer it has not isolated a quantum effect, whatever its headline number. Had we been asked
that question at the outset, this paper would have been much shorter.

The more useful outcome is diagnostic. On the classical benchmark suite, quantum performance is near chance across most cells, leaving the measured edge largely determined by the classical comparator. The proposed complexity vector consequently has little robust quantum variation to predict. Yet the positive control shows that a baseline-robust quantum-sensitive signal can be constructed when the relevant structure is deliberately inaccessible to the generic learning pipeline. The next step is therefore not to add more complexity descriptors to the same selector, but to design experiments in which the quantum-relevant structure, data-access model, representation, and classical comparator are specified together from the outset.


\section*{Data and code availability}

All code, specifications, frozen dataset registry, logged deviations and raw results are released at \url{https://github.com/Christophe-pere/data_complexity_framework}, made public
at publication.

\paragraph{Environment.} Python 3.12 with \texttt{numpy} 1.26.4, \texttt{scipy} 1.13.1, \texttt{scikit-learn} 1.6.1, \texttt{torch} 2.7.1, \texttt{gpytorch} 1.15.2, \texttt{lightgbm} 4.6.0, \texttt{ripser} 0.6.15, \texttt{gudhi} for the pre-specified topological component. Quantum simulation
is exact state-vector in \texttt{numpy}; no quantum hardware was used.


\appendix
\setcounter{equation}{0}
\renewcommand{\theequation}{\thesection\arabic{equation}}
\makeatletter\@addtoreset{equation}{section}\makeatother

\section{The v1 framework as published}
\label{app:v1}

\emph{This appendix reproduces, unchanged, the parts of v1 that the present version supersedes: the composite
definitions of data complexity, the analysis of how complexity was expected to bear on quantum models, and the
discussion of practical boundaries, encoding and dataset diversity. They are kept in full so that the corrected
framework of Sec.~\ref{sec:framework} and the registry of Table~\ref{tab:changes} can be read against the text
they revise, and because v1 remains publicly archived. Nothing here is asserted by this version: the status of
each element, withdrawn before testing, corrected, retained untested, rebuilt, or added, is given in
Table~\ref{tab:changes}, and the evaluative assessment is in Sec.~\ref{sec:ledger}. Forward references to ``Part~II'' in the text below are v1's own, and refer to the empirical work now reported in Secs.~\ref{sec:hypotheses}--\ref{sec:results} of this version.}

\subsection{Composite definitions}
\label{sec:v1composite}

\emph{The four composite quantities below are reproduced as they appeared in v1 of this preprint. Equations~\eqref{eq:classical_comp}, \eqref{eq:v1norm}, \eqref{eq:variance} and \eqref{equation_expressivity} are \textbf{superseded}: they are withdrawn and replaced in Sec.~\ref{sec:framework}, whose first subsection registers every change and the reason for it. They are kept here because the corrected framework, the pre-specified hypotheses and the results in Secs.~\ref{sec:hypotheses}--\ref{sec:results} are all stated against them, and because v1 remains publicly archived.}

We formalize data complexity as a unified quantity that governs trainability and generalization in both classical and quantum machine learning. For a classical dataset $\mathcal{D}=\{x_i\}_{i=1}^N$ with $x_i\in\mathbb{R}^d$, define
\begin{equation}
C_{\text{data}} \;=\; \lambda_1 S(\mathcal{D}) \;+\; \lambda_2 I_{\text{corr}}(\mathcal{D}) \;+\; \lambda_3 K(\mathcal{D})+ \lambda_4 C_{\text{top}}(D),
\label{eq:classical_comp}
\end{equation}

Eq.~\ref{eq:classical_comp} defines the composite classical data complexity as a normalized weighted sum. Here, the weights $\lambda_i$ act as tunable hyperparameters. At present, these weights are heuristic; in Part II we will calibrate them empirically via cross-validation across benchmark datasets. Alternatively, they may be fixed based on axiomatic desiderata (e.g., scale-invariance, monotonicity) or normalized by the variance of each metric across a benchmark set. Eq.~\ref{eq:classical_comp} is composed of: 
\begin{enumerate}
    \item \textbf{Distributional entropy}  
    $S(\mathcal{D})=-\sum_{x\in\mathcal{D}}p(x)\log p(x)$, quantifying diversity and unpredictability of samples.
    \item \textbf{Correlation order / interaction complexity}  
    $I_{\text{corr}}(\mathcal{D})=\sum_{k=2}^d I(x_{i_1};\ldots;x_{i_k})$, capturing higher-order dependencies via multivariate mutual information.
    \item \textbf{Kolmogorov complexity / compressibility}  
    $K(\mathcal{D})\approx L_{\text{compressed}}/L_{\text{raw}}$, estimated via compression ratios, reflecting algorithmic structure, where $L_{\text{raw}}$ is the length (in bits, characters, or bytes) of the original dataset $\mathcal{D}$ in its uncompressed form, and $L_{\text{compressed}}$ is the length of the dataset after applying a lossless compression algorithm. The ratio $L_{\text{compressed}}/L_{\text{raw}}$ estimates how compressible the dataset is, which acts as a proxy for Kolmogorov complexity. 
    \item \textbf{Topological complexity / manifold invariants}  
    $C_{\text{top}}(\mathcal{D}) = \sum_{k=0}^{K} w_k \, \text{Pers}_k(\mathcal{D})$, 
    where $\text{Pers}_k(\mathcal{D})$ is the total persistence of $k$-dimensional homological features 
    (birth–death intervals in persistent homology), and $w_k \geq 0$ are weights. 
    This term captures the global manifold structure of the dataset: 
    datasets with trivial topology (few or short-lived features) have low $C_{\text{top}}(\mathcal{D})$, 
    while those with rich multi-scale homology exhibit higher values, 
    reflecting increased geometric and structural complexity beyond statistical correlations.
\end{enumerate}

To compare across datasets, we normalize:
\begin{equation}
\tilde{C}_{\text{data}} = \frac{C_{\text{data}}}{C_{\max}}, \qquad \tilde{C}_{\text{data}}\in[0,1],
  \label{eq:v1norm}
\end{equation}
where $C_{\text{data}}$ is the complexity measure of a given dataset, $C_{\max}$ is the maximum observed dataset complexity across a benchmark collection, and $\tilde{C}_{\text{data}}$ is the normalized complexity score, always between 0 (simplest dataset) and 1 (most complex dataset in the benchmark). Gradient scaling in variational circuits then follows the form \cite{holmes2022connecting}:
\begin{equation}
\mathrm{Var}[g(\theta)] \;\approx\; \exp\!\big(-\alpha\,n\,d\,\tilde{C}_{\text{data}}\big),
\label{eq:variance}
\end{equation}
where $\mathrm{Var}[g(\theta)]$ is the variance of the gradient of the cost function with respect to parameters $\theta$. This tells how trainable a variational quantum circuit is: if the variance decays exponentially, training becomes infeasible (barren plateau). $\alpha$ is a positive constant (depends on the specific ansatz, cost function type, and noise model). $n$ is the number of qubits. Larger Hilbert spaces accelerate variance decay. $d$ is the circuit depth (number of layers). Deeper are the circuits, faster is the exponential suppression. $\tilde{C}_{\text{data}}$ is the normalized dataset complexity. Higher data complexity increases the effective rate of barren-plateau onset.

This explicitly links data complexity to the barren plateau problem: high-complexity datasets push circuits into barren plateaus faster.

\paragraph{Quantum-native complexity.}  
For a quantum ensemble $\mathcal{E}=\{(p_i,\rho_i)\}_{i=1}^N$ on $n$ qubits, define
\begin{equation}
\begin{split}
C_{\text{quant}}(\mathcal{E})
&=\alpha_1\,\overline{S}_{\text{ent}} \\
& +\alpha_2\,\mathcal{I}_{\text{multi}} \\
& +\alpha_3\,\mathrm{rank}_{\text{eff}}(K_{\mathcal{E}}) \\
& +\alpha_4\,\overline{\mathcal{N}} \\
& +\alpha_5\,\overline{\mathcal{F}} \\
& +\alpha_6\,C_{\text{top}}^{(q)}(E)
\end{split}
  \label{eq:quantum_comp}
\end{equation}
with normalized, dimensionless components:
\begin{enumerate}
    \item \textbf{Average bipartite entanglement entropy}  
    $S_{\text{ent}} = \frac{1}{N}\sum_i S(\rho_A^{(i)})$, quantifying the degree of quantum correlations across bipartitions. 
    Low values indicate nearly product states (classically simulable), while high values signal complex entangled structure.

    \item \textbf{Multipartite total correlation}  
    $I_{\text{multi}} = \sum_j S(\rho_j) - S(\rho_{1\ldots n})$, capturing higher-order dependencies beyond pairwise correlations. 
    \item \textbf{Effective rank of ensemble kernel}  
    $\text{rank}_{\text{eff}}(K_E) = \left(\sum_i \lambda_i\right)^2 / \sum_i \lambda_i^2$, 
    where $\{\lambda_i\}$ are eigenvalues of the kernel Gram matrix $K_E$. 
    A higher rank indicates broader Hilbert-space support and greater sample complexity. 
    \item \textbf{Nonclassicality / magic monotone}  
    $N(\rho)$, quantifying the extent to which states depart from stabilizer (Clifford) structure, 
    reflecting computational resources unavailable to classical simulation. 
    \item \textbf{Quantum Fisher information}  
    $F = \mathrm{Tr}\!\left(\rho L^2\right)$, where $L$ is the symmetric logarithmic derivative, measuring sensitivity of states to parameter variations and encoding capacity for precision. 
    \item \textbf{Topological quantum complexity}  
    $C_{\text{top}}^{(q)}(E) = \gamma_1 S_{\text{topo}} + \gamma_2 \chi(\mathcal{M}_E) + \gamma_3 \sum_{k=0}^K \text{Pers}_k(\mathcal{M}_E)$, 
    where $S_{\text{topo}}$ is the topological entanglement entropy, $\chi(\mathcal{M}_E)$ is the Euler characteristic 
    of the quantum state manifold, and $\text{Pers}_k(\mathcal{M}_E)$ denotes persistent homology contributions of dimension $k$. 
    This term captures long-range order and global manifold structure of the embedded quantum dataset, 
    with higher values indicating topologically rich states requiring more quantum resources to represent and learn. 
\end{enumerate}

\paragraph{Induced quantum complexity.}  
For a classical dataset $\mathcal{D}$ embedded via a quantum feature map $U_\phi:x\mapsto\rho(x)$, we define the induced quantum complexity as
\begin{equation}
C_{\text{ind}}(\mathcal{D};\phi)
=\sum_{j=1}^5 \beta_j\, M_j^{(\phi)}.
  \label{eq:induced_comp}
\end{equation}
Here, $\rho(x)$ is the quantum state encoding of data point $x$, and the quantities $M_j^{(\phi)}$ represent distinct quantum-native complexity measures evaluated on the embedded states $\{\rho(x_i)\}$:

\begin{itemize}
    \item $M_1^{(\phi)}$: \textbf{Hilbert-space support dimension} — the effective dimension of the subspace spanned by $\{\rho(x_i)\}$, capturing how widely data spreads in the quantum state space.
    \item $M_2^{(\phi)}$: \textbf{Quantum Fisher information (QFI) spread} — quantifying the sensitivity of the embedding with respect to variational parameters, linked to gradient concentration.
    \item $M_3^{(\phi)}$: \textbf{Entanglement entropy} — measuring the bipartite entanglement structure across qubits when encoding the dataset, indicative of non-classical correlations.
    \item $M_4^{(\phi)}$: \textbf{Kernel spectrum flatness} — the effective rank of the quantum kernel Gram matrix, connected to sample complexity and generalization.
    \item $M_5^{(\phi)}$: \textbf{Expressibility vs. locality ratio} — balancing how expressive the embedding is compared to its locality, relating to barren plateau onset.
    \item $M_6^{(\phi)}$: \textbf{Topological invariants of the embedding manifold} — 
    persistent homology, Betti numbers, or Euler characteristics extracted from the metric space induced by the embedding $\{\rho(x_i)\}$. This captures global manifold structure beyond local correlations, reflecting 
    whether the embedding induces topologically trivial or highly nontrivial geometry.
\end{itemize}

The coefficients $\beta_j$ are weighting factors that reflect the relative importance of each quantum-native measure in a given application or benchmarking setting. Crucially, this formalism emphasizes that the perceived complexity of the same classical dataset $\mathcal{D}$ is not intrinsic, but depends on the choice of feature map $U_\phi$. A dataset that appears simple under one embedding may induce large entanglement or high kernel rank under another, thereby modifying trainability and sample requirements.

Quantum data complexity is defined as a composite of quantum-native properties such as entanglement entropy, Schmidt rank, multipartite correlations, quantum Fisher information, and topological invariants (TEE, Euler characteristic). Not all of these quantities are efficiently computable; for example, exact TEE estimation is exponential in system size. We therefore propose to rely on efficiently estimable proxies (e.g., Rényi entropies from randomized measurements, classical shadows for QFI, persistent homology on embedding outputs) as practical surrogates.

\paragraph{Spectral connection to sample complexity.}  
Let $K$ be a positive semi-definite kernel Gram matrix with eigenvalues $\{\lambda_i\}$. The kernel effective dimension is
\begin{equation}
d_{\mathrm{eff}}(\lambda) \;=\;\mathrm{Tr}\!\big(K(K+\lambda I)^{-1}\big)\;=\;\sum_i\frac{\lambda_i}{\lambda_i+\lambda}.
\end{equation}
Defining the effective rank $r_{\mathrm{eff}}=\tfrac{(\sum_i \lambda_i)^2}{\sum_i \lambda_i^2}$, we obtain the bound:
\begin{equation}
d_{\mathrm{eff}}(\lambda)\;\ge\;\frac{(\sum_i \lambda_i)^2}{\sum_i \lambda_i^2+\lambda\sum_i\lambda_i}.
\end{equation}

The kernel effective dimension $d_{\mathrm{eff}}(\lambda)$ quantifies how many “directions” in the feature space meaningfully contribute to learning, after accounting for regularization $\lambda$. Intuitively, if the spectrum ${\lambda_i}$ of the kernel matrix is concentrated (i.e., dominated by a few large eigenvalues), then $d_{\mathrm{eff}}$ is small, and only a few principal components matter — leading to lower sample complexity. 

Conversely, a flatter spectrum with many comparable eigenvalues pushes $d_{\mathrm{eff}}$ closer to the effective rank $r_{\mathrm{eff}}$, meaning the model must learn across many directions, thereby requiring more training samples. The inequality highlights that regularization can reduce the effective dimension by dampening the contribution of small eigenvalues, while in the unregularized case $\lambda=0$ the effective dimension lower-bounds to the effective rank. This establishes a direct spectral link between kernel alignment, dataset complexity, and the number of samples needed for generalization \cite{bartlett2021deep}.

\subsection{Effects of data complexity on QML}
\label{sec:v1effects}

\emph{Reproduced as published in v1. Its two load-bearing functional forms, the expressibility--complexity alignment of Eq.~\eqref{equation_expressivity} and the data-gated plateau law of Eqs.~\eqref{eq:variance} and \eqref{eq:variance_topo}, are corrected in Sec.~\ref{sec:framework} and tested in Sec.~\ref{sec:results}; the qualitative observations are assessed in Sec.~\ref{sec:ledger}.}

\subsubsection{Impact on Circuit Size \& Depth}

The complexity of data directly constrains the quantum resources required for faithful representation. 
Data with low structural complexity, such as weak correlations or compressible patterns, can often be modelled with shallow variational circuits, requiring only a few qubits and local entangling gates~\cite{schuld2019quantum}. 
As data complexity increases (e.g., strong multi-way correlations, high intrinsic dimension, or incompressibility), more expressive circuits become necessary. 
This typically entails (i) a larger number of qubits to embed data into an appropriate feature Hilbert space, and (ii) deeper layers of entangling gates to capture higher-order correlations~\cite{sim2019expressibility}.

Let $C_{\text{data}}$ denote a chosen complexity measure (correlation order, entropy, Kolmogorov complexity, etc.). 
We may model circuit requirements as scaling functions of $C_{\text{data}}$:
\begin{equation}
\begin{split}
Q(C_{\text{data}}) &\;\propto\; \log(\mathcal{H}(C_{\text{data}})),\\
D(C_{\text{data}}) &\;\sim\; \mathcal{O}(f(C_{\text{data}})),
\end{split}
\end{equation}
where $Q$ is the qubit count, $D$ the circuit depth, and $\mathcal{H}(C_{\text{data}})$ the effective Hilbert space dimension induced by the dataset. 
The function $f(C_{\text{data}})$ reflects the growth in entangling resources required: polynomial for moderate-order correlations, exponential for highly entangled or incompressible structures.

While richer circuits improve representational capacity, they also introduce challenges: deep circuits exacerbate barren plateaus in optimization~\cite{mcclean2018barren}, amplify noise accumulation, and push beyond the practical limits of near-term devices (NISQ)~\cite{preskill2018quantum}. 
Thus, balancing data complexity with feasible circuit resources remains a central design challenge for quantum machine learning models.

\subsubsection{Impact on Expressibility vs Generalization}

A central challenge in quantum machine learning (QML) lies in balancing the expressibility of a quantum model with its ability to generalize from training data. Expressibility refers to the capacity of a parameterized quantum circuit to generate a sufficiently diverse set of quantum states, thereby enabling it to approximate a wide class of functions \cite{sim2019expressibility,schuld2021quantum}. Generalization, in contrast, measures how well the model performs on unseen data, reflecting its ability to extract meaningful correlations rather than simply memorizing noise \cite{caro2022generalization,abbas2021power}.

When the data complexity is low, yet the quantum circuit possesses high expressibility, the model is prone to overfitting. In such cases, the quantum model risks encoding trivial patterns or spurious correlations, leading to poor generalization on test data \cite{huang2021power}. Conversely, when the data complexity is high but the circuit lacks sufficient expressibility, the model underfits: it cannot represent the higher-order correlations present in the dataset, and its predictive performance stagnates despite additional training. These two extremes illustrate the importance of aligning circuit expressibility with the intrinsic complexity of the dataset.

This alignment can be formalized by considering generalization error as a function of the gap between circuit expressibility and data complexity. If we denote the expressibility of a circuit as $\mathcal{E}(U)$ and the data complexity as $C_{\text{data}}$, then mismatches between $\mathcal{E}(U)$ and $C_{\text{data}}$ increase the expected generalization error. In particular, over-expressible circuits relative to the dataset tend to learn noise, while under-expressible circuits fail to capture meaningful structure. Optimal learning occurs when expressibility and data complexity are of comparable scale, suggesting the existence of a ``sweet spot'' where quantum models achieve their best performance \cite{schuld2021quantum,caro2022generalization}.

Let $\mathcal{E}(U)$ denote the expressibility of a parameterized quantum circuit $U(\theta)$, typically quantified by how uniformly its generated states cover the Hilbert space \cite{sim2019expressibility}. A simple measure of generalization error can be framed as:

\begin{equation}
    \mathbb{E}[\epsilon_{\text{gen}}] \;\approx\; \epsilon_{\text{emp}} + \lambda \, |\mathcal{E}(U) - C_{\text{data}}|,
    \label{equation_expressivity}
\end{equation}

where $\epsilon_{\text{gen}}$ is generalization error, $\epsilon_{\text{emp}}$ is empirical training error, $\mathcal{E}(U)$ is circuit expressibility, $C_{\text{data}}$ is data complexity, and $\lambda$ is a penalty factor for mismatch between data complexity and circuit expressibility.

The implications are twofold. First, data complexity provides a natural criterion for determining the necessary depth and size of parameterized quantum circuits. Second, the study of generalization in QML cannot be disentangled from the structure of the dataset itself: whether quantum circuits outperform their classical counterparts will depend not only on hardware and algorithmic considerations, but also on how well the expressibility of quantum models matches the intrinsic complexity of the data they are applied to \cite{abbas2021power,huang2021power,caro2022generalization}.

This relationship (Eq.~\ref{equation_expressivity}) highlights a fundamental trade-off between expressibility and data complexity. When the expressibility of a quantum circuit is much greater than the complexity of the dataset ($\mathcal{E}(U) \gg C_{\text{data}}$), the model tends to overfit, capturing noise or trivial correlations that do not generalize beyond the training set. In contrast, if the circuit expressibility is far below the data complexity ($\mathcal{E}(U) \ll C_{\text{data}}$), the model underfits, lacking the capacity to represent the higher-order patterns inherent in the dataset. The most effective regime arises when circuit expressibility is commensurate with the data complexity ($\mathcal{E}(U) \approx C_{\text{data}}$), where the quantum model achieves a balance that supports both accurate learning and robust generalization.
    
\subsubsection{Trainability and Barren Plateaus}

While expressibility defines the representational capacity of quantum models, trainability concerns the ability to efficiently optimize their parameters. In practice, even highly expressive quantum circuits may be unusable if their training landscapes are plagued by vanishing gradients. This phenomenon, known as the barren plateau problem, arises when the variance of the gradient of the cost function decreases exponentially with the number of qubits or circuit depth \cite{mcclean2018barren}. As a result, gradient-based optimization becomes ineffective, since updates to the parameters vanish within numerical noise.

The severity of barren plateaus is closely tied to data complexity. For datasets of low complexity, where meaningful correlations are shallow, shallow circuits often suffice, and the optimization landscape remains relatively benign. In such cases, trainability is not the primary bottleneck \cite{cerezo2024does}, and efficient classical simulation may still be possible in the absence of barren plateaus. However, as data complexity increases—particularly when capturing higher-order correlations requires deeper or more entangled circuits—the risk of barren plateaus grows substantially. The exponential suppression of gradients implies that training times scale unfavorably with both the circuit size and the complexity of the data being encoded \cite{cerezo2021cost}.

Formally, if the variance of the gradient scales as Eq.~\ref{eq:variance}, then high-complexity datasets indirectly amplify this scaling by demanding deeper circuits. Thus, even when sufficient expressibility is theoretically available, the optimization becomes practically intractable. This creates a tension between the need to increase circuit complexity to match data complexity and the risk of encountering untrainable barren plateaus.

Mitigation strategies, such as problem-inspired ansätze \cite{grant2019initialization}, layer-wise training \cite{skolik2021layerwise}, or data re-uploading techniques \cite{perez2020data}, can partially alleviate the issue by reducing the effective dimensionality of the optimization space. Nevertheless, the interaction between data complexity and barren plateaus remains an open challenge: as data complexity grows, so too does the need for scalable training strategies that preserve the gradient signal while maintaining the expressive power of the quantum model.

\begin{figure}[t]
  \centering
  \includegraphics[scale=0.45]{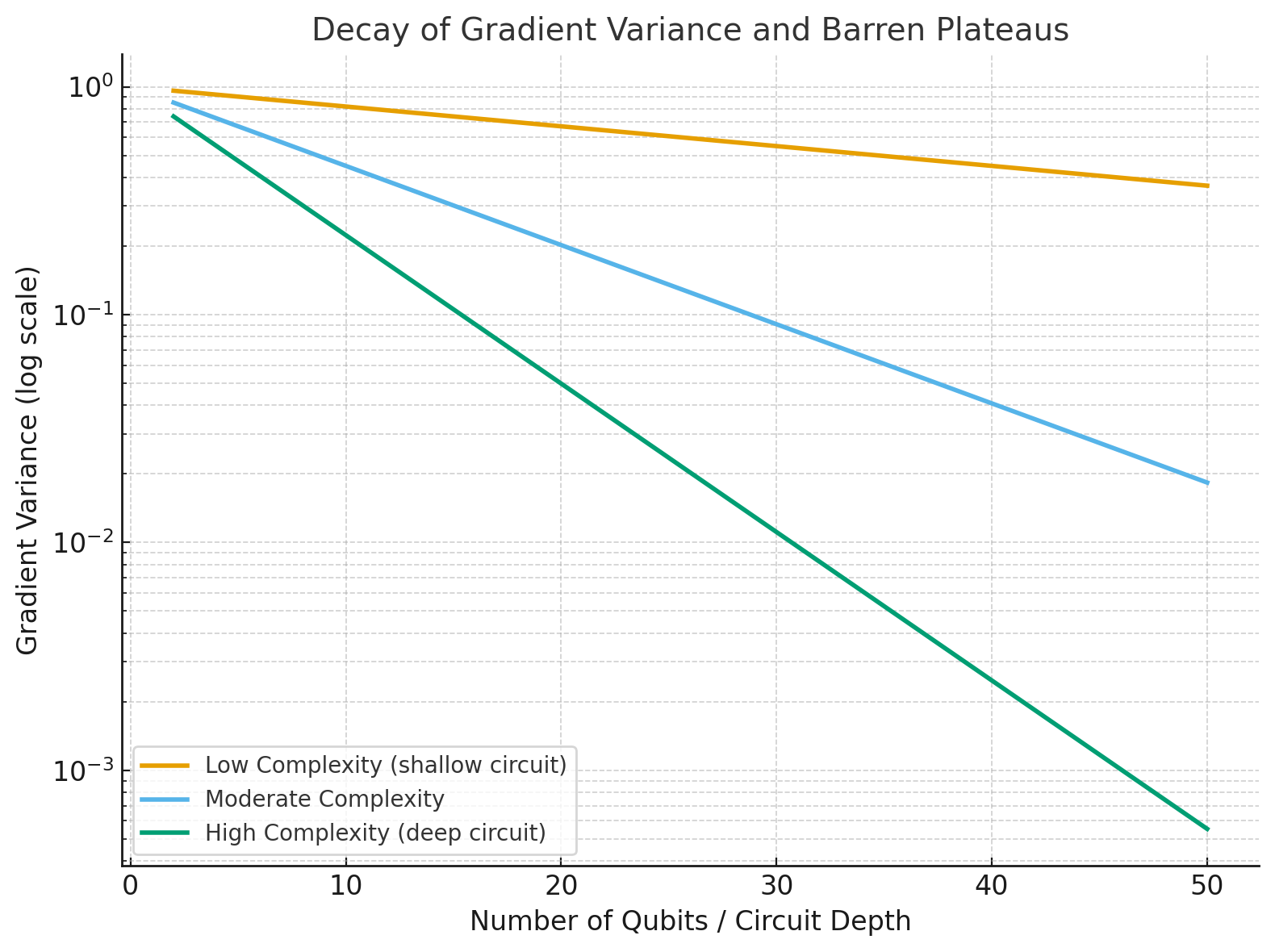}
  \caption{Exponential decay of gradient variance as a function of qubit number and circuit depth, illustrating the barren plateau phenomenon. Shallow circuits (low data complexity) exhibit slower decay and remain trainable, while higher complexity circuits lead to steeper variance decay, making optimization increasingly difficult. Figure is schematic and illustrative only, intended to guide intuition. Quantitative validation will follow in Part II.}
  \label{fig:figure_decay}
\end{figure}

The effect of data complexity on gradient behaviour is closely tied to the barren plateau phenomenon. As shown in Figure~\ref{fig:figure_decay}, the variance of gradients decays exponentially with the number of qubits and circuit depth, with the rate of decay dependent on the interplay between circuit expressibility and data complexity. For shallow circuits aligned with low-complexity datasets, the decay is mild, and optimization remains feasible. However, as data complexity increases, deeper circuits are required, which in turn accelerate the onset of vanishing gradients. This results in optimization difficulties where training becomes effectively impossible. The balanced regime is therefore characterized by a careful alignment between data complexity and circuit design, ensuring sufficient expressibility without incurring the prohibitive costs of barren plateaus.

Formally, let $\theta$ denote a variational parameter in a parameterized quantum circuit (PQC) and $\mathcal{L}(\theta)$ the cost function. The gradient with respect to $\theta$ is
\[
g(\theta) = \frac{\partial \mathcal{L}(\theta)}{\partial \theta}.
\]
It has been shown \cite{mcclean2018barren} that for sufficiently expressive (Haar-random) circuits, the variance of gradients decays exponentially with the number of qubits $n$:
\[
\mathrm{Var}[g(\theta)] \sim \mathcal{O}\!\left(\frac{1}{2^n}\right).
\]
This result highlights that as system size grows, optimization becomes increasingly infeasible due to the vanishing signal.

This scaling makes explicit that deeper circuits processing more complex datasets will experience faster gradient decay. Low data complexity ($C_{\text{data}} \ll 1$) results in slower decay, allowing shallow circuits to be trainable. High data complexity ($C_{\text{data}} \gg 1$) accelerates the onset of barren plateaus, demanding deeper circuits and amplifying optimization difficulties.

A practical trainability condition can therefore be expressed as
\[
\mathrm{Var}[g(\theta)] \gtrsim \epsilon,
\]
where $\epsilon$ is a threshold ensuring that gradients remain above numerical noise. This inequality effectively defines a boundary line between tractable and intractable training regimes, dictated by the joint contributions of qubit count, circuit depth, and dataset complexity.

\subsubsection{Impact of Topological Complexity}

Topological structure introduces an additional axis of difficulty beyond entropy, correlations, or compressibility. Datasets (classical or quantum) with nontrivial homology require models to capture multi-scale connectivity, loops, and voids in the underlying manifold. In the quantum setting, states with nontrivial topological order or embeddings with nonzero Chern numbers force circuits to reproduce long-range entanglement patterns that cannot be localized or truncated without loss of essential information.

From a resource perspective, topological complexity directly affects circuit depth and qubit requirements. Representing homological features of order $k$ generally requires circuits with entangling gates spanning length scales commensurate with the diameter of those features. Persistent homology of data embeddings therefore 
correlates with the minimal entanglement range needed in quantum circuits. Richer topological structures (e.g., multiple long-lived Betti numbers or high topological entanglement entropy) push models toward deeper or more global ansätze.

Trainability is also impacted. When $C_{\text{top}}^{(q)}(E)$ is high, the associated circuits are more expressive but prone to barren plateaus, since gradients vanish rapidly in highly entangled, globally correlated regimes. 
This can be formalized by extending the gradient variance scaling of Eq.~\ref{eq:variance}:
\begin{equation}
\mathrm{Var}[g(\theta)] \approx \exp\!\Big(- \alpha n d \, \big( \tilde{C}_{\text{data}} + \delta \, C_{\text{top}}^{(q)} \big)\Big),
  \label{eq:variance_topo}
\end{equation}
where $\delta$ is a penalty factor quantifying the added optimization difficulty induced by topological complexity. Thus, datasets with nontrivial topology accelerate the onset of barren plateaus, making optimization harder even if entropy and correlation-based measures are moderate.

In summary, topological invariants enrich the definition of data complexity by highlighting global, manifold-level structure that cannot be reduced to local statistics. For classical ML, this manifests as difficulty in approximating decision boundaries on complex manifolds. For QML, it necessitates deeper 
circuits and increases trainability challenges, but also delineates a potential regime of quantum advantage, since classical models typically struggle to capture persistent topological features without exponential overhead.

\subsection{Practical boundaries, encoding, and dataset diversity}
\label{sec:v1bounds}

\emph{Also reproduced as published in v1. The encoding-cost discussion below is superseded by the explicit cost and data-access model of Sec.~\ref{sec:framework:cost}, and the measured consequences of encoding appear in Sec.~\ref{sec:encoder}.}

The previous sections introduced measurable notions of dataset complexity (classical and quantum) and quantified how complexity affects circuit requirements and trainability. This section translates those insights into operational boundaries: concrete conditions under which a quantum machine learning method is expected to outperform classical alternatives in practice. The goal is to produce testable inequalities and a protocol for empirical validation rather than purely asymptotic statements.

\subsubsection{Data Complexity and Error Rates}

A crucial but often overlooked dimension of quantum machine learning (QML) performance is the interaction between data complexity and hardware error rates. Quantum circuits are inherently sensitive to decoherence, gate infidelities, and measurement noise \cite{preskill2018quantum,krinner2022realizing}. As data complexity increases, these errors are not merely additive; they interact with the data representation in ways that can amplify their impact.

For low-complexity datasets, the circuits required to achieve sufficient expressibility are typically shallow, with modest qubit entanglement. In this regime, the error footprint remains manageable, and quantum models may perform competitively with classical counterparts \cite{schuld2019quantum,caro2022generalization}. However, as the complexity of the data increases—whether due to higher-order correlations, low compressibility, or broad distributional entropy—the quantum circuits must grow deeper and more entangled to capture the underlying structure. This results in error accumulation that scales superlinearly with circuit depth \cite{mcclean2018barren,huang2021information}.

Formally, if we denote the circuit error rate as
\begin{equation}
\epsilon_{\text{circ}} \;\approx\; 1 - (1 - \epsilon_{\text{gate}})^{d \cdot W},
\end{equation}
where $d$ is circuit depth, $W$ is the average number of two-qubit gates per layer, and $\epsilon_{\text{gate}}$ is the average gate error rate, then increasing data complexity indirectly drives up $d$ and $W$. As a result, even modest per-gate error rates can lead to dramatic degradation in fidelity when attempting to learn highly complex datasets \cite{murali2019noise}.

This creates a tension: while complex datasets may in principle offer regimes where quantum models hold an advantage, the practical realizability of this advantage is limited by the noise resilience of current hardware. The implication is that error-aware complexity measures are needed—metrics that not only quantify how challenging the data is, but also how feasible it is to encode and process under realistic noise constraints \cite{bharti2022noisy}. Such hybrid metrics could guide algorithm design by balancing theoretical representational power with the error budgets of near-term devices.

\subsubsection{Sample Complexity and Data Availability}

An essential practical constraint for quantum machine learning (QML) lies in \emph{sample complexity}, which determines how many data points—quantum or classical—are needed for a model to generalize effectively. Classical ML theory tells us that low-complexity datasets can often be learned with few samples, but high-complexity datasets (e.g., having high intrinsic dimension or strong higher-order correlations) demand significantly more data. QML holds the promise of representing richer functions. The potential benefit is offset if achieving reliable estimation requires an excessive number of state preparations, thereby rendering the approach experimentally or computationally impractical. 

Caro et al.~\cite{caro2022generalization} provide rigorous generalization bounds for variational QML models, showing that the generalization error scales roughly as $\sqrt{T/N}$, where $T$ is the number of trainable gates and $N$ the number of training examples. Crucially, if only $K \ll T$ gates change substantially during training, the bound improves to $\sqrt{K/N}$, highlighting the importance of effective capacity control for better sample efficiency.

In the PAC learning framework, Cai et al.~\cite{cai2021sample} demonstrate that parametric quantum circuits with at most $n^c$ gates are learnable with $\tilde{O}(n^{c+1})$ samples, providing a polynomial scaling bound that reaffirms QML’s theoretical sample efficiency under structured settings.

On another front, Coopmans and Benedetti~\cite{coopmans2024training} rigorously show that Quantum Boltzmann Machines (QBMs) can be trained using only a polynomial number of Gibbs-state samples, even with expressive models, without encountering barren plateaus.

Nevertheless, these theoretical results rest on favorable conditions. In realistic scenarios, sample efficiency can be undermined by the cost of data generation—including the overhead of embedding classical data into quantum states or the measurement burden when working with quantum-generated data. Encoding a classical dataset into a quantum feature space may require deep circuits, thereby reducing any sample efficiency advantage due to noise and gate errors.

Thus, while structured datasets with aligned embeddings can in principle yield sample-efficient QML, the interplay between data complexity, model capacity, and practical sample cost is delicate. To retain quantum advantage, one must ensure that the model complexity does not outpace data availability and that the cost of preparing or measuring samples does not overwhelm any expressive benefit.

\subsubsection{Hardware Constraints: Qubits, Connectivity, and Coherence}

The pursuit of QML advantage is fundamentally limited by the characteristics of actual quantum hardware—chiefly, qubit count, connectivity, coherence time, gate fidelity, and error rates. These elements shape not only the feasible complexity of QML circuits but also the attainable dataset complexity and model depth.

Current NISQ devices typically support between 50 and 1,000 qubits; yet they remain constrained by decoherence, gate infidelities, and limited qubit connectivity, all of which constrain the depth of circuits that can run reliably. Two-qubit gate fidelities often range from 95\% to 99\%, implying that circuits with more than a few hundred two-qubit operations rapidly become dominated by noise rather than computation \cite{preskill2018quantum,thudumu2025supervisedquantummachinelearning}. 

Moreover, qubit connectivity remains a critical bottleneck. In most hardware architectures (e.g., superconducting qubits), interactions are limited to nearest neighbors, requiring costly SWAP gate insertions to mediate long-range entanglement. This necessity increases both circuit depth and error susceptibility, directly impeding the ability to encode and process high-complexity data efficiently\cite{waring_2024}.

Coherence time further restricts circuit length. Physical decoherence mechanisms, such as amplitude damping and dephasing, degrade quantum states over time. Contemporary superconducting qubits exhibit finite $T_1$ and $T_2$ times—often in the microsecond to millisecond range—requiring circuit execution within tight temporal windows or risking data loss \cite{Krantz2019}.

A promising strategy to circumvent coherence limitations is mid-circuit measurement with qubit reuse\cite{hothem2024, baumer2025, lemelin2025}. The NISQRC algorithm demonstrates this by partitioning temporal inference tasks and recycling qubits, thereby performing inference beyond individual coherence time constraints. Experimental validation on a 7-qubit platform shows that such approaches can substantially extend the effective depth of quantum models without escalating circuit lengths in the traditional sense \cite{Hu_2024}.

Lastly, error characteristics can dramatically impact QML viability. Coherent noise—systematic miscalibrations—aggregates across repeated operations and can be more devastating than stochastic noise, especially in deeper circuits \cite{Resch2021}. Therefore, error-aware circuit design and noise-resilient ansätze are essential for maintaining trainability and fidelity in high-complexity regimes.

\subsubsection{Encoding challenges}

A central aspect of realizing quantum machine learning (QML) advantage lies not only in the algorithmic design but in the process of \textit{data encoding}. As highlighted in Section~4, the complexity of data strongly constrains the ability of quantum models to outperform classical counterparts. Encoding is the interface where this complexity materializes: transforming input data into quantum states entails both resource overhead and potential loss of information. Thus, the very act of encoding may determine whether theoretical advantages translate into practice~\cite{biamonte2017quantum,schuld2018supervised, Mancilla_2022, Deepak_2024}.  

From a complexity-theoretic standpoint, the cost of encoding classical data into quantum states can offset the computational speedups promised by quantum algorithms. For example, \textit{amplitude encoding} offers an exponential compression of input size, yet preparing such states is often intractable without strong assumptions on data access such as \textit{quantum random access memory (QRAM)}~\cite{giovannetti2008quantum,aaronson2015read}. In contrast, \textit{basis} or \textit{angle encoding} is more resource efficient but sacrifices representational capacity, limiting the model’s expressive power~\cite{schuld2019quantum}. This trade-off illustrates a fundamental tension: richer encodings amplify the potential advantage in learning complex patterns, but simultaneously increase the physical and algorithmic costs~\cite{havlicek2019supervised,huang2021power}.  

The nature of the data further deepens this challenge. \textit{Classical datasets} with inherent redundancies or low-rank structure may be more efficiently embedded into quantum states~\cite{tang2019quantum,gilyen2019quantum}, whereas high-dimensional, unstructured data can render encoding prohibitively expensive~\cite{aaronson2015read}. Conversely, for \textit{genuinely quantum data}, arising from quantum experiments, sensors, or simulations, the encoding step is naturally circumvented, as the input is already available in a quantum form~\cite{preskill2018quantum,huang2022quantum}. In such regimes, the bottleneck shifts from encoding to learning and generalization, making QML particularly attractive.  

This interplay between encoding and data complexity underscores that the path to advantage cannot be framed solely in terms of asymptotic algorithmic performance. Instead, encoding constitutes a \textit{critical resource} that must be explicitly accounted for in any analysis of QML scalability~\cite{cerezo2021variational,schuld2021quantum}. In practice, the feasibility of quantum advantage depends on whether the cost of preparing quantum states scales more favorably than the cost of extracting comparable features or representations in classical pipelines. As such, encoding challenges are not merely technical obstacles but rather define the frontier between classical and quantum learning capabilities.  

\begin{figure}[h]
\center 
\includegraphics[scale=0.33]{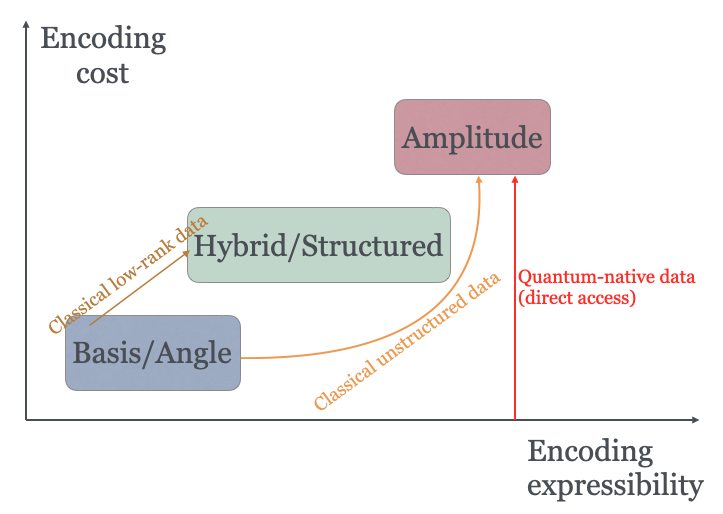}
\caption{Trade-off between encoding cost and expressivity. Classical data increases encoding overhead depending on structure, while quantum-native data bypasses the bottleneck. Figure is schematic and illustrative only, intended to guide intuition. Quantitative validation will follow in Part II.}
\label{fig:encoding}
\end{figure}

Figure~\ref{fig:encoding} illustrates the trade-off between encoding cost and expressivity for different data types and embedding strategies. Simple schemes such as basis or angle encoding occupy the low-cost, low-expressivity regime, making them suitable for low-rank or structured datasets but insufficient for highly complex data. Amplitude encoding, by contrast, provides maximal expressivity but incurs a steep cost, both in state preparation and error sensitivity. Hybrid and structured encodings occupy an intermediate regime, balancing cost and expressivity by exploiting data structure. The arrows indicate how different data types map into this landscape: low-rank classical data can be handled efficiently with simple encodings; unstructured classical data tends to require more expressive (and costly) schemes; while quantum-native data can bypass encoding altogether, avoiding this bottleneck. This highlights encoding as a central challenge in QML, with the feasibility of advantage depending critically on the interplay between data structure, encoding choice, and hardware resources.

\subsubsection{Diversity and dataset difficulty}

The complexity of data does not arise solely from its size or encoding cost, but also from its structural diversity. Homogeneous datasets, characterized by simple correlations or low-dimensional manifolds, often require less expressive models to capture their underlying patterns. In such cases, classical learning methods may achieve competitive performance with limited resources, and the potential advantage of QML is diminished. For instance, linearly separable or low-rank datasets can typically be addressed with classical kernel methods at a cost that scales favorably with data size~\cite{vapnik1995nature,scholkopf2002learning}.

In contrast, diverse datasets exhibit higher manifold dimensionality and richer, often non-linear correlations. This diversity increases dataset difficulty, as the learning algorithm must identify and exploit subtle structures embedded in a high-dimensional space. In classical learning theory, this is captured through notions such as VC-dimension, Rademacher complexity, or covering numbers~\cite{mohri2018foundations}, all of which quantify the richness of a hypothesis class relative to data variability. In the quantum setting, feature maps encoded in Hilbert space offer the possibility of separating data that appears intractable in low-dimensional classical representations~\cite{schuld2019quantum,havlicek2019supervised}. Thus, dataset diversity may shift the balance toward scenarios where QML models achieve nontrivial generalization advantages.

The role of dataset diversity in generalization capacity is particularly relevant for quantum models. While shallow QML architectures may overfit homogeneous datasets due to limited inductive bias, more diverse datasets act as a natural regularizer by requiring the model to capture correlations that are nontrivial to approximate classically. Recent work has shown that quantum kernels can capture data dependencies that are inaccessible to classical kernels without incurring exponential overhead~\cite{huang2021power,schuld2021quantum}. This suggests that the potential for quantum advantage grows not simply with dataset size, but with the intrinsic diversity of the data distribution.

Ultimately, dataset diversity defines a second axis of difficulty—orthogonal to encoding complexity—that determines the feasibility of QML advantage. Homogeneous datasets reduce the necessity of quantum resources, while highly diverse datasets may enhance the expressive gap between classical and quantum hypothesis classes. Identifying and quantifying this diversity, both theoretically and empirically, remains a critical challenge for benchmarking the real-world potential of quantum machine learning.

\subsection{Metric definitions not exercised by this study}
\label{app:v1metrics}

\emph{Also from v1, and removed from Sec.~\ref{sec:defining} for the same reason: these quantities are defined but never estimated here. The quantum interaction complexity and the Schmidt/tensor rank have no estimator in the released code; the quantum Kolmogorov paragraph concerns a quantity that is uncomputable in general; and the classical--quantum analogies are motivational rather than operational. They are kept because they state what the framework intends by ``complexity'' on the quantum side, which the nine measured components only partly cover.}

\subsubsection{High-order dependencies (quantum)}

In the quantum regime, higher-order dependencies are naturally expressed through multipartite entanglement and correlation functions of observables~\cite{horodecki2009quantum, amico2008entanglement}.  
Given a quantum dataset encoded in states $\{\rho_x\}$, correlations among subsystems are captured by reduced density matrices $\rho_{i_1 \dots i_k}$ and their connected correlators:
\begin{equation}
\begin{split}
\small
C_{i_1 \dots i_k} = & \langle O_{i_1} O_{i_2} \cdots O_{i_k} \rangle \\
& - \text{(all lower-order factorizations)},
\end{split}
\end{equation}
where $O_{i}$ are local observables acting on subsystem $i$. This definition parallels classical cumulants: $C_{i_1 \dots i_k}$ vanishes if correlations can be explained by lower-order terms.

The \textit{quantum interaction complexity} is represented as the largest order $k$ such that
\[
\max_{i_1, \dots, i_k} |C_{i_1 i_2 \dots i_k}| > \epsilon,
\]
for some threshold $\epsilon$.  
Datasets requiring high-order entanglement to be faithfully represented exhibit high complexity.  
Tensor network methods exploit such structure by efficiently encoding correlations up to a limited order~\cite{orus2014practical}, while generic quantum circuits can, in principle, represent arbitrarily high-order interactions with polynomial resources~\cite{schuld2021effect}.  
This highlights a potential quantum advantage: classical models struggle with high-$k$ cumulants, whereas quantum systems natively generate and process them.

\subsubsection{Tensor rank / Schmidt rank of data states}

The Schmidt rank measures the number of product states needed to represent a quantum state under a bipartition~\cite{nielsen00}:
\[
|\psi\rangle = \sum_{i=1}^r \alpha_i |a_i\rangle \otimes |b_i\rangle.
\]
Here, $r$ is the Schmidt rank. A higher rank indicates a more complex correlation structure.  
This notion generalizes to tensor rank for multi-partite states and is directly related to the expressivity and resource requirements of variational circuits attempting to model such states~\cite{vidal2003efficient}.

\subsubsection{Quantum Kolmogorov complexity}

Kolmogorov complexity can be applied to the quantum regime where it quantifies the shortest quantum (or hybrid quantum/classical) description necessary to generate a quantum state. In QML, this gives a lower bound for the resources required to represent quantum data or quantum-encoded datasets~\cite{Berthiaume_2001, mueller2007quantumkolmogorovcomplexityquantum, Lemus2024quantumkolmogorov}. Quantum Kolmogorov complexity shares key classical properties — it is uncomputable in general, and, with high probability, most quantum states are incompressible, meaning that a description shorter than exhaustive specification is not possible. Thus, for both data generation and model outputs, only states with low quantum Kolmogorov complexity can be efficiently represented and manipulated\cite{Vitanyi_2001,mora2006quantumkolmogorovcomplexityapplications}.

\subsubsection{Classical--quantum metric analogies}

Classical and quantum metrics exhibit strong analogies that suggest a unified perspective on data complexity. 
For instance, the notion of \emph{intrinsic dimension} in classical datasets, the effective number of degrees of freedom underlying the data distribution~\cite{pope2021intrinsic}, corresponds closely to \emph{entanglement entropy} in quantum states, which quantifies the number of effective quantum correlations~\cite{amico2008entanglement}.  

Similarly, the spectrum of a kernel matrix in classical or quantum machine learning~\cite{schuld2021quantum} plays a role analogous to the \emph{Schmidt rank} in quantum information~\cite{nielsen00}, as both capture how many “directions” of correlation are relevant.  

Finally, \emph{Kolmogorov complexity}, which measures the minimal resources needed to generate a dataset~\cite{kolmogorov1965three}, finds its analogue in the \emph{depth of a quantum circuit}, a measure of the resources required to encode or generate quantum states~\cite{aaronson2004limits,nielsen00}.  

These parallels provide a unified language for comparing dataset difficulty across classical and quantum domains.

\section{Audit and correction of the topological component}
\label{app:topofix}

The pre-specified H2 test (Sec.~\ref{sec:results:h1}) returned a decisive null. Post-hoc inspection of the estimator behind it showed the null was not
interpretable, and this appendix documents what was wrong, what replaced it, and what changes as a result. We report it in full because the defect survived into a confirmatory analysis, and the way it survived is instructive.

\subsection{Four compounding defects}

The frozen component $C_{\text{top}}$ was total Vietoris--Rips persistence, computed with \texttt{gudhi}. Four things were wrong with it at once.

\begin{enumerate}
\item \textbf{$H_0$ entered at full weight.} The deployed call used
  \texttt{max\_dim=2} with default weights $(1,1,1)$, so dimension-0
  persistence was summed into the composite. $H_0$ total persistence is
  $n$ bars each of order an inter-point spacing, a \emph{dispersion}
  statistic. It carried between $50.7\%$ and $99.6\%$ of the composite across
  the suite, median $\approx 90\%$.
\item \textbf{No scale normalisation.} Persistence was accumulated in absolute
  length units, so the component measured how large the point cloud is. Over
  the $27$ classical-track datasets,
  $r(\log \text{diam}, \log C_{\text{top}}) = +0.922$, i.e.\ $R^2 = 0.85$
  (Fig.~\ref{fig:topofix}a).
\item \textbf{The filtration was truncated too early.} The cutoff was
  $0.5\times$ the diameter, but genuine features die later than that, a
  clean circle's $H_1$ class dies at $\sqrt{3}\,r = 0.866 \times$ diameter.
  Real classes therefore never died and were recorded as essential.
\item \textbf{Edge collapse deleted the essential classes.} Edge collapse
  (Boissonnat--Pareek) is persistence-preserving for a \emph{complete} flag
  filtration, not a truncated one. Defects 3 and 4 compound: truncation
  manufactures essential classes and collapse removes exactly those. Directly
  demonstrated on a clean circle, at cutoff $0.5$, $n(H_1) = 0$ with
  collapse and $n(H_1) = 1$ ($P_1 = 0.825$) without; at cutoff $0.9$ the two
  agree exactly.
\end{enumerate}

The consequence is stronger than attenuation. \textbf{The deployed estimator returns $P_1 = 0.000$ for a perfect circle}: it removed topology rather than under-weighting it.

\paragraph{Why this survived validation.} The validation suite exercised $C_{\text{top}}$ with \texttt{max\_dim=1} and weights $(0,1)$, $H_0$ explicitly zeroed, while the feature-extraction path used \texttt{max\_dim=2} with default weights. The validated configuration was never the deployed one. This is the generalizable lesson: validating an
estimator is not the same as validating the estimator \emph{as called}, and only the latter licenses a null result.

\subsection{The corrected estimator}

We rebuilt the component rather than patching it. The backend is \texttt{ripser}, which computes the full untruncated Rips persistence
directly, so defects 3 and 4 cannot recur by construction. $H_0$ is excluded; only dimensions $\ge 1$ count. Every bar is divided by the point cloud's diameter, making the measure scale-invariant. Point clouds are subsampled to $250$ points with a fixed seed.

Because ``how much topology'' has no single canonical summary, we report several and treat the choice as an open specification decision rather than settling it after the fact: $\ell_2$ (sum of squared normalized persistences, threshold-free), the longest single bar, and a count of bars exceeding a
significance cut $\theta$, together with their total persistence.

\subsection{Validation against known shapes}

We validated against point clouds with known Betti numbers (circle, $\beta_1 = 1$; torus, $\beta_1 = 2$, $\beta_2 = 1$; $2$-sphere,
$\beta_2 = 1$) and topologically trivial controls (a $3$D Gaussian blob and a uniform box). A usable summary must separate the two groups.

\begin{table}[t]
\centering
\caption{Known-shape validation. A usable topological summary must place every shape with nontrivial homology above every trivial control.}
\label{tab:topovalid}
\begin{tabular}{lccc}
\toprule
Summary & min(topological) & max(trivial) & separates \\
\midrule
\emph{total} (the pre-specified form) & $0.916$ & $2.602$ & \textbf{no} \\
longest bar                           & $0.318$ & $0.092$ & yes \\
$\ell_2$                              & $0.238$ & $0.107$ & yes \\
count, $\theta = 0.10$                & $1.000$ & $0.000$ & yes \\
thresholded total, $\theta = 0.10$    & $0.792$ & $0.000$ & yes \\
\bottomrule
\end{tabular}
\end{table}

\begin{figure}[t]
\centering
\includegraphics[width=\linewidth]{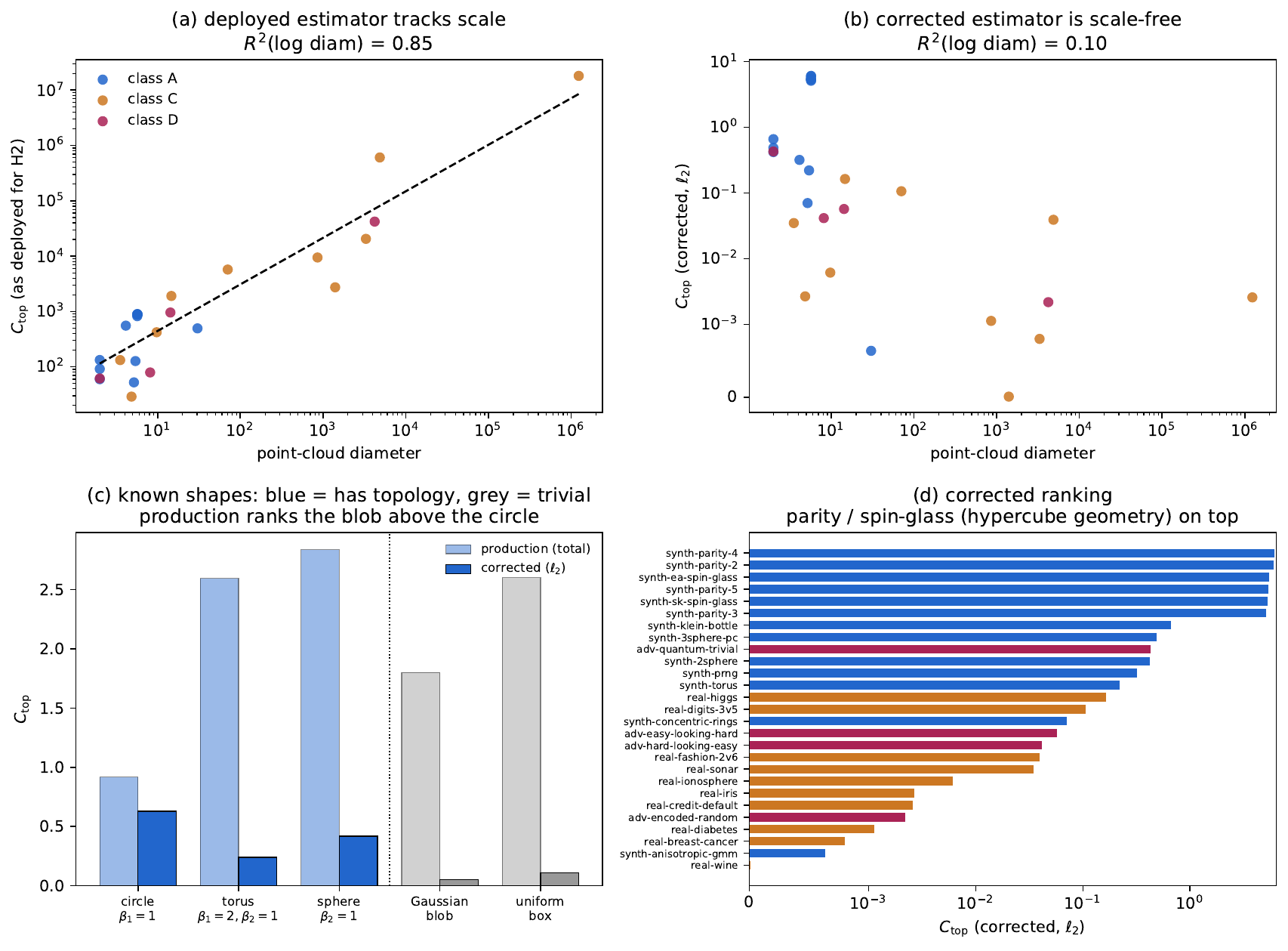}
\caption{Audit of the topological component. (a) The estimator as deployed for H2 tracks point-cloud diameter across five orders of magnitude, $R^2 = 0.85$. (b) The corrected estimator is scale-free, $R^2 = 0.10$. (c) Known-shape validation: blue bars have nontrivial homology, grey are trivial. The pre-specified form ranks the Gaussian blob above the circle; the corrected form does not. (d) The corrected ranking over the $27$ classical-track datasets, coloured by dataset class.}
\label{fig:topofix}
\end{figure}

\textbf{The pre-specified form fails outright} (Table~\ref{tab:topovalid}, Fig.~\ref{fig:topofix}c), and its failure mode explains the downstream damage:
total persistence rewards bar \emph{count}, so a $3$D Gaussian blob's hundreds of tiny sampling-noise cycles outscore a clean circle's single long bar. Every corrected summary passes, with and without Mahalanobis whitening.

The corrected component is invariant where it must be. On a circle with the $\ell_2$ summary it is exact under rescaling ($r = 10^{-2}$ to $10^{3}$: flat at $0.628$) and under ambient-dimension padding ($d = 2$ to $100$: flat), and
stable across sample size ($n = 150$ to $1200$: $0.611$--$0.648$) and subsample seed ($0.601$--$0.654$). Added noise degrades it smoothly and monotonically ($\sigma/r = 0 \to 0.8$: $0.628 \to 0.011$). The scale artifact
is gone: $R^2(\log\text{diam},\cdot)$ falls from $0.85$ to $0.096$
(Fig.~\ref{fig:topofix}b).

Two sanity checks on the suite itself. The corrected ranking is topped by the parity and spin-glass families, binary data on hypercube vertices, which genuinely carries rich Rips topology, together with the designed topological probes; the pre-specified ranking was topped by \texttt{real-credit-default}, which is simply the largest point cloud
(Fig.~\ref{fig:topofix}d). And the four parity datasets share an identical $X$ distribution by construction, differing only in their labels: they score $5.98$, $5.12$, $6.07$ and $5.41$, equal within seed noise, as they must.

\paragraph{A limitation of the correction.} Diameter normalization is a \emph{global} scale, so widely separated clusters inflate the denominator and multiplicity is lost: three circles separated by six radii score \emph{below} one circle. Normalizing by a local scale (median $k$-NN distance) repairs
multiplicity but forfeits sample-size invariance, which matters more when comparing datasets of differing $n$. We took the trade deliberately.

\subsection{Estimation of what changed }

\textbf{H2 becomes inconclusive} (Table~\ref{tab:h2fix}). The threshold-free summaries still show nothing; a significance-thresholded bar count clears the AIC criterion, but only for $\theta \in [0.08, 0.10]$, failing on either side. A binary presence-of-topology indicator gives $\Delta\text{AIC} = +2.8$, also a fail. We decline to select among these after seeing the outcomes, so the reported verdict is that the question is open.

Accordingly, this result should not be read as a universal negative claim about topology-based predictors.  It is a robustness statement limited to the alternative estimators examined here: those alternatives do not recover a stable topology contribution on the present dataset suite.

\textbf{H1 is unaffected.} Pooled LODO $R^2$ stays within $+0.026$ to $+0.031$ across every variant, against a threshold of $0.70$.

\textbf{H5 is unaffected, and provably insensitive to this component.} The obvious objection to a falsification obtained with a broken feature is that a correct feature might have rescued it. We tested that directly rather than arguing it, by varying \emph{only} the topological slot and leaving the other
eight components and the target untouched (Table~\ref{tab:h5ablate}).

\begin{table}[t]
\centering
\caption{Sensitivity of the H5 falsification to the topological component. Only slot $5$ of $\vv$ varies; everything else is held fixed. Pooled leave-one-dataset-out $R^2$ on the budget-matched edge over all $8{,}100$ Phase~B cells, gradient-boosting predictor. The pre-specified pass threshold is $R^2 \ge 0.40$.}
\label{tab:h5ablate}
\begin{tabular}{lcc}
\toprule
Topological slot & Pooled LODO $R^2$ & H5 \\
\midrule
as pre-specified (defective)      & $-1.354$ & fail \\
corrected, $\ell_2$               & $-1.201$ & fail \\
corrected, count $\theta = 0.10$  & $-1.148$ & fail \\
removed entirely ($8$ components) & $-1.292$ & fail \\
replaced by $U(0,1)$ noise        & $-1.254$ & fail \\
permuted across datasets          & $-0.616$ & fail \\
\bottomrule
\end{tabular}
\end{table}

Every treatment fails, and the informative comparison is the spread:
\textbf{the gap between a correctly measured component and one replaced by random noise ($-1.20$ vs $-1.25$) is far smaller than the gap to the decision threshold} ($-1.20$ vs $+0.40$). Deleting the component outright changes little, and the best result of the six comes from \emph{destroying} the signal by permutation, so the slot contributes nothing positive under any treatment. No topological estimator, however well built, could have moved H5 across its threshold.

This is what should be expected once the mechanism behind the H5 failure is identified, because that mechanism sits on the \emph{target} side rather than the feature side: the edge itself is not a stable quantity, collapsing from a $17\%$ to a $1\%$ quantum win rate when the classical baseline is widened from logistic regression to a best-of-six set (Sec.~\ref{sec:results:h5}). No feature vector predicts a target that is largely an artifact of baseline choice, and no repair to $\vv$ addresses that.

The proxy above is a gradient-boosting stand-in for the confirmatory GP pipeline, used because it permits the six-way comparison cheaply. We additionally re-ran the full pre-specified confirmatory protocol, sparse variational GP, leave-one-dataset-out over all 27 datasets with inner $\gamma$ cross-validation and the three-step EM, end-to-end on the corrected component. It returns \textbf{pooled LODO $R^2 = -0.451$} (GBM: $-0.674$), per-fold median $-1.44$, and a fold-bootstrap $95\%$ interval of $[-55.3,\, -8.48]$ whose upper end lies far below the pre-specified threshold of $0.40$. \textbf{H5 fails on the corrected component}, and fails \emph{more} decisively than on the original ($-0.32$): repairing the topological estimator moved the primary result away from its threshold, not toward it.

Two features of that re-run are worth recording. The per-fold \emph{mean} is $-28.8$ against a median of $-1.44$, the same near-constant-edge blow-up discussed for the original run; the pooled figure is the honest headline in both cases. And the folds where the selector does worst are precisely the topology probes: \texttt{synth-klein-bottle} ($-273.6$), \texttt{synth-3sphere-pc} ($-157.1$) and \texttt{synth-2sphere} ($-35.8$), the three datasets on which the corrected component differs most from the
defective one. Whatever the repaired estimator is now measuring, it does not help the selector where topology is by construction the dominant structure.

Software: the defective class is retained unchanged in the released code so the pre-specified numbers stay reproducible; the corrected estimator is a separate class, and the audit is recorded in the project's deviations log.

The audit changes the interpretation of the topology result, but its scope should be stated carefully.  The recomputation is robust to the alternative estimator constructions, normalizations, and summary choices tested in this appendix.  It does not establish robustness to every possible topological estimator or to every representation of the data. Within the tested alternatives, we find no evidence that the topology
term provides a stable contribution to the H2 target on this suite.

\section{Per-dataset complexity vectors}
\label{app:vtables}

Table~\ref{tab:v_raw} gives the raw estimator outputs for the nine implemented components of $\vv(D)$, alongside the corrected topological component $T^{\star}$ of Appendix~\ref{app:topofix}. Table~\ref{tab:p17_values} maps these onto the four v1 pillars and evaluates the v1 composite $C_{\text{data}}$ (Eq.~\eqref{eq:classical_comp}) for each dataset, under both the deployed and the corrected topological estimator. The composite is reported so that the v1 quantities can be reproduced and compared directly; it is \emph{not} used anywhere in the analysis reported here, for the reasons measured in Table~\ref{tab:p17}.

Two further v1 quantities have no entry here, in both cases because they are not properties of a classical dataset. $C_{\text{quant}}(\mathcal{E})$ of Eq.~\eqref{eq:quantum_comp} (Eq.~10 of v1) is defined on a \emph{quantum ensemble} $\mathcal{E} = \{(p_i,\rho_i)\}_{i=1}^N$ on $n$ qubits, combining entanglement entropy, multipartite information, embedded-kernel effective rank, a magic/nonclassicality monotone $N(\rho)$, quantum Fisher information $F=\mathrm{Tr}[\rho L^2]$ and a quantum topological term. It is therefore a quantity of the quantum-native track, which this work runs as a parallel arm rather than as part of the classical-data pathway (Sec.~\ref{sec:positive_control}), and it is not assembled here.

Its six components, with components 5 and 6 respecified as in Eqs.~\eqref{eq:qfi_collective}--\eqref{eq:ctopq_fs}, are evaluated on the ten quantum-native ensembles in Sec.~\ref{sec:framework:cquant}; Table~\ref{tab:cquant} reports the variance decomposition and Table~\ref{tab:cquant_raw} the raw values. The composite is dominated by the effective-rank component, which accounts for $99.97\%$ of its variance. This parallels the degeneracy observed for the classical composite, but here the quantum
components were explicitly evaluated before the aggregation was rejected.

\begin{table}[t]
\centering
\caption{Raw values of the six $C_{\text{quant}}$ components on the quantum-native ensembles ($n=8$, $N=256$). Components 5 and 6 as in Eqs.~\eqref{eq:qfi_collective}--\eqref{eq:ctopq_fs}. \emph{Exploratory.}}
\label{tab:cquant_raw}
\begin{tabular}{lrrrrrr}
\toprule
Ensemble & $\overline S_{\text{ent}}$ & $\mathcal I_{\text{multi}}$ & $\mathrm{rank}_{\text{eff}}$ & $\overline{\mathcal N}$ & $\overline{\mathcal F}$ & $C_{\text{top}}^{(q)}$ \\
\midrule
TFIM ground states        & 0.401 & 3.234 & 1.68   & 1.527 & 0.536 & 0.000 \\
XXZ ground states         & 0.477 & 5.545 & 1.05   & 1.461 & 0.002 & 0.000 \\
Haar-random               & 2.275 & 5.500 & 254.04 & 6.016 & 0.140 & 0.553 \\
Stabilizer with magic     & 0.693 & 5.545 & 2.99   & 1.365 & 0.155 & 1.885 \\
Random MPS                & 0.709 & 3.272 & 252.61 & 4.435 & 0.133 & 0.934 \\
Parity probe $w=1$        & 1.835 & 4.770 & 252.28 & 5.032 & 0.140 & 0.533 \\
Parity probe $w=2$        & 1.835 & 5.468 & 252.28 & 5.032 & 0.149 & 0.533 \\
Parity probe $w=4$        & 1.835 & 5.482 & 252.28 & 5.032 & 0.140 & 0.533 \\
Parity probe $w=6$        & 2.281 & 5.499 & 252.28 & 5.032 & 0.140 & 0.533 \\
Parity probe $w=8$        & 2.281 & 5.515 & 252.28 & 5.032 & 0.141 & 0.533 \\
\bottomrule
\end{tabular}
\end{table}

$C_{\text{ind}}(D;\phi)$ of Eq.~\eqref{eq:induced_comp} (Eq.~11 of v1) has no entry either, and this is deliberate rather than an omission. That quantity is not a
property of a dataset, so there is no per-dataset column it could occupy; it is
replaced by the pair $\big(\vv(D), \qq(D,\phi)\big)$ throughout (Sec.~\ref{sec:framework:align}), whose values are the $\vv$ columns here together with the embedding-interaction components reported per $(D,\phi)$ pair in the results.

\begin{table}[t]
\centering
\caption{Raw estimator outputs for the nine implemented components of $\vv(D)$ (plus the corrected topology column $T^{\star}$), before rank normalization. Reported for reproducibility, and to make concrete the dimensional incoherence motivating Sec.~\ref{sec:framework:predictor}: the columns differ by up to seven orders of magnitude in range and do not share units. $H$ differential entropy (kNN--KL); $d_{\mathrm{pca}}$, $d_{\mathrm{2nn}}$ intrinsic dimension (PCA-effective, TwoNN); $K$ compression ratio (gzip); $T$ topological complexity as pre-specified (Vietoris--Rips, defective --- see Appendix~\ref{app:topofix}); $T^{\star}$ its corrected replacement (untruncated Rips, $H_0$ excluded, diameter-normalised $\ell_2$); $I_2$, $I_h$ second- and higher-order mutual information (kNN, MINE); $R$ kernel effective rank (RBF, median heuristic); $M$ MDL (autoencoder).}
\label{tab:v_raw}
\small
\begin{tabular}{lrrrrrrrrrr}
\toprule
Dataset & $H$ & $d_{\mathrm{pca}}$ & $d_{\mathrm{2nn}}$ & $K$ & $T$ & $T^{\star}$ & $I_2$ & $I_h$ & $R$ & $M$ \\
\midrule
adv-easy-looking-hard & 12.03 & 1.93 & 10.93 & 0.97 & 955 & 0.06 & 0.01 & 0.02 & 4.58 & 80.31 \\
adv-encoded-random & 132 & 1.67 & 5.82 & 0.53 & 4.2e4 & 2.2e-3 & 0.01 & 0.07 & 4.16 & 473 \\
adv-hard-looking-easy & -12.79 & 1.99 & 2.64 & 0.96 & 79.27 & 0.04 & 2.04 & 5.17 & 3.65 & 53.24 \\
adv-quantum-trivial & -11.42 & 3.00 & 2.91 & 0.96 & 61.73 & 0.43 & 0.59 & 3.13 & 3.56 & 54.29 \\
\addlinespace
real-breast-cancer & 101 & 1.04 & 4.91 & 0.53 & 2.1e4 & 8.0e-4 & 0.28 & 1.17 & 3.71 & 429 \\
real-credit-default & 225 & 2.13 & 7.16 & 0.20 & 1.8e7 & 2.6e-3 & 0.29 & 1.10 & 4.96 & 220 \\
real-diabetes & 30.30 & 1.27 & 6.50 & 0.24 & 9456 & 1.1e-3 & 0.07 & 0.34 & 4.59 & 83.69 \\
real-digits-3v5 & 157 & 7.30 & 12.70 & 0.07 & 5728 & 0.11 & 0.04 & 0.26 & 7.21 & 1837 \\
real-fashion-2v6 & 4036 & 5.90 & 20.17 & 0.13 & 6.1e5 & 0.04 & 0.17 & 0.68 & 9.70 & 4.3e4 \\
real-higgs & 31.11 & 15.90 & 18.21 & 0.56 & 1908 & 0.16 & 0.02 & 0.10 & 9.42 & 301 \\
real-ionosphere & -3.32 & 7.78 & 11.80 & 0.43 & 422 & 6.1e-3 & 0.45 & 1.02 & 6.23 & 685 \\
real-iris & 1.68 & 1.43 & 3.70 & 0.22 & 28.99 & 2.7e-3 & 0.36 & 1.27 & 4.18 & 59.98 \\
real-sonar & -41.57 & 5.93 & 9.03 & 0.37 & 132 & 0.03 & 0.07 & 0.47 & 6.77 & 2582 \\
real-wine & 46.25 & 1.00 & 3.10 & 0.31 & 2737 & 1.2e-5 & 0.12 & 0.75 & 2.75 & 269 \\
\addlinespace
synth-2sphere & 0.41 & 2.99 & 2.89 & 0.96 & 59.79 & 0.42 & 0.40 & 0.96 & 3.56 & 21.68 \\
synth-3sphere-pc & 1.62 & 3.99 & 4.37 & 0.96 & 133 & 0.49 & 0.13 & 0.56 & 4.18 & 38.91 \\
synth-anisotropic-gmm & 6.55 & 4.46 & 10.17 & 0.97 & 487 & 6.4e-4 & 0.79 & 2.12 & 4.34 & 66.96 \\
synth-concentric-rings & -13.99 & 2.00 & 2.86 & 0.96 & 52.43 & 0.07 & 1.61 & 4.90 & 3.45 & 54.15 \\
synth-ea-spin-glass & -14.14 & 7.97 & 7.64 & 0.90 & 823 & 5.48 & 8.7e-3 & 4.1e-3 & 5.94 & 79.38 \\
synth-klein-bottle & -0.15 & 3.99 & 2.91 & 0.96 & 91.66 & 0.66 & 0.37 & 0.62 & 4.16 & 39.03 \\
synth-parity-2 & -14.52 & 7.97 & 7.22 & 0.90 & 811 & 5.98 & 0.01 & 3.9e-3 & 5.95 & 80.45 \\
synth-parity-3 & -14.98 & 7.97 & 7.61 & 0.90 & 955 & 5.12 & 6.7e-3 & 2.2e-3 & 5.94 & 80.21 \\
synth-parity-4 & -14.51 & 7.98 & 7.93 & 0.90 & 894 & 6.07 & 0.01 & 2.1e-3 & 5.95 & 80.18 \\
synth-parity-5 & -14.87 & 7.97 & 7.78 & 0.90 & 897 & 5.41 & 8.2e-3 & 3.4e-3 & 5.94 & 79.27 \\
synth-prng & 6.73 & 7.96 & 9.99 & 0.95 & 555 & 0.32 & 7.5e-3 & 0.02 & 6.40 & 80.33 \\
synth-sk-spin-glass & -14.21 & 7.97 & 7.88 & 0.90 & 878 & 5.28 & 5.5e-3 & 3.3e-3 & 5.94 & 79.86 \\
synth-torus & 2.58 & 2.22 & 3.04 & 0.96 & 128 & 0.22 & 0.30 & 0.16 & 3.53 & 21.65 \\
\bottomrule
\end{tabular}
\end{table}

\begin{table}[t]
\centering
\caption{Part~I's four complexity pillars and the composite $C_{\text{data}}$ (Eq.~\eqref{eq:classical_comp}, equal weights) for each classical-track dataset. $C_{\text{top}}$ is the topological component as deployed for the pre-specified tests; $C_{\text{top}}^{\star}$ and $C_{\text{data}}^{\star}$ use the corrected estimator of Appendix~\ref{app:topofix}. Provided so the Part~I quantities can be reproduced and compared; Part~II does not use the composite (Sec.~\ref{sec:framework:predictor}, Table~\ref{tab:p17}). \emph{Exploratory.}}
\label{tab:p17_values}
\small
\begin{tabular}{lrrrrrrr}
\toprule
Dataset & $S$ & $I_{\text{corr}}$ & $K$ & $C_{\text{top}}$ & $C_{\text{top}}^{\star}$ & $C_{\text{data}}$ & $C_{\text{data}}^{\star}$ \\
\midrule
adv-easy-looking-hard & 12.03 & 0.03 & 0.97 & 955 & 0.06 & 968 & 13.08 \\
adv-encoded-random & 132 & 0.09 & 0.53 & 4.2e4 & 2.2e-3 & 4.2e4 & 132 \\
adv-hard-looking-easy & -12.79 & 7.21 & 0.96 & 79.27 & 0.04 & 74.66 & -4.57 \\
adv-quantum-trivial & -11.42 & 3.72 & 0.96 & 61.73 & 0.43 & 55.00 & -6.31 \\
\addlinespace
real-breast-cancer & 101 & 1.45 & 0.53 & 2.1e4 & 8.0e-4 & 2.1e4 & 103 \\
real-credit-default & 225 & 1.40 & 0.20 & 1.8e7 & 2.6e-3 & 1.8e7 & 226 \\
real-diabetes & 30.30 & 0.41 & 0.24 & 9456 & 1.1e-3 & 9487 & 30.94 \\
real-digits-3v5 & 157 & 0.31 & 0.07 & 5728 & 0.11 & 5885 & 157 \\
real-fashion-2v6 & 4036 & 0.85 & 0.13 & 6.1e5 & 0.04 & 6.1e5 & 4037 \\
real-higgs & 31.11 & 0.13 & 0.56 & 1908 & 0.16 & 1940 & 31.96 \\
real-ionosphere & -3.32 & 1.47 & 0.43 & 422 & 6.1e-3 & 420 & -1.42 \\
real-iris & 1.68 & 1.63 & 0.22 & 28.99 & 2.7e-3 & 32.53 & 3.54 \\
real-sonar & -41.57 & 0.54 & 0.37 & 132 & 0.03 & 91.79 & -40.62 \\
real-wine & 46.25 & 0.88 & 0.31 & 2737 & 1.2e-5 & 2784 & 47.44 \\
\addlinespace
synth-2sphere & 0.41 & 1.37 & 0.96 & 59.79 & 0.42 & 62.52 & 3.15 \\
synth-3sphere-pc & 1.62 & 0.69 & 0.96 & 133 & 0.49 & 136 & 3.76 \\
synth-anisotropic-gmm & 6.55 & 2.91 & 0.97 & 487 & 6.4e-4 & 497 & 10.43 \\
synth-concentric-rings & -13.99 & 6.51 & 0.96 & 52.43 & 0.07 & 45.92 & -6.44 \\
synth-ea-spin-glass & -14.14 & 0.01 & 0.90 & 823 & 5.48 & 810 & -7.74 \\
synth-klein-bottle & -0.15 & 0.99 & 0.96 & 91.66 & 0.66 & 93.46 & 2.46 \\
synth-parity-2 & -14.52 & 0.01 & 0.90 & 811 & 5.98 & 797 & -7.62 \\
synth-parity-3 & -14.98 & 9.0e-3 & 0.90 & 955 & 5.12 & 940 & -8.96 \\
synth-parity-4 & -14.51 & 0.01 & 0.90 & 894 & 6.07 & 880 & -7.53 \\
synth-parity-5 & -14.87 & 0.01 & 0.90 & 897 & 5.41 & 883 & -8.55 \\
synth-prng & 6.73 & 0.02 & 0.95 & 555 & 0.32 & 563 & 8.02 \\
synth-sk-spin-glass & -14.21 & 8.8e-3 & 0.90 & 878 & 5.28 & 865 & -8.01 \\
synth-torus & 2.58 & 0.46 & 0.96 & 128 & 0.22 & 132 & 4.23 \\
\bottomrule
\end{tabular}
\end{table}


\section*{Acknowledgements}

This research was supported by Institut transdisciplinaire d’information quantique (INTRIQ), a Strategic Cluster funded by Fonds de recherche du Québec – secteur Nature et technologies. I would like to thank Steven Rayan, Olivier Landon-Cardinal, Yogendran Boniface, Javier Mancilla Montero and Antoine Lemelin for their helpful comments and suggestions. 

\section*{AI statement}

This study used  Claude Opus 5 and 5.5 (Anthropic) as an engineering collaborator, under the author's direction. The AI was used as a tool to write code and improve text smoothness. All hypotheses, decision rules, direction, reviews, analysis and conclusions were human-made.

\bibliography{refs}

\end{document}